\documentclass[10pt]{scrbook} 
\makeatletter
\long\def\@makecaption#1#2{%
  \vskip\abovecaptionskip
  \sbox\@tempboxa{{\caplabelfont #1\captionformat}{\capfont #2}}%
  \ifdim \wd\@tempboxa >\hsize
    {\caplabelfont #1\captionformat}{\capfont #2\par}
  \else
    \global \@minipagefalse
    \hb@xt@\hsize{\hfil\box\@tempboxa\hfil}%
  \fi
  \vskip\belowcaptionskip}
\makeatother
\typearea[12mm]{7} 
\usepackage{latexsym,longtable,amssymb,cite,color}
\usepackage{epsfig}
\usepackage[francais,italian,swedish,german,english]{babel}
\renewcommand{\caplabelfont}{\sffamily\bfseries} 
\renewcommand{\capfont}{\slshape}

\begin{document}
\bibliographystyle{unsrt}
\thispagestyle{empty}
\vspace*{\fill}\vfill
\hspace*{\fill}
Torsten Henning~$\bullet$
~PhD thesis 1999
\vfill
\noindent
\textcolor{red}{
\textbf{\Large cond-mat version note:}\\[1.5ex]
Some  images have been downsampled to comply with
the total file size limitation of the cond-mat e-print archive.
\\[1ex]
A hardcopy  of this thesis can be requested 
by mail from
\begin{quote}
  Prof. Per Delsing\\
  MINA-ASSP, GU/CTH AB\\
  SE-41296 G\"oteborg, Sweden.
\end{quote}
}
\vfill\vfill
\clearpage
\thispagestyle{empty}
\cleardoublepage
\thispagestyle{empty}
\begin{center}
\vspace*{2ex}
{\large Akademisk avhandling f\"or teknisk doktorsexamen}\\
\vspace*{\fill}
\textbf{\textsf{\huge Charging effects\\[0.5ex]
in niobium nanostructures}}\\
\vspace*{5mm}
\textbf{\textsf{\Large ~}}\\
\vspace{8ex}
{\LARGE Torsten Henning}
\vfill
\epsfig{file=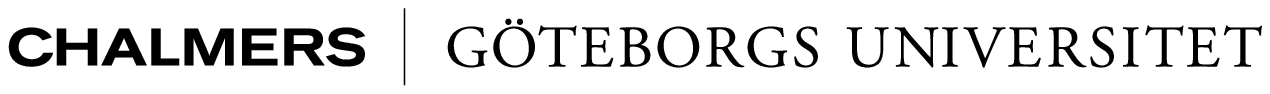,width=0.7\textwidth}
\vfill
Mikroelektronik och Nanovetenskap\\
Chalmers Tekniska H\"ogskola AB och G\"oteborgs Universitet\\
G\"oteborg 1999
\end{center}
\vspace*{12mm} 
\clearpage
\thispagestyle{empty}
\begin{center}
\vspace*{\fill}
ISBN 91-7197-746-5\\
ISSN 0346-718X\\
cond-mat/9901308\\
\vspace*{3ex}
Chalmersbibliotekets reproservice\\
G\"oteborg 1999\\
\end{center}
\vspace*{5ex}
\cleardoublepage

\selectlanguage{english}
\thispagestyle{plain}
\centerline{\textbf{\textsf{Abstract}}}
\vspace*{2ex}
Three types of metallic nanostructures comprising niobium were
investigated experimentally; in all three types, electric transport at
very low temperatures was governed by Coulomb blockade effects.
\begin{enumerate}
\item Thin film strips of niobium could be tuned into resistor strips by
an electrochemical anodisation process, using microfabricated masks and
in situ resistance monitoring. These resistors showed a transition from
superconducting to insulating behaviour with increasing sheet
resistance, occurring at a value approximately equal to the quantum
resistance for Cooper pairs, $h/(4e^2)$.
\item Combining the anodisation technique with lateral size minimisation
by shadow evaporation, devices in a single electron transistor-like
configuration with two weak links and a small island between these were
made. Direct evidence for the Coulomb blockade in the anodisation
thinned niobium films was found when the transport characteristics could
be modulated periodically by sweeping the voltage applied to a gate
electrode placed on top of the structure.
\item `Conventional' single electron transistors with Al base electrodes,
AlO$_x$ barriers formed in situ by oxidation, and Nb top electrodes were
made by angular evaporation. The output current noise of such a
transistor was measured as a function of bias voltage, gate
voltage, and temperature. The low frequency noise was found to be 
dominated by charge input noise. The dependence of the noise on the bias
voltage is consistent with self-heating of the transistor activating the
noise sources.
\end{enumerate}
\clearpage
\selectlanguage{swedish}
\thispagestyle{plain}
\centerline{\textbf{\textsf{Sammanfattning}}}
\vspace*{2ex}
Tre typer av metalliska nanostrukturer inneh\aa{}llande niob har
un\-der\-s\"okts experimentellt. I alla tre typerna
best\"amdes den elektriska
transporten vid mycket l\aa{}ga temperaturer av
Coulomb-blockad-effekter.
\begin{enumerate}
\item Tunnfilmstr\aa{}dar av niob kan f\"orvandlas till
resistortr\aa{}dar genom en elektrokemisk anodiseringsprocess med
mikrotillverkade masker och \"overvakning av resistansv\"ardet in
situ. Dessa resistorer visar en \"overg\aa{}ng fr\aa{}n supraledande
till isolerande beteende  
n\"ar ytresistansen \"okar. \"Overg\aa{}ngen upptr\"ader vid
ett v\"arde i n\"arheten av kvantresistansen f\"or Cooperpar,
$h/(4e^2)$. 
\item Genom att kombinera anodiseringsteknik med lateral
f\"or\-minsk\-ning genom skuggf\"or\aa{}ngning,
tillverkades element med
en\-elek\-tron\-trans\-istor-lik 
geometri med tv\aa{} tunna l\"ankar och en \"o
d\"aremellan. Bevis f\"or Coulomb-blockad i dessa filmer
som uttunnats
genom anodisering,
kunde direkt p\aa{}visas eftersom
transportkarakteristikerna
kunde moduleras periodiskt genom att svepa sp\"anningen som lades
p\aa{} en grindelektrod placerad ovanp\aa{} strukturen.
\item 'Konventionella' enelektrontransistorer med Al-baselektroder,
AlO$_x$-barri\"arer framst\"allda genom oxidering in situ, och
Nb-toppelektroder har tillverkats med
skuggf\"or\aa{}ngning. Utg\aa{}ngsstr\"ombruset av en s\aa{}dan
transistor har uppm\"ats som funktion av biassp\"anning, 
grind\-sp\"an\-ning
och temperatur. L\aa{}gfrekvensbruset var dominerat av
laddningsing\aa{}ngsbrus. Brusets beroende av biassp\"anningen kan
f\"orklaras
med en aktivering av brusk\"allorna genom transistorns
sj\"alv\-upp\-v\"arm\-ning. 
\end{enumerate}
\clearpage

\selectlanguage{german}
\thispagestyle{plain}
\centerline{\textbf{\textsf{Zusammenfassung}}}
\vspace*{2ex}
Drei Typen von 
metallischen Nanostrukturen, die jeweils Niob enthielten, wurden
experimentell untersucht; in allen drei Typen wurde der elektrische
Transport bei sehr tiefen Temperaturen durch Coulomb-Blockade-Effekte
bestimmt. 
\begin{enumerate}
\item D\"unnfilmstreifen aus Niob konnten mit einem elektrochemischen
Anodisierungsproze\ss{} zu Widerstandsstreifen getrimmt werden; dabei
wurden mikrofabrizierte Masken verwendet und der Widerstand in situ
kontrolliert. Diese Widerst\"ande zeigten einen \"Ubergang von
supraleitendem zu isolierendem Verhalten bei zunehmendem
Fl\"achen\-wider\-stand, der bei einem Wert nahe des Quantenwiderstandes
f\"ur Cooperpaare, $h/(4e^2)$, auftrat.
\item Durch Kombination der Anodisierungstechnik mit Mimimierung der
lateralen Abmessungen mittels Schattenverdampfung wurden Elemente in
Einzelelektronentransistor-Geometrie, mit zwei Schwachstellen und einer
Insel dazwischen, hergestellt. Ein direkter Beleg f\"ur die
Coulomb-Blockade in den anodisations-ausged\"unnten Niob-Filmen wurde
gefunden als es gelang, die Transportkennlinien periodisch mit einer an
eine Gatterelektrode, die sich auf der Struktur befand, angelegten
Spannung zu ver\"andern.
\item 'Konventionelle' Einzelelektronentransistoren mit
unteren Elektroden aus Al, durch in situ-Oxidation  erzeugten 
AlO$_x$-Barrieren, und oberen Elektroden aus Nb wurden mittels
Schattenbedampfung hergestellt. Der Ausgangsrauschstrom eines solchen
Transistors wurde als Funktion der Vorspannung, der Gatterspannung und
der Temperatur vermessen. Das niederfrequente Rauschen wurde dominiert
von Eingangsladungsrauschen. Die Abh\"angigkeit des Rauschens von der
Vorspannung ist konsistent mit einer Aktivierung der Rauschquellen durch
die Selbstaufw\"armung des Transistors.
\end{enumerate}
\clearpage
\selectlanguage{francais}
\thispagestyle{plain}
\centerline{\textbf{\textsf{R\'esum\'e}}}
\vspace*{2ex}
Trois types de nanostructures m\'etalliques comprenant du niobium ont
\'et\'e \'etudi\'es. 
Dans toutes les structures, le transport \'electrique aux
tr\`es basses temp\'eratures est gouvern\'e par les effets de blocage de
Coulomb.
\begin{enumerate}
\item Des 
  lignes tr\`es fines 
  de niobium ont \'et\'e obtenues par lithographie
  \'electronique. Par la suite elles ont \'et\'e affin\'ees par un
  proc\'ed\'e d'anodisation 
  \'electro\-chimique en contr\^olant in situ la
  r\'esistance des lignes. Leur comportement pr\'esente une transition
  entre un \'etat supraconducteur et isolant lorsque leur \flqq{}
  r\'esistance carr\'e \frqq{} augmente. Cette transition se produit
  \`a une valeur environ \'egale \`a la r\'esistance quantique par
  paire de Cooper,
  $h/(4e^2)$.
\item La combinaison des techniques d'anodisation et de r\'eduction
  lat\'erale de la ligne par \'evaporation inclin\'ee a permis de
  r\'ealiser des dispositifs comparables aux transistors \`a un seul
  \'electron ayant une petite \^\i{}le reli\'ee \`a l'ext\'erieur
  par deux liaisons faibles.
  La modulation de la tension appliqu\'ee \`a une grille plac\'ee au
  sommet de la structure g\'en\`ere une modulation des
  caract\'eristiques de transport. Cet effet met en \'evidence l'effet
  de blocage de Coulomb dans ces circuits de niobium obtenus par
  anodisation. 
\item Des transistors \flqq{} classiques \frqq{} \`a un seul \'electron,
  avec des
  \'electrodes de base en Al, des barri\`eres 
  en AlO$_x$ form\'ees par
  oxydation in situ, et des \'electrodes sup\'erieures en
  Nb ont \'et\'e
  r\'ealis\'es par \'evaporation inclin\'ee. Le bruit du courant \`a la
  sortie de ces transistors a \'et\'e mesur\'e en fonction de la
  tension de polarisation, de la tension de la porte et de la
  temp\'erature. Le mesures ont montr\'e que le bruit \`a basse
  fr\'equence est domin\'e par le bruit d'entr\'ee de la charge.
  La d\'ependance
  du bruit  de la tension de polarisation est compatible
  avec l'auto-\'echauffement du transistor 
  qui active les sources de
  bruit. 
\end{enumerate}
\clearpage

\selectlanguage{english}
\tableofcontents
\cleardoublepage
\addcontentsline{toc}{chapter}{List of Figures}
\listoffigures
\chapter*{Preface}
\addcontentsline{toc}{chapter}{Preface}
\markboth{}{}

This thesis is the result of work that I have done in the years 1994 to
1998 in the Applied Solid State Physics group at Chalmers and G\"oteborg
University, and at the Swedish Nanometre Laboratory.\par
Experimental results have been published in
the following conference contributions and papers:
\begin{itemize}
\item 
 Torsten Henning, D.~B. Haviland, and P.~Delsing.
 \newblock Transition from supercurrent to {Coulomb} blockade tuned by
  anodization of {Nb} wires.
 \newblock {\em Czech. J. Phys.}, 46(Suppl. S4):2341--2342, 1996.
 \newblock Proc. 21st Int. Conf. on Low Temperature Physics, Prague, August
  8--14, 1996.
\item
 Torsten Henning, D.~B. Haviland, and P.~Delsing.
 \newblock Fabrication of {Coulomb} blockade elements with an electrolytic
  anodization process.
 \newblock {\em Electrochemical Society Meeting Abstracts}, 96-2:561, 1996.
 \newblock Fall Meeting San Antonio, Texas, October 6-11.
\item
 Torsten Henning, D.~B. Haviland, and P.~Delsing.
 \newblock Charging effects and superconductivity in anodised niobium
  nanostructures.
 \newblock In H.~Koch and S.~Knappe, editors, {\em ISEC'97. 6th International
  Superconductive Electronics Conference. Extended Abstracts}, volume~2, pages
  227--229, Berlin, June 1997. Physikalisch-Technische Bundesanstalt.
\item
 Torsten Henning, D.~B. Haviland, and P.~Delsing.
 \newblock Coulomb blockade effects in anodized niobium nanostructures.
 \newblock {\em Supercond. Sci. Technol.}, 10(9):727--732, September 1997.
 \newblock \texttt{cond-mat/9706302}.
\item
 Torsten Henning, B.~Starmark, T.~Claeson, and P.~Delsing.
 \newblock Bias and temperature dependence of the noise in a single electron
  transistor.
 \newblock accepted by Eur. Phys. J. B 1998-10-12, 
 \texttt{cond-mat/9810103}.
\item
 B.~Starmark, Torsten Henning, A.~N. Korotkov, T.~Claeson, and P.~Delsing.
 \newblock Gain dependence of the noise in the single electron transistor.
 \newblock \texttt{cond-mat/9806354}.
\end{itemize}
The \emph{monografiavhandling} style was chosen with the
best intentions of reader friendlyness \cite{granqvist:74:thesis}.\par
There are so many people that I am indebted to after these years, and I
will try to name them here with no particularly deep thoughts about the
order.
First of all (and here order is intentional), I would like 
to thank Tord Claeson
for giving me the chance to work in his 
excellent \cite{tolles:96:nanoreview}
group, and for his continued
interest in the progress. And for paying me for having all this fun, of
course. Another big Thank You to Per Delsing, whose Single Electron
Tunnelling group I joined and who has supported me in every way from day
one, and to David Haviland, who got me on the niobium anodisation
track. I fondly remember the extensive 
and inspiring lunch time
discussions with Per and David
and Yuichi Harada, ChiiDong Chen, Magnus Persson,
Joakim Pettersson and
Peter Wahlgren. From Per Davidsson, I learned a lot about low
temperature physics in general and Old Dil in particular. The work
during the last year has profited very much from the cooperation with
Bj\"orn Starmark, whose amplifier was not only well-designed, but 
absolutely
physicist-proof. My other fellow PhD students contributed in one way or
another, not least by creating a very pleasant atmosphere at the
workplace: Tobias Bergsten, Karin Andersson, Denis Chouvaev, Linda
Olofsson, our \emph{exjobbare} P\aa{}l Dahle,
and all the High-$T_\mathrm{c}$ students that I will not
enumerate more for the fear of forgetting one than out of space
concerns. Edgar H\"urfeld and Oliver Kuhn were welcome additions to our
Little Germany in the SET group, and always good for a discussion or
two. \par
Experimental physics is twice the fun if you are blessed with a technical
staff like ours: Staffan Pehrson can build anything, and Henrik Fredriksen
installed the niobium system and kept it alive at working hours that
would be considered odd even for a student. Alex Bogdanov did the same
thing for the JEOL lithography, and I learned much about nanofabrication
from him. Of the people I met in the nanolab, Julie Gold was
particularly helpful.
Ann-Marie Frykestig tought
me that less bureaucracy is better, something you can only appreciate
fully when you have experienced the opposite. \par
Bengt Nilsson deserves a paragraph of his own.\par
On second thought, make that two. We all know why, and I won't say it
here, because we would not want another laboratory to hire him away from
us, right?\par
During these five years overseas, I have made some very dear new
friends, and kept old ones. I think that these acknowledgements
should be restricted to contributions that are directly related to this
thesis, so sorry, being a friend gets you a place in my heart, but not
on these pages. Andreas Klinkm\"uller might meet the contribution
criterion, though, 
for helping me when he found me clueless at something that was
obviously called a ``tc-shell prompt''. Things have developed a
little, and this thesis is proudly made with Linux.\par
Starting in 1995, I have had much fun during some incredibly active (and
productive) weeks in the nanolab, digging trenches in semiconductor
heterostructures with my best friend, Peter Klar. Where is the
connection with this thesis, one may ask? 
There is appendix E.\par
Finally, let's talk about money again. My first year here was financed
by the German Academic Exchange Service (DAAD), and I am very grateful
to the German taxpayers for risking this investment. The taxpayers of
Europe have subsequently fed me via several programmes, and deserve my
heartfelt thanks as well. \par
\begin{flushright}
G\"oteborg, 24th of January, 1999\\Torsten Henning
\end{flushright}
\chapter{Overview}
This chapter contains, first, a brief orientation on where the subject
of this thesis is located within physics. This may be skipped
immediately by anyone who can interpret the thesis' title without
help. Secondly, there is a summarising overview that casts light on the
motivation for and the connections between the different parts of this
work. 
\section{What is this thesis about?}
It is about \textbf{solid state physics}, predominantly about the
electronic properties of the solid state. This encompasses phenomena
which can be understood in terms of the Fermi Liquid Theory, which tells
us that the behaviour of interacting electrons can be understood in
terms of independent electrons with slightly modified properties, as
well as phenomena that can only be understood as the consequence of
collective effects (e.\,g. superconductivity).
\par
It is about \textbf{electronics}, the science and art of designing solid
state systems in which the laws governing the motion of electrons are
exploited to produce or manipulate useful electrical, usually time
dependent,  quantities that we call signals.
\par
It is about \textbf{three terminal devices}, the bread and butter of
electronics. In a three terminal device, an electrical current 
between two terminals is influenced by the potential applied to, or the
current sent into, the third terminal. In digital electronics, three
terminal devices are used as switches opening or closing a current path.
The classical three terminal devices are the thermionic tube, the
bipolar transistor and the field effect transistor. A lot of novel three
terminal devices are under development
\cite{claeson:96:threeterminaldevices}. 
\par
It is about \textbf{miniaturisation}. Electronics have experienced a
series of booms when new three terminal devices were introduced and
quickly became dominant in most applications. This progress is
intimately linked to the continuing miniaturisation of devices and the
consequent integration and complexity that these devices make possible.
\par
It is about \textbf{single electronics}. In each new generation of three
terminal devices, fewer and fewer electrons 
are needed per switching process. On
the horizon, we now see a limit where only a single electron is needed
to switch a device completely. The device still contains bazillions of
electrons, but the presence or absence of just one electron makes all
the diference! We will have to come back to that.
\par
It is about \textbf{tunnel junctions}. One of the first applications of
quantum mechanics, tunnelling has been understood since the 1920s. 
Basically, tunnelling means that a particle that has been on one side of
a region that it cannot enter (because it does not have enough energy)
can, after a certain time and with a certain probability,
be found on the other side of that forbidden
region (\emph{barrier}).
\par
It is about \textbf{very small capacitors} with very small
capacitances. Tunnel junctions can be made with such small capacitances
that the energy associated with charging them with a single elementary
charge (the \emph{characteristic charging energy})
gives rise to a substantial voltage over the capacitor.
\par
It is about \textbf{the Coulomb blockade}. Said 'substantial voltage'
can be so high that it prevents the tunnelling of further electrons
until the tunnelled charge has been removed to  the environment of
the junction. 
\par
It is about \textbf{very low temperature physics}. To observe the
Coulomb blockade, the thermal energy of the electrons must be
considerably below the characteristic charging energy. We will do the
calculations later, but suffice it to say that contemporary single
electronics usually involves advanced cryogenics, typically with a
dilution refrigerator that can  cool samples  to millikelvin
temperatures continuously. 
\par
It is about \textbf{nanostructures}. Calculations will show that the
critical sample dimensions for the Coulomb blockade to be observable are
in the range of one hundred nanometres, where deep submicron technology
gets dubbed \emph{nanotechnology}.
\par
It is about \textbf{mesoscopic physics}. On this length scale, not all
of the laws of solid state physics, often derived assuming infinite
system size, are valid without modification. On the other hand, the
behaviour of such small systems is still different from that of single
atoms and molecules. The twilight zone between solid state physics
(\emph{macroscopic physics}) and atomic and molecular physics
(\emph{microscopic physics}) is where 
\emph{mesoscopic physics} happens. This
is one of the most interesting areas of solid state physics because
almost everything we know about the solid state in bulk may need to be
reexamined at small length scales.
\par
It is about \textbf{single electron transistors}, the first useful
single electronics device invented. 
Their current-voltage characteristics depend on
the number of electrons on a small island separated from two terminals
by very small tunnel junctions, and are switched between extremal
characteristics  when the charge of the island is varied by half an
elementary charge.
\par
It is about \textbf{electrometry}. With another small capacitor as the
third terminal, the single electron transistor can be used as a very
sensitive measurement instrument for electrical charges. 
\par
It is about \textbf{noise}. Noise, that is unwanted signals following
along with and adding to the useful signals, limits the performance of any
electronic device. In single electronics, noise can limit the accuracy
of the current standard, or the possibility to store bits of information
represented by single electrons. An understanding of the noise is
slowly emerging.
\par
It is about \textbf{superconductivity}. Operating single electronics
devices in the superconducting state is interesting from the physical
basic research point of view, and may even be beneficial for some
applications. Most single electronics materials become
superconducting at the typical operation temperatures.
\par
It is about \textbf{niobium}. Niobium is an interesting material because
it has the highest superconducting transition temperature of all
elements, which, together with its durability,
made it the standard material of established low
temperature electronics (which is practically synonymous to Josephson
electronics). It is hard to handle, though, due to its metallurgical
properties that we will examine in detail later.
\par
It is about \textbf{a superconductor-insulator transition}. Increasing
the disorder in thin superconducting films, an indicator of which is the
resistance, can cause a transition between superconducting and
insulating behaviour at low temperatures. 
\par
And last, but not least, it is about a fair amount of 
\textbf{nanofabrication
techniques}, or ``technology''. There
are no industrial standard techniques for single electronics and no
commercial foundries, so if one wants to examine the properties of a
single electronics device, one has to fabricate it first. Especially
with the introduction of relatively new materials like niobium, figuring
out the right fabrication process may be more time and resource
consuming than the actual measurements on the device.
\par
\section{History of and motivation for this work}
The work documented in this thesis started out with a project aiming at
developing a new method for the fabrication of very small, high ohmic
resistors. Such resistors would have a very small stray capacitance and
should  be of interest for experiments on the effect of the
electromagnetic environment on the behaviour of single electron
transistors or single ultrasmall junctions
\cite{haviland:91:zfpb,kuzmin:91:cpprl}. 
They were expected to play a role in
the attempt to raise the operating frequency of the SET by integrating it
with a HEMT \cite[and references therein]{pettersson:96:prb}. \par
The fabrication method we chose to explore was the controlled anodic
oxidation of nanofabricated niobium wires. We later found that the idea
of raising and tuning the resistance value of thin film resistors by
anodic oxidation was  not exactly brand new
\cite{westernelectric:60:resipat}. \par
The technological challenge could be broken up into three parts:
\begin{enumerate}
\item Definition of deep-submicron niobium wires and electrical
  connections by a suitable process,
\item Definition of a microfabricated anodisation mask 
  laterally confining the  region to be anodised, and
\item Implementation of a resistance monitoring setup 
  and anodisation voltage control procedure.
\end{enumerate}
Since some previous knowledge about niobium nanofabrication by liftoff
processing was available locally
\cite{harada:94:technotes}, and it was obvious that
this technique would be useful for tunnel junction device fabrication
\cite{harada:94:nbset}, we chose to employ liftoff processing already at
this stage (as opposed to an etching process starting from predeposited
niobium). Chapter~\ref{chap:nanofab} will treat the nanofabrication
issues in detail as far as they are relevant for all parts of this
work.\par
The anodisation mask and processing details are the subject of section
\ref{sect:microanodi}. The samples produced were characterised by very
low temperature transport measurements, described in
section~\ref{sect:lowtempsetup}. This section will also be important for
the subsequent chapter~\ref{chap:noise} on noise measurements, since the
same measurement equipment was used for all very low temperature
experiments. \par
While 
characterising the resistor samples, we found that they showed a
Coulomb blockade when the sheet resistances exceeded a few kiloohm (see
\ref{sect:resistorsamples}). We then tried to modulate this blockade
like in the experiments of Chandrasekhar and Webb
\cite{chandrasekhar:91:prl,chandrasekhar:94:jltp}, to demonstrate its
charging effect nature, and in the hope of building some kind of
device. It turned out, however, that the anodised niobium wires were
such a fine grained system that gating them failed to produce a
modulation of their current-voltage characteristics. It was not until we
modified the deposition process into a shadow evaporation process,
producing sub-100\,nm lateral structures, that we were able to show the
electric field effect in an anodised niobium variable thickness wire
system. Section \ref{sect:setlikesamples} elaborates on these processes
and measurements. \par
By this time, the interest in the resistors as such had
declined. Resistors consisting simply of arrays of relatively large
tunnel junctions worked nicely for the purposes of the SET-HEMT
integration project \cite{pettersson:96:prb}. \par
Since the resistors now seemed only useful in combination with niobium
based junctions, it was decided to proceed with fabrication and
characterisation of the latter. Special attention was paid to their low
frequency noise properties, since there were (and still are)
considerable gaps in our understanding of this noise that is severely
limiting the operation of single electronics
devices. Chapter~\ref{chap:noise}, elaborating these investigations, is
outlined as follows. The specific details on the fabrication process, not
covered in the general fabrication background of chapter
\ref{chap:nanofab}, are given in section \ref{sect:noisefabrication}. We
were in the fortunate situation of having a very low noise current
amplifier (see \ref{sect:bjoernsamplifier}), which allowed us not only
to corroborate previous findings about the gain dependence of the low
frequency noise 
\cite{starmark:98:lic,starmark:98:condmat} (see
\ref{sect:gaindependence}), but also gave new insight into its bias and
temperature dependence (\ref{sect:nonidealnoise}).
\chapter{Background topics}
\section{Tunnelling and superconductivity}
  \label{sect:tunnsup}
Effects associated with tunnelling have been observed early in
the twentieth century, and tunnelling has been one of the first
cases where quantum mechanics was 
essential for the understanding of
experimental findings. Superconductivity had already been discovered by
that time, but it had not yet
been recognized as a new thermodynamic state. Both phenomena are closely
linked not only because both can only be explained in quantum mechanical
terms, but also because much of our knowledge about superconductivity
comes from tunnelling experiments.
\subsection{Tunnelling}
Consider a conduction electron with energy $E$ near the
Fermi surface in a region that we call `left electrode', and an adjacent
region called `barrier' where the conduction band edge lies at a higher
energy. On the other side of the barrier we have the `right
electrode'. Even if the Fermi energy in the right electrode is 
different from that in the left electrode, classical physics  would
prohibit the transmission of the electron. Quantum mechanics, however,
predicts a nonvanishing transmission through the barrier as long as it is
finite in height (energetic) and width (in real space). \par
A  voltage applied between left and right electrode results in a
current flow. For small voltages, the
current-voltage characteristics are linear and define a tunnelling
resistance $R_\mathrm{T}$.
Tunnelling processes are classified as elastic, if the energy of the
tunnelling particle is conserved, or inelastic. In the latter case,
dissipation occurs through excitations in the barrier, the electrodes,
or the electrode-barrier interfaces. 
\begin{figure}
\centering
\epsfig{file=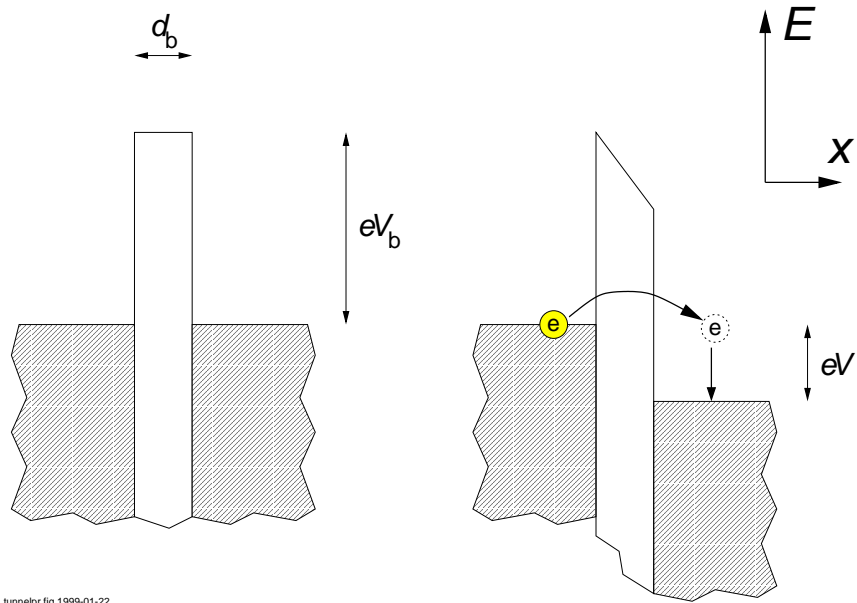,width=\textwidth}
\caption[Tunnelling between two metallic
conductors]%
{Tunnelling between two metallic
conductors separated by a barrier of height $eV_\mathrm{b}$ 
and width
$d_\mathrm{b}$. In the right picture, the junction is biased with a
voltage $V$, favouring tunnelling of electrons from the left
to the right electrode. Dissipation will occur in the right electrode
after the tunnelling, bringing the tunnelled electron closer to the
Fermi level again.}
\label{fig:tunnelpr}
\end{figure}
Figure~\ref{fig:tunnelpr} is a
schematic representation of the tunnelling between two metallic
conductors.
\par
Two examples of systems showing tunnelling are metallic tips producing
field emission, and  structures with a thin insulator separating
two conductors. In the first case, the `left electrode' is the metal tip,
and the barrier is created by the work function, i.\,e. the energy needed
to move an electron from the conductor to infinity. An electric field
tilts the vacuum level, resulting in a triangular shaped barrier that
allows the emission of electrons
into vacuum (`right electrode'). This effect is not only of historical
interest as an early verification of quantum mechanics (observed by
Lilienfeld in 1922  and explained by Fowler and
Nordheim in 1928), it is also of technological
importance as the working principle of advanced electron guns in
electron beam lithography machines. The second example,
conductor-insulator-conductor structures in different varieties, is the
main subject of this thesis.
\subsection{Superconductivity}
Below a critical temperature $T_\mathrm{c}$ and critical magnetic 
field $H_\mathrm{c}$,
certain materials are in a thermodynamic state called the
`superconducting state'. It manifests itself in a number of effects.
The first effect discovered was the vanishing of the electrical
resistance, which gave the phenomenon its name
\cite{royalsociety:35:disc}.
Of at least equal importance is the fact that superconductors 
(type I, the only superconductors known at that time) expel
magnetic fields, the Meissner-Ochsenfeld effect
\cite{meissner:33:nw}. \par 
We will in this report only be dealing with so-called `low temperature'
superconductors. These materials are rather well understood
theoretically, much better than the `high temperature' superconductors
that were discovered just more than a decade ago
\cite{bednorz:87:epl}. A microscopic theory that is
well confirmed experimentally is that of Bardeen, Cooper and Schrieffer
(BCS, \cite{claeson:74:supcon}). Pairs of electrons in time reversed
states (opposite spin and wave vector) interact by exchange of virtual
phonons, leading to an attractive interaction and the formation of
so-called `Cooper pairs'. A collective effect results in the formation
of a ground state that has a lower energy than the 
electron-filled Fermi sphere, and a
gap in the density of states 
opening up around the Fermi energy. The energetic
width of this gap is $2\Delta$, the energy required to break up a Cooper
pair and create an excitation. \par
The density of states
has a singularity at the gap edges.
It was measured by Giaever 
\cite{giaever:61:physrev} in tunnelling experiments 
which gave a direct verification of the BCS theory.\par
Characteristic material parameters are the gap 
(at zero temperature) $\Delta(0)$, the London
penetration depth $\lambda$, the characteristic length over which a
magnetic field drops at the superconductor surface, and the coherence
length $\xi$, over which the Cooper pair density varies. The critical
temperature is related to
the gap at zero temperature, and the gap energy
vanishes at the critical temperature. Niobium is the element with the
highest critical temperature
(at ambient pressure), 
$T_\mathrm{c}\approx 9.2\,$K in bulk, corresponding
to a gap of $2\Delta\approx 3$\,mV. 
The gap is reduced in thin films, considerably below 50\,nm thickness
\cite{lehnert:94:nbgap}. 
The London
penetration depth of niobium
is $\lambda=32$\,nm \cite{auer:73:prb}, 
and the coherence
length $\xi=$39\,nm \cite{auer:73:prb}.\par
\subsection{Superconductive tunnelling}
By \emph{superconductive tunnelling}, we mean tunnelling between two
electrodes, of which at least one is in the superconducting state, via
an insulating barrier. This type of system is usually denoted as an SIS
junction (both electrodes consist of the same superconducting material),
SIN junction (one electrode is normal conducting), SIS$^\prime$
junction (two different superconductors), and so on. Two types of
tunnelling are of particular importance, namely Giaever tunnelling 
\cite{giaever:61:physrev} and
Josephson tunnelling \cite{josephson:64:rmp}.\par
\subsubsection{Giaever tunnelling}
The net quasiparticle (electron) current $I$ between the two electrodes
under consideration is proportional to the densities of states
$\varrho_i$ on both sides, a transition matrix element $M$, and the
difference in occupation of the Fermi distributions. Assuming a constant
$M$, a sharp edge of the Fermi distributions $f_i$ (that is, low
temperature), and assuming that the difference in Fermi levels is equal
to the voltage $V$ between the electrodes, one can probe the densities
of states on both sides. If one of the electrodes is normal conducting,
one can assume $\varrho_\mathrm{N}=\mathrm{const.}$, and the
current-voltage characteristics directly reflect the superconducting
energy gap. The IVC of an NIS junction has a zero current branch between
the voltages $\pm\frac{\Delta}{e}$, and current flow sets in at these
voltages. Farther away from the origin, the IVC approach an ohmic
asymptote.\par
For an SIS$^\prime$ junction, there is a zero current branch between
$\pm\left(\frac{\Delta_1+\Delta_2}{e}\right)$, except for small current
peaks at $\pm\left(\frac{\Delta_2-\Delta_1}{e}\right)$. At
$\pm\left(\frac{\Delta_1+\Delta_2}{e}\right)$, strong  current flow
sets in, again with asymptotically ohmic behaviour. We use this feature
in \ref{subsubsec:dccharsup}, where we consider SIS$^\prime$ junctions
with S$=$Nb and S$^\prime=$Al.\par
\subsubsection{Josephson tunnelling}
While in the processes considered so far, Cooper pairs are first broken
by applying some finite voltage, before their constituting electrons
tunnel, Cooper pairs can also tunnel without breaking. This results in a
current-voltage characteristic with a zero voltage branch, called the
\emph{supercurrent}. Every junction has a characteristic maximum
supercurrent, and once this maximum is exceeded, the IVC jump to a
stable state at finite voltage and current. Josephson tunnelling is only
observed in junctions with sufficiently transparent barriers.\par
The maximally possible supercurrent 
(Josephson pair current) has
the following dependence on the phase difference 
$\gamma=\phi_r-\phi_l$
between the wavefunctions of the condensates in the two electrodes
\cite{tinkham:80:intro}:
\begin{equation}
I_\mathrm{c}=I_\mathrm{c0}\,\sin\gamma.
\end{equation}
The coupling between the superconductors is often expressed in terms of
a coupling energy
\begin{equation}
E_\mathrm{J}=\frac{\displaystyle \hbar}{\displaystyle 2e}\,I_\mathrm{c0}.
\end{equation}\par
In many respects, the described (DC) Josephson effect is a dual analog
to the Coulomb blockade \cite{delsing:90:thesis} that will be discussed
below. By combining two SIS Josephson junctions in parallel, forming a
loop, one can build a device called a DC SQUID, whose critical current is
very sensitively dependent on the magnetic flux penetrating the loop,
with a periodicity of only one flux quantum. The DC SQUID is the basis
for the most advanced magnetometers available today.\par
Another Josephson effect is the generation of high frequency
electromagnetic radiation in a junction. By locking external microwaves
to the radiation in an array of junctions, this AC Josephson effect is
exploited metrologically by linking the representation of the voltage
unit to a frequency standard.\par
\section{Charging effects}
In this section, we will give some background on a group of phenomena
known as charging effects, the most important of which is the Coulomb
blockade. Though some charging effects have been suggested almost half
a century ago \cite{gorter:51:filmres}, it was not after they had been
observed in purposely designed systems that the word \emph{Coulomb
blockade} was coined
\cite{averin:86:osc,grabert:92:sctbook}
(a close predecessor is \emph{Coulomb suppression}, 
\cite{cavicchi:84:coulprl}). 
We will not proceed historically, but roughly
reverse chronologically, starting with the relatively well understood
single and few junction systems and ending with thin films that are
still much more of an open field.
\subsection{Charging effects in very small tunnel junctions}
In the previous section \ref{sect:tunnsup}, we introduced
metal-insulator (oxide)-metal junctions with a tunnel resistance
$R_\mathrm{T}$. This structure is reminiscent of a parallel plate
capacitor with the barrier oxide as dielectric, and in fact,
investigations have shown that the formula for the parallel plate
capacitor's capacitance,
\begin{equation}
  C=\frac{1}{\varepsilon_0\varepsilon}\frac{A}{d},
\end{equation}
where $A$ is the area and $d$ the distance 
between the plates (that is the
thickness of the oxide), holds down to very small dimensions.\par
Whenever an electron tunnels through the junction, the capacitor's
charge changes by the amount of an elementary charge, and accordingly,
its charging energy
\begin{equation}
  E_\mathrm{ch}(Q)=\frac{Q^2}{2C}
\end{equation}
changes by 
\begin{equation}
  \Delta{}E_\mathrm{ch}=
  E_\mathrm{ch}(Q)-E_\mathrm{ch}(Q\pm e)
  \in \left[-E_\mathrm{c},+E_\mathrm{c}\right],
\end{equation}
with the characteristic charging energy
\begin{equation}
  E_\mathrm{c}=\frac{e^2}{2C}.
\end{equation}
Suppose we miniaturise the junction so much that $C\lesssim 1\,$fF, then
the characteristic charging energy will be of the order of 100\,$\mu$eV,
which in turn corresponds to a characteristic temperature
$T_\mathrm{c}=E_\mathrm{c}/k_\mathrm{B}$ of about 1\,K. At very low
temperatures, those attainable with a dilution refrigerator for example,
this characteristic charging energy is non-negligible.\par
Making such junctions, however, was a technological challenge not solved
until well into the 1980's, because the lateral dimensions have to be
rather small. Suppose we have a dielectric permeability
$\varepsilon=10$ and an oxide thickness of 1\,nm, then the area for the
junction to give $C\lesssim 1\,$fF must be less than approximately
100\,nm squared. Nowadays, making such small junctions is a known
technology thanks to the progress in deep submicron processing,
especially in electron beam lithography.\par
If a junction of the kind described above, which we will refer to as a
\emph{very small junction}, is charge neutral initially, tunnelling of an
electron would increase the charging energy by the characteristic
charging energy and is thus prohibited; this is the phenomenon called
\emph{Coulomb blockade}. If we could increase the charge on the electrodes
steadily by moving the electron cloud in the circuit with respect to the
lattice, we would reach a point where tunnelling would suddenly decrease
the charging energy at which an electron would actually
tunnel. Continuing to displace the charge, we would reach the initial
state again half a cycle later, and everything would start all over
again. The charging energy would  cause a time correlation of the
tunnelling events \cite{delsing:89:octprl}
and a voltage oscillation (\emph{SET-oscillation}) 
\cite{averin:86:osc} with
a frequency
\begin{equation}
  \label{eq:fgleichidurche}
  f=\frac{I}{e}.
\end{equation}
Metrologists love an equation like (\ref{eq:fgleichidurche}), because it
relates a quantity that is defined in a cumbersome way and hard to
measure ($I$) to a quantity that is relatively easily defined in the
laboratory ($f$) via nothing but a natural constant.\par
Nature is not that kind, though.\par
Steadily displacing the charge means biasing the junction with a
current. Unlike an electrochemical potential, a current is not a
thermodynamic variable of state. Fixing a potential, i.\,e. voltage
biasing a sample, is much more natural than current biasing.\par
In this special case, the key to understanding lies in the environment
of the junction. The minimalistic circuit involving a very small tunnel
junction contains at least a source of some kind and two leads from the
source's ports to the junction electrodes.\par
Speaking in electrotechnical terms now: even if we had a perfect current
source, the stray capacitance of the leads would still act as a shunt
and introduce a voltage bias component, and since the junction
capacitance we are talking about is so  small, even a few
millimeters of lead are enough to destroy the charging effects.\par
The effect of the electromagnetic environment on very small tunnel
junctions has been an important part of the research ever since the
first such junctions were made 
\cite{delsing:89:sepprl,ingold:91:envzfpb,ingold:92:sct,%
cleland:92:envprb,wahlgren:98:envprb}, and it
is a field still evolving both experimentally and theoretically. 
A proposed type I perpetuum mobile 
\cite{nakashima:98:perpetuummobile} demonstrates the importance of always
considering the environment together with a very small junction.\par
To observe the Coulomb blockade, and SET oscillations, one has to
protect the very small tunnel junctions against the shunting influence
of the environment. This can be done by surrounding it with thin film
resistors \cite{haviland:91:zfpb,kuzmin:91:blochprl,wong:93:gecu}, 
or even easier, with
other junctions, producing a one or two dimensional array. The special
(and simplest) case of the two junction one dimensional array leads us
to the device called the single electron transistor, discussed in
\ref{sect:se-transistor}. 
\subsection{Charging effects in films and arrays}
The other extreme way of surrounding a junction with other junctions is
to include it in a large, possibly irregular, two dimensional
array. Lithographically made arrays of this kind have been investigated
for some time now 
\cite[and references therein]{mooij:92:sct}. Small
superconducting arrays consisting of just a few junctions and holes have
given new insight into the interplay between charges (Cooper pairs) and
vortices 
\cite{geerligs:89:jjaprl,elion:93:aceprl,vanderzant:94:ringpb}. 
Large arrays, however, still pose
considerable lithographical problems, namely in getting them
homogeneous, and there are gaps in our understanding of phenomena such
as the Hall effect 
\cite[and references therein]{delsing:sienaxxx}.\par
\subsubsection{Granular films: discovery of the charging effect}
Already in 1951, Gorter \cite{gorter:51:filmres} suggested that the
observed increase of the resistance in thin films at low temperatures
and low bias might be due to a granular structure of these films and to
charging effects, by virtue of which the charge transfer (by tunnelling)
between the grains is impeded. This paper is generally regarded as
the beginning of single electronics.\par
A decade later, Neugebauer and Weller \cite{neugebauer:62:films} made
transport measurements on ultrathin metal films prepared by
evaporation. These films had a granular structure, as was demonstrated
by TEM imaging. An Arrhenius type dependence of the resistance on
temperature was found, and explained within a model of the film as a
planar array of small islands connected by tunnel junctions and affected
by charging effects.\par
Intentionally created small particles, made with the aim of studying
superconductivity at small dimensions, were used in another classical
experiment by Giaever and  Zeller
\cite{giaever:68:part,zeller:69:tunn}. They made Al-AlO$_x$-Al junctions
($T_\mathrm{c,Al}=1.2\,\mathrm{K}$) and embedded tin particles
($T_\mathrm{c,Sn}=3.7\,\mathrm{K}$ in bulk) in the oxide. The finite size
of the grains caused a finite spacing of energy levels in them, so that
the Fermi levels in the grains and in the electrodes would not line
up. Nonlinear current-voltage characteristics, with a feature we today
call the Coulomb blockade, were observed and could be explained assuming
a distribution of grain sizes. As a byproduct, these measurements
indicated that superconductivity in small Sn particles persisted down to
the smallest particle sizes of 5\,nm.\par
The third experiment that lay the foundation to single electronics was
made by Lambe and Jaklevic \cite{lambe:69:tunn,jaklevic:75:prb}. They
made a sandwich structure consisting of a base electrode, an oxide layer
transparent for tunnelling electrons, a layer of small particles,
another oxide layer, but much thicker than the first one and opaque to
tunnelling electrons, and a top electrode. The top electrode provided a
capacitive coupling to the grains, and in modern terminology, we would
describe their system as a parallel coupling of (inhomogeneous) 
\emph{single
electron boxes} 
\cite{lafarge:91:boxzfpb,lafarge:92:cr,lafarge:93:thesis}. 
Capacitance
measurements showed that with increasing voltage, the small particles
were charged stepwise by tunnelling through the thinner barrier. By
applying an external field, the energy levels in the grains could be
modulated. This field effect device is the ancestor of modern three
terminal charging effect devices like multiple tunnel junction systems
(see \ref{subsec:granular_techniques}) or the 
single electron transistor (see \ref{sect:se-transistor}).\par
\subsubsection{Development after 1975: from theory to devices}
The theory of the charging effect started with a 1975 paper by Kulik and
Shekhter \cite{kulik:75:gran}, giving a quantum mechanical treatment
with a charging Hamiltonian and a tunnelling Hamiltonian. They calculated
the current-voltage characteristics for tunnelling through a small
grain, that is, for a double junction system, and predicted a step
structure in the characteristics in the case of asymmetric
junctions. This effect has been verified experimentally,
e.\,g. by Wilkins et al. \cite{wilkins:89:stmprl} using an STM,
and is known as the 
\emph{Coulomb staircase} \cite{amman:89:chet}.\par
In 1982, Widom et al. \cite{widom:82:duality} pointed out the duality
between the Josephson effect and the \emph{current bias frequency effect}
we know today by the name of \emph{Bloch oscillations}. The theory of these
oscillations was worked out by Likharev and Zorin in 1984/1985
\cite{likharev:lt17,likharev:85:jltp}. 
\par
Oscillations occurring even in junctions with normal conducting
electrodes were predicted by Ben-Jacob et al. \cite{benjacob:85:osc},
and a theory of the \emph{SET oscillations} was presented soon thereafter
by Averin and Likharev \cite{averin:86:osc}.\par
The single electron transistor was introduced as a dual analog to the
DC SQUID by Likharev in 1987 \cite{likharev:87:squidieee}
(the more general name \emph{charging effect transistor} or CHET
\cite{amman:89:chet} never got popular). Kuzmin and
Likharev found single electron charging experimentally in granular
junctions \cite{kuzmin:87:jetp}. With the observation of charging in
lithographically made junctions by Fulton and Dolan
\cite{fulton:87:charprl}, the science of single electron devices took
off in 1987, moving away from granular films to more well-controlled
systems with a few junctions.\par
\subsection{Superconductor-insulator transition in thin films}
\label{subsec:sitransition}
While the entry into the superconducting state below a critical
temperature $T_\mathrm{c}$ is a thermodynamic phase transition, the
superconductor-insulator transition (S-IT) we will consider in this
subsection is
(or `may be interpreted as')
an example of a quantum phase transition (QPT)
\cite{sondhi:97:qptrmp}.
QPT take (in principle) place at zero temperature, and the S-I phase
boundary is crossed by varying a parameter other than the temperature in
the system's Hamiltonian. This can be the charging energy in a
Josephson-junction array \cite{geerligs:89:jjaprl,chen:92:ps} or the
amount of disorder in a metal undergoing a metal-insulator transition
(M-IT). Superconducting thin films have similarities with both
disordered metallic films and arrays of Josephson junctions
\cite{jaeger:89:scprb}. The small grain size generally means that even
charging effects are important in these films.\par
Experimental studies have been performed on quench condensed films, that
are films deposited from the vapour phase onto a very cold surface. This
technique allows to grow amorphous or nanocrystalline
films. Experimental studies on a variety of materials show that the
parameter governing the behaviour seems to be the sheet resistance
of the thin films. White et al. \cite{white:86:quenchprb} have measured
the superconducting 
energy gap with tunnelling experiments and found that the
broadening of the gap edges became comparable to the gap itself, and
hence superconductivity disappeared, when the sheet resistance at high
temperature reached (10\dots 20)\,k$\Omega/\Box$. \par
Jaeger et al.
\cite{jaeger:86:gallprb} found that gallium films became globally
superconducting when the sheet resistance was below about
6\,k$\Omega/\Box$. Experiments
\cite{haviland:89:2dscprl,liu:93:sitprb}
suggest that the threshold for the
superconductor-insulator transition is a sheet resistance of one-fourth
the Klitzing resistance, the so-called quantum resistance for pairs
\begin{equation}
R_\mathrm{Q}=\frac{\displaystyle h}{\displaystyle (2e)^2}=
\frac{\displaystyle R_\mathrm{K}}{\displaystyle 4}\approx
6.45\,\mbox{k}\Omega.
\end{equation}
There is no conclusive
agreement to date \cite{katsumoto:95:gran} on whether this value is a
universal sheet resistance for the S-IT \cite{belitz:94:rmp}
or not \cite{lee:90:mocprl},
and whether thin films actually better be modelled as junction arrays
or disordered metals \cite{lee:85:disorder,altshuler:88:phystoday}.\par
In all these experiments, the sheet resistance depended very sensitively
on the film thickness; in most cases, a difference of one nominal
monolayer can change the sheet resistance over the entire range of the
S-IT \cite{jaeger:89:scprb}. This problem was addressed by Wu
and Adams
\cite{wu:94:sitprl}, who used the same technique we used for the
experiments described in this thesis
(chapter \ref{chap:anodi}): the films
(Al in their case, Nb in our experiments) were deposited with a
certain (relatively high) thickness and then thinned by controlled
anodic oxidation \cite{wolf:77:granarr}.\par
\section{The single electron transistor (SET)}
\label{sect:se-transistor}
The simplest device that bases its operation entirely on the charging effect
is the double junction system known as the 
\emph{single electron [tunnelling]
transistor} or SET. It consists of two very small tunnel junctions,
preferably with approximately equal tunnel resistances $R_1\approx R_2
\gtrsim R_\mathrm{K}$, and a metallic island between them. The island
carries an integer number of (excess) electric charge carriers, and is
furtherfore polarised with a certain displacement charge. In the
simplest case, the island is capacitively coupled to a gate
electrode. The gate voltage $V_\mathrm{g}$ then induces the displacement
charge 
\begin{equation}
  \label{eq:gatecharge}
  Q_\mathrm{g}=V_\mathrm{g} C_\mathrm{g},
\end{equation}
which is a continuous quantity.\par
The operation of the SET is based on the fact that on one hand, the
gate charge $Q_\mathrm{g}$ can be varied continuously, whereas
on the other hand, transport through the island via the tunnel barriers
always involves an integer number of electrons.\par
In addition to the gate charge $Q_\mathrm{g}$, 
there is an offset charge
charge 
$Q_0$ which results from other sources than the gate coupling
capacitively to the island, namely background charges in the vicinity
of the island. 
For the following discussions, we can absorb
the offset charge into the gate charge. The distinction will become
important later, especially when we consider the noise properties of the
SET in chapter~\ref{chap:noise}.\par
\begin{figure}\centering
\epsfig{file=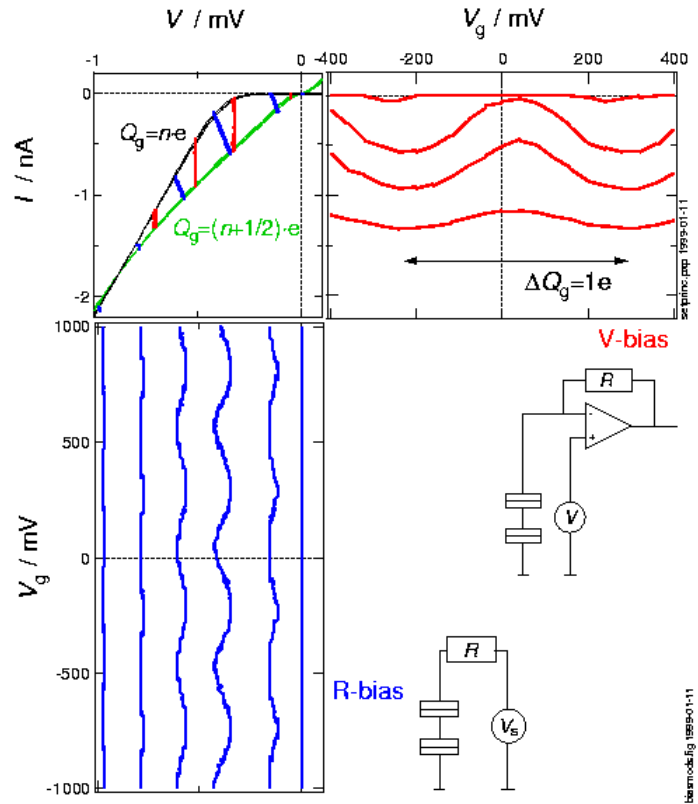,width=0.82\textwidth}
\caption[SET: IVC and modulation 
curves; R-bias and V-bias]{Current-voltage characteristics of
a single electron transistor showing maximum and minimum blockade 
(top left), and modulation curves for two different bias schemes.
In R-bias (left bottom), the SET is biased from a voltage source via a
high ohmic resistor, whose value determines the inclination of the trace
in the IVC (the \emph{load line}) as the gate voltage is varied. In V-bias
(top right), the voltage across the SET is held constant by an
operational amplifier setup (causing a
vertical trace in the IVC), and the current
is modulated with a periodicity in the gate voltage corresponding to a
difference of one elementary charge induced on the gate.}
\label{fig:biasmods}
\end{figure}

If the initial gate charge 
$Q_\mathrm{g}$ is an integer multiple of the
elementary charge $e$, tunnelling of an electron onto the
island increases the electrostatic energy of the system. Therefore, a
threshold voltage
\begin{equation}
  V_\mathrm{th}=\frac{\displaystyle e}{%
    \displaystyle C_\Sigma}
\end{equation}
has to be overcome before the tunnelling process can occur and current
can flow through the transistor. This is the state of maximum Coulomb
blockade, shown in the current-voltage characteristic (IVC) of
fig.~\ref{fig:biasmods} (top left) as the trace labelled
``$Q_\mathrm{g}=n\cdot e$''. \par
The other extremal IVC is obtained when the gate charge is changed by
half an elementary charge. Here,
the initial state and the state after tunnelling of an electron are
energetically equal, and no threshold voltage for
current flow exists. The current-voltage characteristic in this case is
not ohmic, though, as it might appear from the IVC in the limited range
shown in fig.~\ref{fig:biasmods}. Rather, there is an offset voltage
from the ohmic characteristic which, at high bias, becomes independent
of the island charge. We will re-address the topic of the offset voltage
again later. Suffice it here to say that the offset voltage has been the
subject of extensive study in our group 
\cite{wahlgren:94:xjobb,wahlgren:95:lic,%
wahlgren:95:prb,wahlgren:98:envprb}.
\par
Figure~\ref{fig:biasmods} also shows how to exploit the characteristics
of the single electron transistor for purposes of electrometry. In the
simpler case (circuitwise) of resistive bias or \emph{R-bias}, shown in the
lower part of the figure, the double junction system is biased via a
high ohmic resistor from a voltage source. The gate has been omitted
from the sketches. As the gate voltage $V_\mathrm{g}$ and thus the
induced polarisation charge according to eq.~(\ref{eq:gatecharge}) are
varied, the voltage drop across the SET is modulated periodically. One
period corresponds to a difference of the gate charge 
of exactly one elementary charge. Also the current, which can be
measured via the voltage drop over the bias resistor, is modulated
e-periodically. In the current-voltage diagram, the trace of
$(I,V)(V_\mathrm{g})$ for fixed source voltage $V_\mathrm{s}$ falls on a
line whose inclination is given by the bias resistance, the so-called
\emph{load line}. \par
A perfectly vertical load line is attained in the voltage bias or
\emph{V-bias} mode depicted on the right in fig.~\ref{fig:biasmods}. Here,
an operational amplifier circuit keeps the voltage across the double
junction constant, and the current, which can be measured via the
voltage drop over the feedback resistor, is e-periodic with the gate
charge.\par
In either bias mode, electrometry is done by biasing the SET at such a
transport voltage $V$ or
source voltage $V_\mathrm{s}$ and gate voltage $V_\mathrm{g}$ that the gain
\begin{equation}
  \eta=\frac{\displaystyle dI}{\displaystyle dV_\mathrm{g}}
\end{equation}
is maximised to optimise the sensitivity (in chapter~\ref{chap:noise} we
will relativise this statement slightly). Small variations of the gate
charge $\Delta Q_\mathrm{g}$ will then result in measurable variations of the
current $\Delta I$
or voltage $\Delta V$ (in R-bias), 
and the resolution limit of the SET electrometer is
set by its noise properties 
\cite{zimmerli:92:noiseapl,krech:92:sensi,korotkov:94:noiseprb}
(see also the following section). We will
discuss this in more detail in chapter~\ref{chap:noise}.\par
\section{Noise}
The purpose of electronics is the handling and processing of some time
dependent electrical quantity (e.\,g. a current) whose
variation we call
\emph{signal}. Unfortunately, in a real world scenario, not only does the
signal get distorted, but it may also get drowned in a mess of other,
unwanted signals that we call \emph{noise}. In the time domain, noise
manifests itself in fluctuations of the signal quantity. Since most
electronics involve some kind of frequency selection, it is useful and
customary to think about noise in the frequency domain. The noise
spectral density of the current $I$, for example, is the Fourier
transform 
\begin{equation}
  \label{eq:fourier}
  S_I(f)=\int_{-\infty}^{+\infty}dt\,
  R_I(t)\mathrm{e}^{\mathrm{i}\omega t}
\end{equation}
of the current autocorrelation function
\begin{equation}
  \label{eq:autocorr}
  R_I(t)=\lim_{T\to\infty}
  \frac{1}{T} \int_{-T/2}^{+T/2}dt^\prime
  I(t^\prime) I(t^\prime-t),
\end{equation}
that is the properly normalised
convolution of the current with itself.
$S_I(f)$ gives us a measure of the noise power in a certain
bandwidth. The larger the bandwidth in the measurement, the more
noise power will be present.\par 
The noise spectral densities of uncorrelated noise sources add up:
\begin{equation}
  S_{I,\Sigma}(\omega)=\sum_j
  S_{I,j}(\omega).
\end{equation}
Noise sources are uncorrelated when the processes causing the respective
fluctuations are stochastically independent.\par
\subsection{Noise nomenclature}
When considering a device, we can attribute the noise measured at the
output of the device as coming from two groups of sources. First, there
is noise already present at the device's inputs that proliferates
through the device and eventually gets transformed 
in the way the output quantity depends on this input quantity
(e.\,g. amplified). 
The result of this transformation of the noise present at the input of
the system is called \emph{input noise}.
Observe
the unfortunate use of the word \emph{input} here: input noise is the noise
at the output of the device that is the direct result of the
noise present at the input.\par
Second, the device itself will add some noise to the output
quantity. This contribution is called \emph{output noise}, 
another unlucky
word coining. Output noise should better be called 
\emph{internal noise},
but since it cannot be detected before it has reached the output, it
received its name. Since input noise and output noise are uncorrelated
by definition, the measurable noise spectral density at the output is
the arithmetic sum of the spectral densities of input noise and output
noise. \par
If the noise present at the input triggers some process in the device
that adds noise, we absorb this contribution into the input noise by
stating that the transfer function for the noise present at the input
need not be identical to the transfer function for the input quantity.\par
Instead of working with separate numbers for input noise and output
noise, in engineering these two are often combined into yet another
quantity, the \emph{input referred noise}. If the transfer function is
known, e.\,g. as a simple gain $\eta$, one can take the measured 
noise at the output, which is the sum of input noise and output noise,
and divide it by the 
transfer function to calculate the input referred noise. More
general, the input referred noise is the noise at the input of a system
that would cause the observed noise at the output of the system under
the assumption that the system just transforms the input quantity
without adding noise by itself. In chapter~\ref{chap:noise}, we will
encounter such a quantity in our discussion of the noise in the single
electron transistor, where all measured noise is referred to the input
quantity charge.\par
Another real life complication is that we often cannot measure the
output quantity of the device directly, but have to use an
amplifier. This amplifier itself is a device that adds (output) noise,
which we call \emph{amplifier noise}. The noise measured at the output of
the whole system, consisting of the device under test (DUT) and the
amplifier, is then the sum of amplifier noise, amplified input noise of
the DUT, and amplified output noise of the DUT.\par
Noise with a frequency independent spectral density is called \emph{white
noise}. Of course, white noise must be limited by some cutoff frequency
since the energy of the system is finite. Typical white noise
contributions in electrical systems are the thermal noise of the
currents in resistors, and shot noise. Thermal current noise in a
resistor is known as 
\emph{Nyquist}  or \emph{Johnson noise}, its
spectral density is
\begin{equation}
  S_{I,\mathrm{th}}=\frac{4k_\mathrm{B}T}{R},
\end{equation}
where $R$ is the resistance value and $T$ the resistor's
temperature.\par
Since electric current is carried by discrete electrons, statistics of
the number of electrons per unit time lead to an inevitable noise term
known as shot noise \cite{schottky:18:schrot}, with a spectral density
of 
\begin{equation}
  S_{I,\mathrm{e}}=2eI
\end{equation}
in a single junction.
\subsection{Low frequency noise}
\label{subsec:lfnbackground}
The noise spectral density of many systems shows a pronounced increase
at low frequencies, often with a dependence of the form
\begin{equation}
  S_X(f)=S_X(f_0) \left(\frac{f}{f_0}\right)^{-\alpha},
  \quad \alpha\approx 1.
\end{equation} 
This noise is known as \emph{$1/f$ noise}, 
\emph{flicker noise} or \emph{excess
noise}, however, we prefer the more general term \emph{low frequency
noise} (LFN). Among the systems that show this LFN are such distinct
ones as metal films \cite{voss:76:flickerprb},
nerve fibres, the human heartbeat period, 
semiconductor field effect transistors, SQUID
magnetometers, and the SET electrometers described in
\ref{sect:se-transistor}. Since single electrons transistors are in many
respects analogous to SQUIDs
\cite{likharev:87:squidieee,delsing:90:thesis}, this parallel deserves
some more attention in chapter~\ref{chap:noise}.\par
Despite years of effort, the physics of LFN is not completely
understood. A model that seems to reach quite far, at least for
semiconductor structures, is that of charge traps. These charge traps,
which are located in the silicon oxides covering devices, and in gate
oxides, can catch an electron for an average lifetime $\tau$. A single
such trap would then give rise to a current spectral noise density with
a Lorentzian shape \cite{machlup:54:noise}, 
$S_I\propto (1+\beta f^2)^{-1}$, and an ensemble of many such traps with
lifetimes distributed evenly over a certain interval 
$\left[\tau_1,\tau_2\right]$
would lead to an approximate $1/f$ dependence of the resulting spectral
noise density over the frequency interval 
$\left[(2\pi\tau_2)^{-1},(2\pi\tau_1)^{-1}\right]$ 
\cite{klem-artikel}.\par
At very low frequencies, the noise becomes very hard to measure
accurately because of the long integration times required. On the high
frequency side, there appears a \emph{corner frequency} where the LFN
crosses over to a noise floor set by some kind of more or less white
noise, e.\,g. by amplifier noise. Stating the frequency range of the LFN
is, therefore, quite difficult. \par
Low frequency noise is detrimental to all potential applications of
single electron tunnelling. In logic devices, it can cause computing
(bit) errors. Since realistic logic applications seem to lay in the more
distant future, this may not be a top priority concern.
Proposed analog-to-digital converters (ADC) \cite{sareen:exjobb}
certainly work better without LFN and the need for background charge
re-trimming it imposes.\par
In metrological devices, LFN limits the accuracy 
\cite{zimmerman:97:cpem}
by causing random errors. However, even with quite noisy devices
measured by the state of the art, metrologically sufficient accuracy could
be achieved 
\cite{keller:97:cpem,keller:98:pumpprl}.\par
In single electron memory (SETRAM), the primary industrial scale
application envisioned today, LFN will limit the lifetime of a single
bit. \par
With the recent invention of the radio frequency single electron
transistor (RF-SET) \cite{schoelkopf:98:science,wahlgren:98:thesis},
it has become possible to conduct
electrometry at such high frequencies that the contribution from low
frequency noise becomes negligible.
\section{Niobium}
Niobium was discovered by Charles Hatchett in the year 1801
and originally named Columbium. In the
following years, it became confused with Tantalum, discovered 1802,
with which it occurs mostly in nature,
and was finally isolated and rediscovered
in 1844 by Rose and named
Niobium (after Niobe, the daughter of Tantalos). 
Both names were used until element 41 was officially named
Niobium by IUPAC in 1950, but the name Columbium is
to date still used 
occasionally by the American metallurgical community and e.\,g.
the United States Geological Survey. In metallic form, niobium was
isolated for the first time by Bolton in 1905.\par
Early work on niobium anodisation was inspired by potential applications
in electrolytic capacitors
\cite{guentherschulze:37:elko,johansen:57:anox}.
Today, niobium oxides are often studied because they form the surface
of superconducting accelerator cavities, and since acceleration is a
high frequency application, the surface is very important to the cavity
quality. Such cavities used to be made of sheet niobium, but are
nowadays also produced from copper covered with sputter deposited niobium.
\subsection{Properties}
The superconducting critical temperature $T_\mathrm{c}$ of BCS
superconductors is known to depend on the density of states
$N\left(E_\mathrm{F}\right)$ at the Fermi energy. For niobium,
$E_\mathrm{F}$ lies on the flank of a narrow peak in the density of
states \cite{elyashar:77:apw}. This explains the pronounced sensitivity
of niobium's critical temperature to crystal lattice imperfections,
since these tend to smear out the DOS peak, thus lowering
$N\left(E_\mathrm{F}\right)$ and $T_\mathrm{c}$. \par
The influence of gaseous impurities on the superconducting properties of
niobium has been investigated by DeSorbo at General Electric in 1963
\cite{desorbo:63:dissgas}. He found that the addition of oxygen reduced
the $T_\mathrm{c}$ of wire samples to as low as 5.8\,K for an oxygen
content slightly below four atomic percent. Interestingly enough, the
addition of more oxygen raised $T_\mathrm{c}$ again, and finally lead to
the inclusion of oxide phases as the solubility limit was
exceeded. Nitrogen showed the same tendency of first lowering, then
raising $T_\mathrm{c}$, but its effect was much weaker. Note that this
refers to the solution of N in Nb, not to the stoichiometric compound NbN.
\subsection{Niobium and its oxides}
\label{subsec:nboxidechemistry}
There exist three stable oxides, niobium pentoxide Nb$_2$O$_5$, niobium
dioxide NbO$_2$, and niobium monoxide NbO, and  the solution of oxygen
\cite{seybolt:54:solu}
in niobium notated as Nb(O), with up to one weight percent of oxygen at
high temperatures. Niobium pentoxide occurs as NbO$_x$ with $x\in
[2.4\dots 2.5]$, the dioxide and monoxide only in narrower
stoichiometry \cite{gray:75:iss}. Nb$_2$O$_5$ is the principal
constituent of anodic oxide films on niobium
\cite{bakish:60:anox,young:60:anox}. Its density in bulk amorphous form
is $\varrho=4360$\,kg/m$^3$, and its dielectric constant 
$\varepsilon\approx 41$ \cite{young:60:anox}; values for thin films
might deviate from this value, though.
Nb$_2$O$_5$ is an insulator, NbO a superconductor with 
$T_\mathrm{c}\approx 1.4$\,K.\par
The microstructure of an anodic oxide film on niobium is rather
complicated. The outermost layer is Nb$_2$O$_5$, followed by a thin
layer of NbO$_2$ followed in turn by NbO. This sequence was determined
by Gray et al. using ion scattering spectroscopy \cite{gray:75:iss}. It
is noteworthy that they found a more gradual falloff in stoichiometric
oxygen content from Nb$_2$O$_5$ to Nb on anodised foils than in natural
oxide layers. A thin layer of NbO seems to 
always appear between the metal and
the pentoxide, independent of the preparation details
\cite{grundner:80:japplphys}.
Halbritter \cite{halbritter:84:habil,halbritter:87:nbapa}
points out that the interface between niobium
and its oxides is not even but serrated
\cite{halbritter:89:arxpseca}. 
This serration is stronger for `bad'
niobium as measured by the residual resistance ratio (RRR), the ratio of
the resistivities at room temperature and at 4.2\,K
or just above the transition. Niobium deposited by
thermal evaporation is, compared to sputter deposited material, always
worse, but evaporation in conjunction with a liftoff mask offers
more flexible patterning techniques.\par 
The reason for the serration of the interface is the volume expansion from
Nb to Nb$_2$O$_5$ by a factor of about 3 \cite{kroger:81:anodiapl} in
combination with the mechanical properties of the compounds involved. Nb
(density $\varrho = 8570$\,kg/m$^3$ at room temperature)
is relatively soft, and niobium pentoxide microcrystallites cut into the
metal. This serration does not occur on the metals NbN and NbC that are
harder than Nb; carbon inclusion in the interface is known to improve
the quality of Nb based tunnel junctions \cite{kuan:82:nbox}.\par
As a consequence of the complicated interface structure, niobium oxide
on niobium is a bad choice for a tunnel junction material
\cite{aponte:87:barr}. To form good
homogeneous junctions with relatively high transparency, one uses other
insulators to create barriers between niobium electrodes. A good choice
is aluminium oxide \cite{fisher:61:tunnjap}.\par

\chapter{Nanofabrication with niobium}
  \label{chap:nanofab}
\section{Definition of nanosize patterns}
Compared to the processes encountered in modern industrial semiconductor
processing, producing metallic nanostructures is in principle very
simple indeed. There are basically two kinds of process steps,
viz. lithographic steps and pattern transfer steps, which we will
discuss in \ref{subsec:lithotechniques}. After that, we will give an
overview over the lithographic techniques used for the various types of
samples presented in this thesis (\ref{subsec:lithooverview}) and
mention how patterns are produced with the help of computers
(\ref{subsec:lithocad}). \par
\subsection{Lithographic and pattern transfer techniques}
  \label{subsec:lithotechniques}
Lithography is the creation of a physical pattern on some  surface
with the aim of transferring this pattern to another structure that
could not have been patterned directly, i.\,e. without the lithography
in between. The oldest application of lithography is the (mass)
reproduction of drawings by creating the pattern on the surface of a
flat stone (hence the name, ``lithos'' is greek for stone), and then
copying it by inking the stone, and transferring the graphic to
paper. Some time after the invention of silver halogenide photography,
photosensitive organic materials were found and used to reproduce
photographic images for printing. Such an organic material, a
\emph{resist}, is at the heart of any lithographic process. Apart from
visible light, any sufficiently energetic 
radiation source can be used to
modify a suitable resist. The semiconductor industry uses (deep
ultraviolet) light for all mass fabricated circuits, because X-ray
techniques
are too expensive in almost every respect. The masks for this
photolithography, in turn, as well as nanopatterns for research
purposes, are usually produced by electron beam lithography (EBL).\par
Resists come in two basic flavours, called 
\emph{positive} and
\emph{negative}, depending on their behaviour under exposure and
development. \par 
Positive resists consist of some kind of macromolecules that are broken
down by the radiation during exposure. The development is done with a
suitable kind of weak solvent. Provided that the cracked (exposed) and
unexposed resist molecules differ sufficiently in their respective
solubilities, the former will be removed completely from the exposed
regions of the sample. \par
Negative resists, on the other hand, consist of
small molecules with functional groups. The irradiation removes some
atoms from these groups, making them reactive and the molecules
susceptible to chemical bonding, so-called crosslinking. Upon
development, once again in a weak solvent, the unlinked molecules are
removed, and the resist stays where the sample has been exposed.\par
The choice of resist is governed by two considerations, speed and
required profile. Speed is dependent on the area where resist is needed
to be, or be removed, and on the sensitivity of the available resists. It
used to be that positive resists were much less sensitive than negative
resists and thus needed higher doses, resulting in lower resolution
(higher beam diameter) and/or longer exposure times. With the
introduction of ZEP\,7000B, however, one now
has a high resolution positive e-beam resist with a sensitivity
comparable to negative resists. A recipe for photomask making with
ZEP\,7000B can be found in \ref{rec:photomaskpositive}.\par
\begin{figure}\centering
\epsfig{file=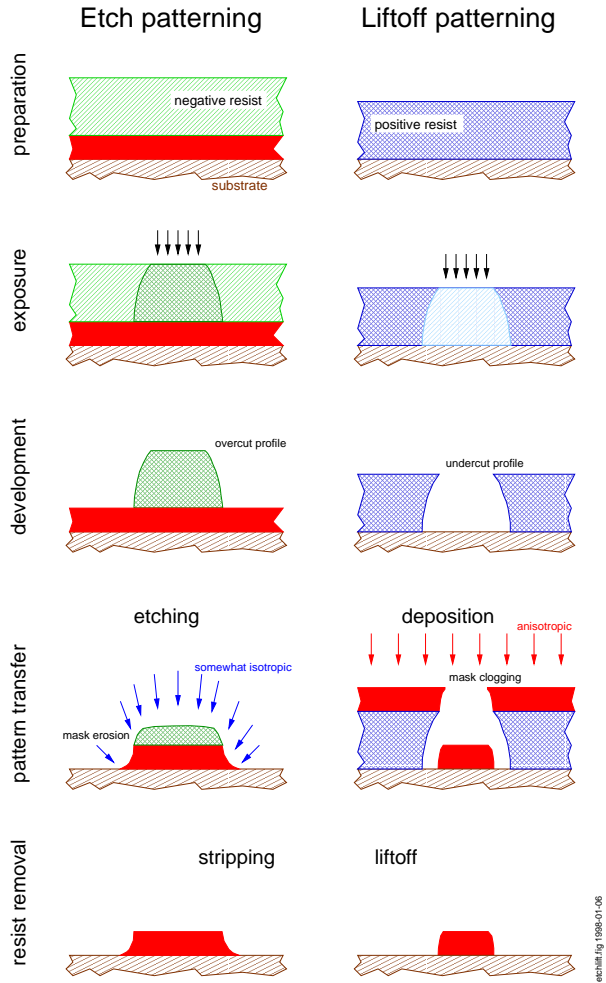,width=0.7\textwidth}
\caption[Etch patterning vs. liftoff patterning]{%
Patterning process types relevant for nanofabrication: etching and
liftoff. Note the widening of the exposure profile due to scattering of
the irradiating particles and the resulting resist profiles, overcut in
negative and undercut in positive resists. Isotropy of etching and mask
erosion lead to a concave, clogging of the liftoff mask on the other
hand to a convex profile of the patterned material in the respective
processes. 
}
\label{fig:etchlift}
\end{figure}

Before discussing the resist profile issue, we should look at the
techniques used for the pattern transfer steps, so that we can
understand the demands. The two fundamental types of pattern transfer
processes are the etching process and the liftoff technique, both
illustrated in fig.~\ref{fig:etchlift}.
Etching processes are dominating for the metallisation in contemporary
industrial lithography, and came long before the liftoff processing was
introduced \cite{hatzakis:67:liftoff}. In a pattern transfer by etching,
the material that is to be patterned is deposited on the substrate
as a first step. 
Since no organic materials are present at this stage, rather
extreme conditions regarding temperature, vacuum etc. can be created,
allowing the optimisation of material properties to a certain
degree. Then, a layer of resist is prepared on top of the material and
patterned by lithography. After development, the material not covered by
resist is removed by some kind of etching process, be it reactive ion
etching (RIE), ion beam etching (IBE), or wet chemical etching
(WCE).\par 
A resist profile with perfectly vertical sidewalls is very, very hard to
attain. Under this premise, for the etching process one will prefer to
err on the side of the bell shape or \emph{overcut profile} indicated in
fig.~\ref{fig:etchlift}, to prevent etchant from sipping under the
resist. Such an overcut profile is almost automatically generated in
negative resist, either by light diffraction behind a photomask in the
case of a photo resist, or, in the case of EBL, by scattering of the
electron beam in the resist (forward scattering) and the materials below
(backscattering). Hence, etching processes normally use negative
resists.\par 
Liftoff processing, on the other hand, does not start with pre-deposited
material, but with resist prepared directly on the substrate. The resist
is then lithographed into a stencil, and the material is deposited on
the substrate through the holes in the stencil. This obviously limits
the ranges of suitable materials, available deposition techniques, and
possible environmental parameters, but on the other hand, allows to
deposit material on substrates that are not suitable for etching. The
biggest advantages of liftoff processes, however, are the resolution and
the ability to make more complex patterns by nonvertical evaporation
techniques. 
Liftoff deposition cannot create structures with aspect ratios as high
as strongly anisotropic etch processes can. High energetic IBE or
chemically and energetically well balanced RIE can be used to make
structures with aspect ratios of 10 in suitable materials, while liftoff
masks clog and thereby limit the amount of material that can be deposited.
After the deposition of the material on and through the stencil, the
resist with the stencil and superfluous material on top is removed
(\emph{lifted off}) in a strong solvent. To allow this liftoff, the metal
film must not cover the resist sidewalls, but tear cleanly at the edges
of the stencil holes. This is achieved by creating an undercut profile
as drawn in fig.~\ref{fig:etchlift}. By virtue of the same diffraction
or scattering mechanisms mentioned above, such an undercut profile is
almost automatically created in positive resists. The effect can be
enhanced by using multilayer resist systems, and the extreme form of
this line of evolution are the resist systems for shadow evaporation
described in \ref{subsec:niemdol}.\par
\subsection{Lithography and processing overview}
  \label{subsec:lithooverview}
\begin{figure}\centering
\epsfig{file=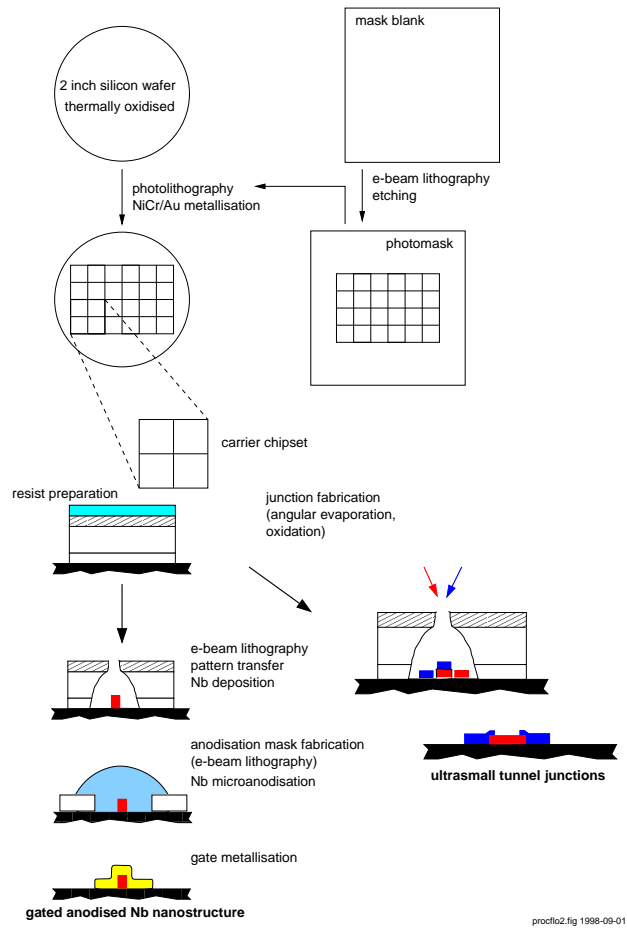,width=0.75\textwidth}
\caption[Nanofabrication process steps]{%
Nanofabrication process steps for anodised niobium nano\-structures
(chapter~\ref{chap:anodi}) and ultrasmall tunnel junctions
(chapter~\ref{chap:noise}).}
\label{fig:procflow}
\end{figure}

For the various parts of work documented in this thesis, a number of
lithographic or lithography related techniques were used and will be
discussed in the remainder of this chapter and in specific sections of
chapters \ref{chap:anodi} and \ref{chap:noise}. To provide an
orientation, fig.~\ref{fig:procflow} gives an illustrative flow
scheme. \par 
Any sample fabrication started with the preparation of the substrate
chips. We used two inch wafers in (100) orientation with a thickness of
about 0.25\,mm that had been thermally oxidised to a depth of
approximately 1\,$\mu$m. Oxidised silicon has the advantages that it
allows the testing of samples at room temperature, and anodisation
without current leakage through the substrate. Its disadvantages are
that it makes the samples more prone to damage from electrostatic
discharges, and possibly, that it makes SET electrometers more noisy
(see chapter~\ref{chap:noise}).\par 
To save exposure time, only an area of $160\times 160\,\mu\mathrm{m}^2$
on each chip was actually patterned with electron beam lithography (this
area corresponds to four \emph{fields} of the system in highest resolution
mode). From this EBL writing field in the centre of the chips, sixteen
gold leads went to contact pads. Electrical contact to the samples
was made by means of spring-loaded probes (\emph{pogo pins}) in a distance
of about 2.5\,mm from the chip centre. This gold lead and pad pattern
was most economically produced by photolithography, the photomask in
turn by electron beam lithography (see recipes 
\ref{rec:photomaskpositive} and \ref{rec:photomasknegative} in the
recipe appendix \ref{chap:recipes}).\par
A documented process for the lead and pad pattern photolithography
\cite[appendix to chapter 4]{chen:94:thesis} had to be abandoned to
comply with environmental regulations restricting the use of toxic and
carcinogenic chemicals. Therefore, a new process, described in
\ref{rec:chiplitho}, was developed. This process not only eliminated the
use of chlorobenzene, but also required fewer chemicals and treatment
steps and less time while producing good resolution. This issue was
especially important to ensure proper detection of fiducials integrated
in the pattern that were later used to align patterns in various layers
of electron beam lithography.\par
After the photolithography, the wafers were scribed by pre-sawing them
from the back. Resist was prepared on the whole wafer, and subsequently,
the wafer was divided into typically six carrier chipsets of two by two
chips, each chip measuring $7\times 7$\,mm$^2$. Processing started with
these sets of four, later continued with individual chips. \par
Figure~\ref{fig:procflow} shows the two different fabrication lines for
the anodised niobium nanostructures described in
chapter~\ref{chap:anodi}, and for the ultrasmall tunnel junction systems
used for the measurements that are the subject of
chapter~\ref{chap:noise}. Both lines started with the preparation and
patterning of four layer resist (see
\ref{subsec:fourlayerresist}). Niobium was then deposited by vertical
evaporation for the anodised wire samples (\ref{sect:resistorsamples}),
or by a special kind of angular evaporation for the anodised samples in
SET-like geometry (\ref{sect:setlikesamples}), followed by
liftoff. On both
kinds of samples for anodisation, a mask was produced that defined the
anodised areas (\ref{subsubsec:anodimask}). The anodisation mask could
later serve as a liftoff mask for the fabrication of a top gate on the
anodised structures, especially on the SET-like samples
(\ref{subsec:gatefailures}). \par
For the fabrication of the more conventional single electron
transistors, that is the noise measurement samples, aluminium and
niobium were used in a shadow evaporation \`a la Niemeyer
\cite{niemeyer:74:mitt} and Dolan \cite{dolan:77:masks} as
described in \ref{subsec:niemdol}.\par
\subsection{Computer aided design of nanostructures}
  \label{subsec:lithocad}
Computer aided design (CAD) is an excellent tool for rapid prototyping,
and in nanostructure research, each and every device is a prototype.\par
Since there are no complex design rules to be observed, CAD of
nanostructure is mainly a drawing and programming aid for the researcher
with whom the responsibility for the design rests. It is not comparable
to the design of complex semiconductor electronics that would be
completely impossible with paper and pencil.\par
\begin{figure}
\centering
\epsfig{file=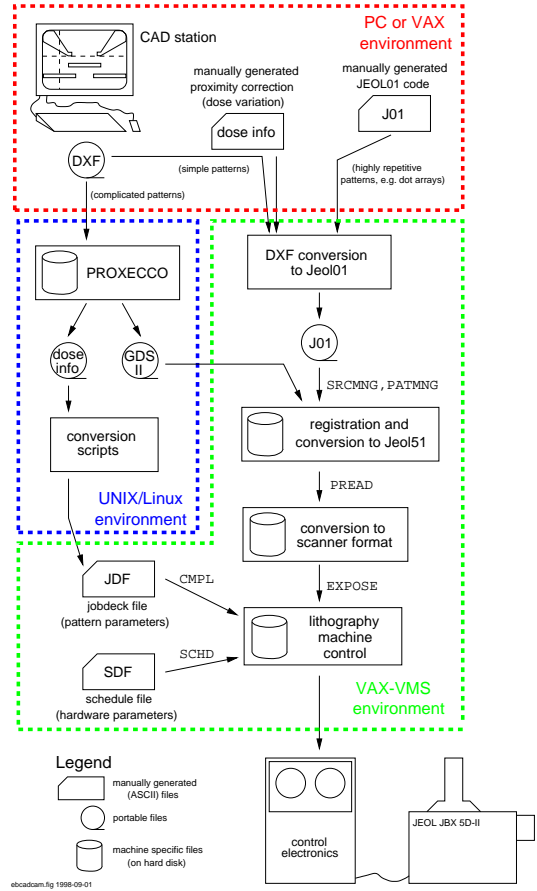,height=0.7\textheight}
\caption[CAD/CAM for nanostructure EBL]{%
CAD/CAM for nanostructure EBL. Patterns were designed on a CAD station
(mini- or microcomputer) and rendered as DXF files. Depending on the
necessity of elaborate proximity correction, they were either processed
with PROXECCO
\cite{eisenmann:93:proxecco}, 
or a shot modulation was entered manually as a variation
of elevation in the drawing. PROXECCO-corrected patterns were delivered
in stream format (GDS\,II), manually corrected patterns in JEOL\,01
format. For simple, highly repetitive patterns, 
JEOL\,01 code was generated manually.}
\label{fig:ebcadcam}
\end{figure}
Figure~\ref{fig:ebcadcam} gives an overview of the pattern data
flow. The nanostructures described in this thesis were designed (drawn)
using AutoCAD version 10 on an old, slow and very reliable VAXSTATION P3100
minicomputer, not least because the DXF output of
this old programme was compatible with all following handling stages.
\par
AutoCAD 10 has a rudimentary support for three dimensions that allows to
assign an \emph{elevation} to every drawing object. This ability was used
to implement a manual proximity correction by trial and error. Features
of the design were drawn with different elevations, increasing with
decreasing feature size. During compilation of the lithography machine
job, an assignment of doses (\emph{shot modulation}) to the elevations
(\emph{shot ranks}) was made, iterating from initial dead reckoning to a
rather stable set of values.\par
This method works for single electronics systems with few junctions,
whose drawings consist mainly of scarce lines and rectangles. It fails
for complex structures like two dimensional arrays of tunnel
junctions. For these patterns (which were made, though they did not
enter this thesis), proximity correction 
\cite{waas:95:cdprox} was done using the commercial
software PROXECCO \cite{eisenmann:93:proxecco}. Its output was, on one
hand, a pattern file in the industry standard GDS~II (Calma stream)
format. On the other hand, PROXECCO generates a dose variation table
that was converted to a JEOL jobdeck file shot modulation table by a
homebrewn script that was written as an exercise in tclsh,
but should be easily portable to other languages.\par
Manually proximity corrected patterns were converted to the JEOL\,01
format by a noncommercial programme and  subsequently underwent a number
of conversions, as did the PROXECCO-generated patterns in GDS II
format.\par
In addition to the pure pattern data, two files completed the
information required by the lithography system. The jobdeck file (JDF)
detailed the placement of various patterns and the specifics of the
fiducials needed for this placement. It also contained the shot
modulation table. The schedule file (SDF) gave hardware parameters like
autoloader levels and origins of the coordinates for the JDF in
absolute (machine) coordinates, and the base dose (resist sensitivity)
that was then multiplied by the factors in the modulation table.

\section{Fabrication methods}
In this section, we will first state what we expect from a fabrication
technique for niobium nanojunctions. After an introduction to the state
of the art in metallic junction and multigranular system
fabrication techniques, we will take a closer look at the
shadow evaporation technique we used in this thesis, and compare it to
its main competitors. The wide field of semiconductor-based single
electronics, including all systems involving gated two-dimensional
electron gases, is intentionally left out, although there are some very
interesting developments concerning memory applications based on
semiconductor technology 
\cite{guo:97:apl,nakajima:97:floatdot}.
\subsection{Criteria for a Nb nanofabrication technique}
Techniques for nanostructure fabrication will, in the academic research
environment, be judged by the quality of the product, while factors like
simpleness and cost effectiveness play hardly any role. In the present
stage of very low scale integration in single electronics, flexibility
of design is also a criterion of minor relevance. This leaves us with
the following criteria for a good nanofabrication technique, in the
special case of single electronics devices based on tunnel junctions:
\begin{enumerate}
\item The junctions should be as small as possible, to minimise their
  capacitances and thus maximise their characteristic charging energy
  and the operating temperature of the device. 
\item In the case of niobium, the quality of the metal should be such
  that a maximal superconducting energy gap $\Delta$ is attained, as
  close as possible to the bulk value of 1.5\,meV.
\item The barrier resistances should be tunable to a value that
  maximises the signal and, consequently, optimises the signal-to-noise
  ratio. For a single electron transistor, this means a target value of
  the order of 25\,k$\Omega$ combined junction resistance, depending
  on the biasing scheme.
\item The junctions should be as homogeneous and equal as possible, both
  in capacitance as in resistance values. The latter criterion is more
  stringent, since the tunnel resistance of an oxide barrier depends
  more strongly (viz. exponentially) on the barrier thickness than
  the capacitance.
\end{enumerate}
No single technique known today fulfills all of these criteria equally
well, so, depending on the intended application, certain trade-offs will
have to be made in choosing a technique.\par
Table \ref{table:strongpoints} compares the three techniques
that can be regarded as so far established that they are presently
competing in the field of fabrication of ultrasmall single junctions
involving niobium:
\begin{enumerate}
\item the Niemeyer-Dolan or shadow evaporation technique, which will be
  described in \ref{subsec:niemdol} since it is the
  technique we chose for the work described here,
\item the various trilayer processes developed at PTB, represented by
 the one that makes use of a planarising spin-on glass
  (SOG), and
\item the self-aligned in-line technique (SAIL).
\end{enumerate}
\begin{table} 
\caption{Criteria for good single electronics
  niobium junctions and strong points 
  ($\blacksquare$) of the competing fabrication
  techniques. ND: Niemeyer-Dolan, SOG: Spin-on glass, SAIL: Self-aligned
  in-line technique. $T_\mathrm{dil}$ means a typical dilution
  refrigerator attainable temperature. References in the text
  (\ref{subsec:competing_techniques}).}
\label{table:strongpoints}
\begin{center}
\begin{tabular}{c||c|c|c|c}
~ & $E_\mathrm{C}\gg k_\mathrm{B}T_\mathrm{dil}$ &
   $\Delta_\mathrm{Nb}\approx 1.5\,\mathrm{meV}$ &
   $R_i\approx R_\mathrm{K}$ & $R_1\approx R_2$ \\ \hline\hline
ND & $\blacksquare$ & ~ & $\blacksquare$ & $\blacksquare$ \\
   \hline 
SOG & ~ & $\blacksquare$ & $\blacksquare$ & $\blacksquare$ \\
   \hline 
SAIL & $\blacksquare$ & $\blacksquare$ & ~ & ? \\ \hline 
\end{tabular}\end{center}
\end{table}
Before comparing these techniques in more detail
in \ref{subsec:competing_techniques}, 
where  references will be given, we will review other
techniques for the fabrication of charging effect devices that were
excluded from this comparison, either because they are not suitable for
niobium or because they cannot fulfill more than two of the criteria
enumerated. Techniques for the fabrication of more or less controlled
multigranular systems will be treated extra in
\ref{subsec:granular_techniques}, and chapter \ref{chap:anodi} deals
with our own special technique in this field.
\subsection{Other fabrication techniques in single electronics}
Over the last decade, numerous techniques for single electron charging
effect devices have been conceived and developed. Many of these aim
primarily at maximising the charging energy in order to achieve a
Coulomb blockade at 4.2\,K, 77\,K, or even room temperature. A few even
allow for operation at these elevated temperatures (as opposed to those
mereley producing a blockade that cannot be modulated). Often, though,
the high charging energy is bought at the expense of a certain degree of
randomness in the structure, and very often a high resistance and
consequently low signal levels.
\paragraph{Direct writing} A thin, lithographically defined metal line
is deposited by a liftoff process. During and after the liftoff process,
the metal is exposed to atmosphere, and a native oxide coating is
formed. Oxide formation may be enhanced by heating or a plasma
treatment. A second thin line crossing the first one is then fabricated
in a second lithography process, requiring a certain alignment precision
in the lithography. This process worked for chromium
\cite{kuzmin:96:crapl}, but is applicable neither to aluminium, which
would oxidise too heavily for reasonable resistance values, nor to
niobium due to the complicated structure of niobium oxides formed in
atmosphere (see \ref{subsec:nboxidechemistry}).
\paragraph{Membrane-with-hole} To study superconductivity in ultrasmall
alu\-mi\-n\-ium particles, Ralph et
al. \cite{ralph:97:alxxx,braun:98:grainxxx} etched a groove into a
silicon nitride membrane, ending in a tiny hole to the other
side. Covering both sides of the membrane with aluminium, using a
discontinuous film on the flat side first and oxidising before
evaporating a continuous film, they were able to obtain structures in
which tunnelling occurred through a single, very small grain.
This technique can be combined with non-vertical evaporation on the
grooved side and vertical evaporation through the hole from the flat
side \cite{aassime:98:holeapl}, creating a single particle with the cross
section of the hole and a length determined by the membrane thickness
and the evaporation angle on the grooved side. This type of structure
has been used to study tunnelling in a very low impedance environment
\cite{aassime:98:holeapl}. 
\paragraph{Nanoparticle assembly} A variety of methods exploit the fact
that it is per se not so difficult to produce fairly homogeneous
nanoparticles or clusters by themselves (especially for
chemists). Then, a pair of electrodes is made by lithography, and one or
more of the prefabricated particles are assembled in the gap between the
electrodes. Electrostatic trapping can be used to place single particles
\cite{bezryadin:97:trapapl} or assemble chains of particles
\cite{bezryadin:98:chainxxx} between the electrodes. Other
schemes use bifunctional linker molecules 
\cite{klein:96:assemblyapl,sato:97:colloapl}
to chemically \emph{self-assemble}
\cite{persson:submitted} metal clusters in the gap. Also, some
SPM techniques (see below) fall into this category.
\paragraph{SPM techniques} The methods involving scanning probe
microscopy can be divided into STM measuring techniques on single grains
in granular systems on one side, and STM/AFM fabrication techniques on
the other side.\par
In STM measurements
\cite{wilkins:89:stmprl,schoenenberger:92:rtepl}, the tip is
brought close to a grain that is separated from the conducting substrate
by an oxide coating. This grain then forms the island of an SET. Of
course, the junctions are quite dissimilar, one consisting of a
dielectric and the other of a vacuum gap. By actually scanning the tip
over the sample, locally resolved Coulomb blockade spectroscopy can be
performed, e.\,g. near the percolation threshold of a granular film
\cite{barsadeh:94:granprb}. \par
SPM-based fabrication techniques come in a large number of
variations. An AFM tip can be used to assemble metallic disks with
atomic precision, forming an SET with 
at least partly vacuum gap tunnel junctions
\cite{junno:98:spmapl,carlsson:98:spmsest}. 
Such devices are of potential
interest for comparing noise characteristics, helping to assess the
barriers' contribution to transistor noise. These devices are limited to
resistance values above 1\,M$\Omega$, though, because of the limited tip
placement accuracy. Other approaches use SPM to create barriers in thin
films, be it by nanoanodisation of thin films with an STM
\cite{matsumoto:96:nanoanodiapl}
or by scratching grooves
(\emph{ploughing}) with an AFM tip 
\cite{snow:95:afmscience,irmer:98:ploughxxx}.
Even an AFM can be used to induce local oxidation 
\cite{irmer:98:afmoxixxx}.
\subsection{Techniques for the fabrication of multigranular systems}
\label{subsec:granular_techniques}
Following the discovery of the charging effect by Giaever and Zeller
\cite{giaever:68:part,zeller:69:tunn},
using an
aluminium oxide matrix with embedded tin particles,
researchers have striven to produce granular systems in which the grain
size is much smaller than typical lithographic resolutions. Ideally, the
grains are so small that they show charging effects at room
temperature. In recent years, the ambition has again been to laterally
nanostructure these systems, to reduce the number of parallel current
paths that contribute to the measured characteristics, as
in the 1987 experiments by Kuzmin et al. \cite{kuzmin:87:jetp}.\par
Since every metal thin film has granular structure, ultrathin films
below the threshold for complete coverage 
(but above the electric transport percolation threshold)
can be suitable systems. The incrementally evaporated quench-condensed
films in which Haviland et al. \cite{haviland:89:2dscprl} observed a
superconductor-insulator transition (see \ref{subsec:sitransition}) are an
example. Recently, Kubatkin et al. \cite{kubatkin:98:rtapl} have
combined quench condensation with lateral miniaturisation by angular
evaporation, and they were able to observe the Coulomb blockade 
at room temperature and its gate
modulation \cite{kubatkin:posteranderwand}.\par
The step edge cut-off
technique, SECO \cite[and references therein]{altmeyer:97:thesis} should
be classified as a granular system fabrication method. In SECO, grooves
with sharp edges, etched into the substrate, are filled with metal by
evaporation under normal incidence almost up to the brim. The idea is
that then the metal in the groove and the film on top of the unetched
substrate will come so close that a tunnel junction is created, with a
barrier consisting partly of the substrate dielectric, partly of
vacuum. It
must be assumed that metal granularity, step rounding and other
imperfections in general lead to the creation of multigranular
systems. Charging effects have been observed in these structures at
77\,K, and gate modulation was possible.\par
Multigranular systems can be combined with SPM methods, as in the
experiments of Nejoh et al. \cite{nejoh:93:roomjap}, who observed
Coulomb blockade at room temperature in a system of metal grains
connected by liquid crystals.\par
Choosing a material other than metal has been the approach of many
groups. One of the classic experiments, and the one we tried to follow
with our niobium anodisation experiments, was the observation of
charging effects in short, highly resistive wires of indium oxide,
In$_2$O$_3$, by Chandrasekhar and Webb 
\cite{chandrasekhar:91:prl,chandrasekhar:94:jltp}.
Indium oxide is a strange material behaving in some respects like a
semiconductor. For example, irradiation with light can change the
conductivity significantly \cite{benchorin:93:hoppprb}.\par
Another superconductor material used in a number of experiments is
niobium nitride, NbN. Coulomb blockade has been observed in nanobridges
made by ion beam etching and using the shadow of a silicon dioxide step
\cite{chen:93:mbieee}, and the gate effect has been demonstrated in NbN
nanobridges at 4.2\,K \cite{miura:95:apl}.\par
Other semiconductor multiple tunnel junction (MTJ) systems have been
made by ionised beam deposition of particles in a gap between
electrodes \cite{chen:95:islandapl}, and by side gating of semiconductor
mesa narrows \cite{tsukagoshi:97:mtjpump} and wires
\cite{smith:97:siwirejap}. Such MTJ have been discussed as a possible
element of single electron tunnelling RAM applications
\cite{nakazato:93:multiadv,nakazato:94:memjap}. \par
Our own niobium nanoanodisation process for MTJ fabrication is described
in detail in chapter~\ref{chap:anodi}.\par
\subsection{The Niemeyer-Dolan technique}
\label{subsec:niemdol}
Getting it small is not even half the battle in nanoelectronics. Making
electrical contacts that allow measurements on a structure is often as
complicated as making the objects themselves. In multilayer structures,
there is the additional problem of properly placing the different
layers, a process known as \emph{alignment}. The angular evaporation
technique, pioneered by Niemeyer \cite{niemeyer:74:mitt} and implemented
in the form described below by Dolan
\cite{dolan:77:masks}, addresses not only the problem of aligning the
two electrodes forming a junction. It also permits forming the barrier
in situ, i.\,e. without breaking  vacuum and moving the sample to
atmosphere (which would produce too thick and opaque barriers).\par
The essential feature in an angular evaporation process is a narrow
structure suspended above and shadowing part of the substrate. In the
original technique of 1973, a thin glass wire was placed over a groove
in the substrate. We will here describe the angular evaporation
technique as it was introduced by Dolan, using masks lithographically
patterned first by photolithography, now regularly by electron beam
lithography. 
\begin{figure}\centering
\epsfig{file=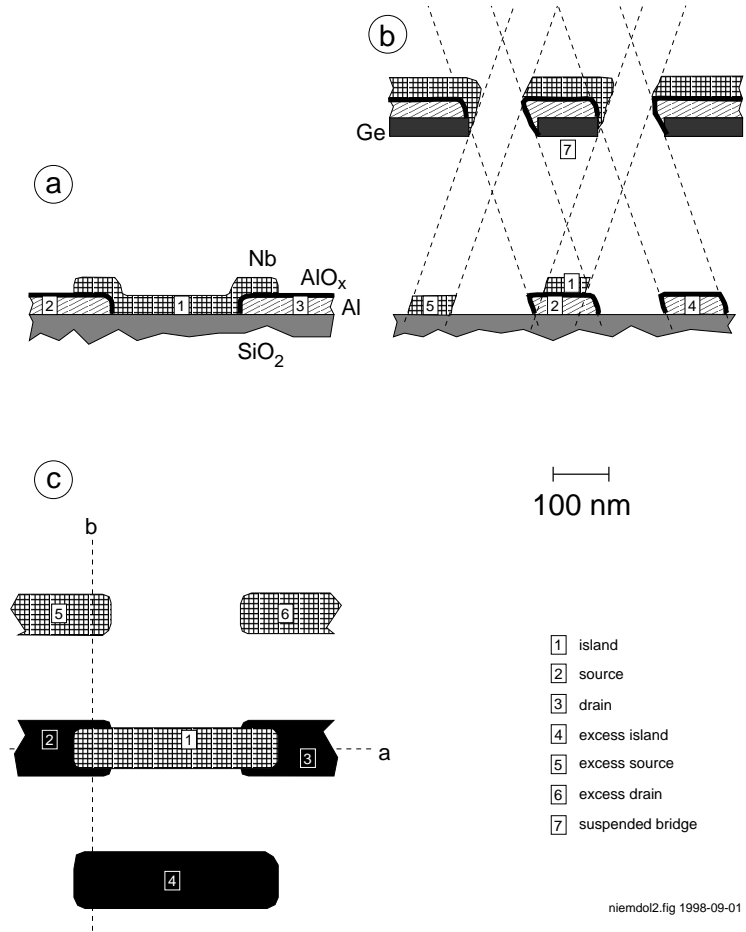,width=0.9\textwidth}
\caption[Niemeyer-Dolan technique 
for the fabrication of SET]{%
Niemeyer-Dolan (angular evaporation) technique for the fabrication of
single electron transistors. \textbf{a}: Side view, cut along the
current path. \textbf{b}: Side view, cut perpendicular to the current
path, and showing the resist mask and the layers deposited on it during
the evaporations. Note how these depositions change the cross section of
the openings in the mask. \textbf{c}: Top view, indicating the cut
planes for views a and b. Drawings are approximately to scale in all
dimensions.
}
\label{fig:niemdol2}
\end{figure}
Figure~\ref{fig:niemdol2} gives three views of a double junction made by
angular evaporation, as we fabricated them for the noise measurements
described in chapter~\ref{chap:noise}.\par
To produce the suspended bridge, a resist system with at least two,
often three layers is used that can generate the extreme undercut
profile required. We will refer the discussion of resists to
\ref{subsec:nd-resists}.\par
The first step after development and processing of the mask is the
deposition of the base electrode material, aluminium in our case shown
in fig.~\ref{fig:niemdol2}. The oxide layer is then created by exposing
the sample to oxygen under low pressure, either inside the evaporation
system chamber or, in more advanced systems, in a separately pumped load
lock. After the oxidation, and once again under UHV conditions, the top
electrode material (here: niobium) is deposited under another tilt angle
between the source direction and the substrate normal. In practice, both
evaporations will be carried out under approximately oppositely equal
tilt angles.\par 
The tilting causes the mask pattern to be shifted, as we see in the
bottom drawing of fig.~\ref{fig:niemdol2}. Designing the pattern
accordingly, one can create overlap regions between the island and the
source and drain electrodes, respectively, where the junctions are
formed. \par
If the barrier material requires a separate evaporation, care must be
taken to avoid any electrical shorts between base and top
electrode. This requires much greater precision in angle control than
the shown process involving just two evaporations, and a fine grained
structure of the barrier material. \par
The deposition of excess material is an unavoidable drawback of the
angular evaporation technique, as is the fact that design flexibility
is limited by defining one fixed direction and amount of pattern shift
for the whole chip. Since single electronics today, however, is still in
the stage of very low scale integration, we felt that adapting the
Niemeyer-Dolan technique according to our needs was preferable over
developing a completely new process.\par
A variation of the shadow evaporation technique that we used for the
definition of very short thin weak links in a niobium wire is described
in \ref{subsec:shadowmin}.
\subsection{Competing Nb nanojunction fabrication techniques}
\label{subsec:competing_techniques}
As described in the last subsection, the Niemeyer-Dolan technique is a
self-aligned technique for the fabrication of planar (overlap)
junctions. Its design flexibility is limited by the definition of one
shift direction for the whole chip, and by the deposition of superfluous
material. Addressing these limitations while pursuing a more ``natural''
way to size reduction is the self-aligned in-line technique (SAIL)
\cite{bluethner:96:sailjdp}.\par
Instead of planar junctions, SAIL creates vertical edge junctions. Since
one of the junction dimensions is a film thickness, which is usually
much smaller than lithographically defined line sizes, SAIL made
junctions tend to have nice large characteristic charging energies.\par
SAIL consists of two lithography steps. A thin line of the island
material is deposited first. An etch mask then defines the island
position. The metal line is removed in the vicinity of the island, and
the barrier layer generated by oxidation or sputtering. The same mask
then serves as a liftoff mask for the deposition of both electrodes.\par
Since both island and source/drain materials can be sputter deposited,
junctions with niobium electrodes can be made very small and still have
a critical temperature and superconducting gap close to the bulk
optimum. Junction capacitances of the order of $10^{-16}$\,F have been
reported \cite{bluethner:97:nbsetieee}. Though there does not seem to be
an inherent limitation, junction resistances have been rather high so
far ($\approx 1$\,M$\Omega$), and the future will have to show how far
this can be reduced.\par
The three layer techniques under development by the collaboration
between PTB Braunschweig and Moscow SU exploit the fact that no in situ
fabricated junction has ever been as good as those made from a
prefabricated sandwich. The electrodes can be made quite thick with this
technique, up to 200\,nm \cite{pavolotsky:98:nbtechxxx}, which helps to
improve the superconducting properties even more. This improvement is
bought at the price of larger junctions than made with SAIL or shadow
evaporation. The latest reported values \cite{zorin:estocolmo} are
$(0.4\times 0.4)\,\mu\mathrm{m}^2$ for the ``cross-strip'' and
$(0.3\times 0.3)\,\mu\mathrm{m}^2$ for the ``etching-anodisation'' and
``SOG'' techniques, respectively. The SOG or spin-on-glass technique
appears to have come farthest in the development of charging device
fabrication techniques that are compatible with the old Josephson
junction processes and facilities known from SQUID technology and the
voltage standard. It uses multiple electron beam lithography and etching
steps. After an etching step that leaves pillars with the cross section
of the junctions, the sample is covered with a planarising resist called
spin-on glass. An island between two junctions is then formed by
covering the whole sample with niobium, which again can be done by
sputtering or evaporation and to sizeable thicknesses, and by etching
the niobium after the island has been defined by EBL with negative
resist. A more detailed description can be found in
\cite{pavolotsky:98:nbtechxxx}. \par
Deviations of the etch profile from perfect verticality seem to produce
effective junction sizes down to $(0.2\times 0.2)\,\mu\mathrm{m}^2$
\cite{zorin:estocolmo}, but at the same time point to a limitation of
the cross sections ultimately attainable with this technique. If one is
satisfied with a charging energy corresponding to a few hundred mK, the
SOG technique fulfills all the criteria for a good niobium nanojunction
technique. \par
An aspect that we have not mentioned so far is the quality of the
AlO$_x$ barrier. The main reason is that we know very little about our
barriers, for example, how homogeneous they are and how much of the
conductance actually is via pinholes or other kinds of defects. One can
only assume that prefabricated sandwich barriers will be at least as
good as in situ produced ones since much larger parameter spaces can be
explored in their making.\par
\section{Shadow evaporation patterning of niobium}
\subsection{Resists for the Niemeyer-Dolan technique}
\label{subsec:nd-resists}
The suspended bridge is usually generated in two or three layer resist
systems patterned by electron beam lithography.
\begin{table}
\begin{center}
\caption{Some resist systems for EBL and shadow evaporation. References
  in the text.}
\label{tab:resistsystems}
\vspace*{1ex}
\begin{tabular}{|c|p{0.5\textwidth}|}\hline
\multicolumn{2}{|c|}{\rule{0in}{3ex}Two layer systems} \\
\rule[-0.5ex]{0in}{2ex}layer sequence & 
  undercut mechanism \\ \hline
\rule[-0.5ex]{0in}{3ex}PMMA / P(MMA-MAA) & 
  sensitivity difference, non- or moderately
  selective developers \\
\rule[-0.5ex]{0in}{2ex}ZEP 520 / PMGI & 
  sensitivity difference, 
  highly selective developers
  \\ \hline \hline
\multicolumn{2}{|c|}{\rule{0in}{3ex}Three layer systems} \\
\rule[-0.5ex]{0in}{2ex} layer sequence & 
  pattern transfer; remarks \\ \hline
\rule[-0.5ex]{0in}{3ex}PMMA / Ge / P(MMA-MAA) & 
  development of e-beam sensitive bottom layer
resist \\
\rule[-0.5ex]{0in}{2ex}PMMA / Ge / S1813,PMMA & 
  reactive ion etching; called ``four layer
resist'' here \\
\rule[-0.5ex]{0in}{2ex}PMMA(?) / Si$_3$N$_4$ / SiO$_2$ & 
   wet chemical etching with HF; no liftoff
   \\
PMMA / Si or Ge / PES & solvent  \\ \hline
\end{tabular}
\end{center}
\end{table}
Table \ref{tab:resistsystems} gives an overview of some resist systems
that can be used to produce nanosize patterns with the shadow
evaporation technique:
\begin{itemize}
\item Two layer resists consist of an e-beam sensitive top layer and
  usually a more sensitive bottom layer. The difference in sensitivity,
  ideally in combination with selective developers, leads to the removal
  of bottom layer material at larger distances from the exposing beam
  than in the top layer and to the creation of the undercut profile with
  the Dolan bridge.
\item Three layer resists have an e-beam patterned top layer, a mask or
  stencil layer (usually inorganic), and a supporting bottom layer. The
  pattern is transferred from the top layer to the mask by some kind of
  etching process (often RIE or IBE, sometimes WCE). The four layer
  resist we used (\ref{subsec:fourlayerresist}) 
  is really a special form of three layer resist.
\end{itemize}
\subsubsection{Conventional multilayer resist systems}
The combination of PMMA and CP, copolymer P(MMA-MAA),
has been the working
horse of (aluminium based) single electronics for the last dozen
years. Exploiting the sensitivity difference between both resists, that
can be enhanced by preexposing the CP prior to PMMA coating, its
disadvantage is that there are no highly selective developers for the
different layers available. This narrows the parameter ranges for doses,
developer concentrations, and development time(s), and can make the
fabrication of high density patterns  
\cite{klar:96:zmtdot,griffin:98:zmtjcg,klar:98:zmsqdprb}
quite arduous. A recently introduced combination of a ZEP\,520 top layer
and a polyimide bottom layer \cite{harada:unpublishedresist} 
allows for the use
of highly selective developers (xylene and Shipley MF322,
respectively), gives good resolution, and can even be baked to higher
temperatures than PMMA and CP.
\subsubsection{Resist considerations for Nb evaporation}
The problems in the deposition of niobium by evaporation are twofold:
\begin{enumerate}
\item The intense heat in the evaporation system tends to melt organic
  resists, resulting in distorted patterns or seriously sagging or
  damaged Dolan bridges. PMMA for example starts flowing at about
  110$^\circ$C, and the substrate temperature should not exceed
  100$^\circ$C \cite{hatzakis:88:fabr}.
\item The deposited niobium generally shows a suppression of the
  critical temperature and residual resistance ratio. Inclusion of
  gaseous impurities emanating from the resist is generally held
  responsible for this problem, which is affectionately known as the
  ``stinking resist'' problem.
\end{enumerate}
The first complex has been addressed by the introduction of three layer
resist systems with sturdy inorganic (metallic) stencils supported by a
resist layer. Table~\ref{tab:resistsystems} shows a selection of
these.\par
Inserting a germanium layer between the PMMA top and P(MMA-MAA) bottom
layer of conventional two layer resist was an early approach to creating
a three layer mask for liftoff nanopatterning. The pattern can in
principle be transferred to the bottom layer by development, though
reactive ion etching seems to be more common. This system is still not
completely satisfactory for niobium deposition, where substrate
temperatures can reach 150$^\circ$C \cite{wang:87:middlesys}. The
stability is improved by depositing the germanium on a layer of hard
baked photo resist, which brings us to the four layer resist 
\cite{harada:94:nbset} described
in \ref{subsec:fourlayerresist}.\par
At present, the focus of interest is shifting from the first problem,
that is getting niobium deposited in nanopatterns at all, to the second,
improving the quality. Two parallel trends can be noted. One trend is to
find organic resists that can be baked to higher temperatures, to outgas
as much as possible before entering a deposition system. These resists
are then also expected to outgas less under the actual evaporation
conditions. An example for such an resist is polyethersulfone (PES),
known in France under the band name Vitrex. In collaboration with
CNRS-CRTBT Grenoble, we have patterned niobium wires with this
resist. Due to a very limited number of samples, however, statistics
about the resulting Nb quality are not conclusive yet.\par
The second trend is that towards completely inorganic resists. Hoss et
al. have very recently reported \cite{hoss:99:nitridxxx} using a
Si$_3$N$_4$ mask supported by an SiO$_2$ layer over a silicon
substrate. Since the undercut in this system is generated by wet
chemical etching with hydrofluoric acid, the choice of materials is
somewhat limited, and liftoff after the deposition is hardly an
option. 
The remaining metal would  cause a certain shunting
by introducing additional stray capacitances, but it is
possible to measure on single electronics devices without doing
liftoff \cite{wahlgren:94:xjobb}.
Therefore, such
inorganic resists definitely are an interesting development for certain
applications demanding good superconducting niobium.\par
For the preparation of very thin niobium films, a suitable surface
cleaning must be allowed prior to the deposition 
\cite{park:86:ultrathin}. 
\subsection{Four layer resist}
\label{subsec:fourlayerresist}
\begin{figure}\centering
\epsfig{file=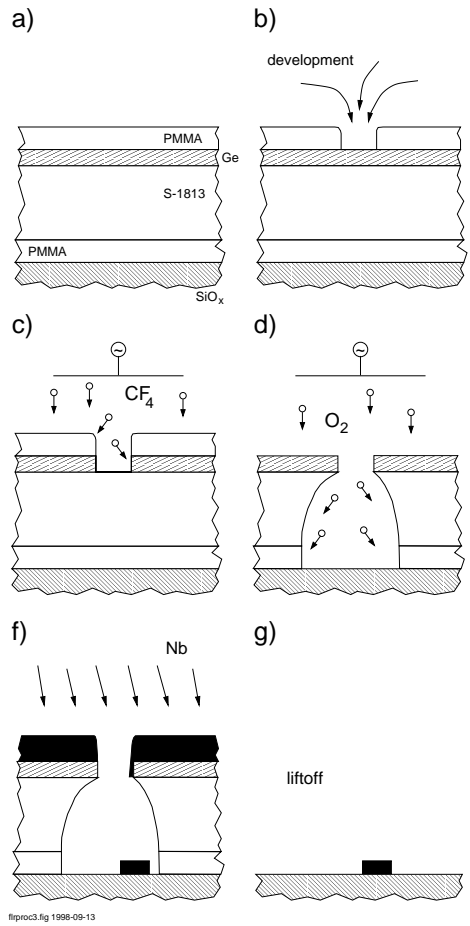,width=0.56\textwidth}
\caption[Four layer resist processing]{%
Four layer resist processing: overview of processing steps for Nb
liftoff nanopatterning. After exposure and development of the PMMA top
layer, the pattern is transferred to the germanium mask by reactive ion
etching (RIE) with carbon tetrafluoride CF$_4$. Subsequent oxygen RIE
creates the undercut profile necessary for a liftoff mask, especially if
intended for the angular evaporation technique.}
\label{fig:flrprocx}
\end{figure}
\subsubsection{Preparation and pattern transfer}
The structure of four layer resist we used can be seen in the top left
of fig.~\ref{fig:flrprocx}. From the top it consists of a 50\,nm PMMA
layer that is patterned by electron beam lithography, a 20\,nm germanium
mask, and the support layers. These support layers in turn are 200\,nm
of hard baked Shipley S-1813 photo resist and 50\,nm of PMMA. Details on
the process are given in \ref{rec:fourlayerprep}. It should be
emphasized here that keeping the right baking temperatures is very
important. Baking a higher layer at a too high temperature can lead to
warped masks, cracks due to tensile stress in the germanium layer, or
problems with liftoff.\par
A simple developer for PMMA is a mixture of isopropanole and water. As
soon as possible after exposure, the patterns were developed using the
concentration, times etc. given in \ref{sect:flpprcapp}.
The openings in the EBL-patterned PMMA top layer were transferred to the
germanium layer by reactive ion etching (RIE). RIE is a combination of
chemical etching and physical sputtering. The sample is placed inside a
low pressure reaction chamber on an insulated electrode, and a reactive gas
(`process gas') is let into the reaction chamber.
A radio
frequency cold plasma discharge is then ignited in the chamber.
The process gas becomes
partially cracked by the discharge, creating highly reactive radicals
that can reach the sample because their mean free path is long enough in
the low pressure. These radicals provide the chemical etching, which is
essentially isotropic. Additionally, molecules accelerated in
the electric field have a sputtering effect. The lower the chamber
pressure, the higher is the anisotropy of this sputtering. Near the
electrode on which the sample is placed, a DC bias voltage occurs
between the electrode and the plasma potential due to different
mobilities of positive and negative ions. The anisotropy of the
etching increases with this DC bias.\par
A suitable reactive gas for germanium etching is carbon tetrafluoride
CF$_4$. Our RIE system was not equipped with an etch end detection, so
that the required etching time had to be estimated
based on experience,
allowing some extra margin since reactive ion etching processes tend to
be somewhat irreproducible. Etch rates may vary depending on
contaminations present in the chamber,
or on the size of the sample areas, just to name a
few factors. Overdoing this etch step resulted in a slight increase in
linewidths, but this increase was considered tolerable.\par
Successful etching of the germanium layer is clearly visible by optical
microscope inspection. While developed PMMA areas appear just slightly
brighter than their surroundings in an optical microscope, the etched Ge
areas are much darker.
\par
Both the hardbaked photoresist and
the PMMA bottom layers were etched with RIE
using oxygen as reactive gas. A higher pressure than in the Ge etch step
was used to increase the anisotropy of the etch rate to create the
desired undercut profile. Another advantage of higher pressure is that
the physical sputtering, leading to damage of the suspended Ge mask
parts, is reduced.\par
Other than in the Ge etch step, the lack of an etch end detection is
rather
unfavourable here, because unnecessarily
long etching causes damage
of the Ge mask
that could otherwise have been avoided.
Figure~\ref{fig:wealimsk} 
on page \pageref{fig:wealimsk} is a scanning electron
micrograph of an etched four layer resist mask. The undercut is clearly
visible if one uses acceleration voltages of (5\dots 8)\,kV. The
suspended bridge in this picture is damaged in the form of a tiny
crack. More careful timing of the oxygen etch step and use of Teflon
piedestals to adjust the position of the sample in the reaction chamber
can reduce the risk of such damage.\par
\subsubsection{Niobium deposition and quality problems}
For the niobium resistor samples, we deposited the metal in a
non-bakable multipurpose HV system. During evaporation, the pressure
rose to typically $(3\pm 1)\cdot 10^{-5}$\,Pa, which gave niobium films
of 20\,nm thickness with a $T_\mathrm{c}$ of about 1.5\,K. Later we used
a bakable UHV system with a usual background pressure of $3\cdot
10^{-7}$\,Pa. In films of 40\,nm thickness deposited with four layer
resist masks, we obtained residual resistance ratios between 1.3 and
1.9, and critical temperatures between 1.5\,K and just below 4\,K,
depending on the individual conditions.\par
At a deposition rate of between 2\,nm/s and 4\,nm/s, and without
either substrate heating or cooling, we had a niobium grain size so
small that we could not distinguish grains in an SEM.
\label{comm:grainsize}
From an AFM analysis, we obtained a niobium surface roughness of between
5\,nm and 7\,nm (root-mean-square), which should indicate a grain size
of a few nanometres.\par
\begin{figure}\centering
\epsfig{file=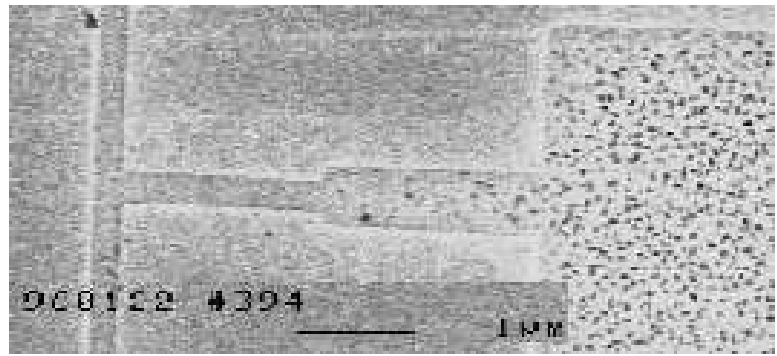,width=0.9\textwidth}
\caption[Surface contamination after etch processing]{%
Scanning electron micrograph of niobium patterned by the four layer
process, showing grainy contaminants whose density increases with the
width of the structure.}
\label{fig:contami}
\end{figure}
Figure~\ref{fig:contami} illustrates another quality problem with four
layer resist: on large open areas, grainy contaminants occurred after
the reactive ion etching pattern transfer steps. Their density depended
sensitively and not completely reproducibly on the conditions during the
oxygen etching step. Since the fine structures were spared from these
contaminations, we decided to tolerate them.\par

\chapter{Anodised niobium nanostructures}
\label{chap:anodi}
\section{Anodic oxidation}
\subsection{Principle and History}
In an anodic oxidation or \emph{anodisation}, a metal is brought into
contact with an electrolyte, and negatively charged ions containing
oxygen are driven into the metal by an externally applied electrical
field. The ions react with the metal to form an oxide. The process is
not reversible, i.\,e., the oxide layer does not decrease when the
polarity is reversed. This ``one way only'' behaviour in mind, the
metals suitable for such a process were called \emph{valve metals}.\par
Transforming metal into an oxide, anodisation can be viewed either as a
technique to create an oxide layer, or to remove metal. The former
application is dominant, and anodic oxidation is used at an industrial
scale today for the surface treatment of metals, e.\,g. aluminium. While
in this example, a certain porosity of the surface that allows the
inclusion of colour pigments is desirable, the application most
important for the development of anodisation techniques had just the
opposite aim, namely the creation of pinhole-free oxide layers in
electrolytic capacitors 
\cite{guentherschulze:37:elko,jackson:76:elcap}. A fair amount of
development, especially regarding electrolytes, was conducted behind
closed doors in industry, and since many decades have passed and modern
analysis techniques had not been invented at that time, much know-how
associated with electrolyte recipes today gives an impression of black
magic. A standard reference for early developments in electrolytic
capacitors and anodisation is the 1937 book by G\"untherschulze and Betz
\cite{guentherschulze:37:elko}, in which numerous metals are
treated. The most important metal for capacitors, still today, is
tantalum, to which niobium is closely related in its physicochemical
properties. \par
A standard reference on anodisation is the 1961 book by Young
\cite{young:61:anox}. The behaviour of niobium has been covered in a
1993 series of review papers by d'Alkaine 
\cite{alkaine:93:anodi1,alkaine:93:anodi2,alkaine:93:anodi3}.
\subsection{Micro- and nanofabrication by anodisation}
For purposes of micro- and nanofabrication, anodic oxidation basically
serves as a means to remove metal. Since in most cases, however, the
oxide is not dissolved in the electrolyte or at all removable without
affecting the other structures, the oxide layers usually become an
integral part of the device design.\par
Commercially, anodisation of niobium is used for the production of
small Josephson junctions from prefabricated three layer 
(Nb/AlO$_x$/Nb) sandwiches
through a process called selective niobium anodisation
process (SNAP) \cite{kroger:81:anodiapl}. The top layer metal is oxidised
where a photolithographically defined resist mask exposes it to the
electrolyte, and then suitable contacts are made to the bottom layer and
the unoxidised top layer regions. Sandwiches for such a \emph{three
layer technique} can be fabricated under extreme conditions and give
the best niobium and interface quality (for their application in single
electronics, see \ref{subsec:competing_techniques}). The top layer is
anodised all the way through, and the proper etch end can be detected by
driving a constant current through the anodisation cell and monitoring
the voltage; a discontinuity in the voltage increase rate indicates that
the AlO$_x$ layer has been reached.\par
Anodisation can also be used to not completely remove a metal, but just
thin it out very much. Apart from the resistor fabrication process,
which was patented \cite{westernelectric:60:resipat} long before SNAP
was devised, several processes were developed in the last decades.\par
Ohta et al. \cite{ohta:87:wljj} used anodisation to thin out a weak link
of niobium between two thick and large niobium pads. In the temperature
dependence of the critical current, they observed a transition from weak
link Josephson junction behaviour to tunnel Josephson junction behaviour
with decreasing residual thickness of the linking metal. Goto et
al. \cite{goto:79:jap,goto:81:vtb} produced a silicon monoxide mask over
a niobium strip with a long ($\approx 10\,\mu$m) and narrow ($\approx
200$\,nm) slit, created by shadow evaporation at the edge of a resist
strip. They then anodised the niobium through this small slit and found a
transition from the behaviour of a continuous superconductor to that of
a Josephson junction, manifesting itself in the occurrence of Shapiro
steps under microwave irradiation.\par
In neither experiment was the resistance or thickness of the remaining
metal monitored during the anodisation. Instead, the assumption of a
niobium consumption directly proportional to the anodisation voltage
\cite{kroger:81:anodiapl} was made, which is an oversimplification
\cite{chiou:71:timedep}, as we will see in \ref{subsec:anodidyn}.\par
Nakamura et al. \cite{nakamura:96:acme} used anodisation to thin out the
top electrodes, and thereby the cross sections, of Al-AlO$_x$-Al single
electron transistors made by the conventional shadow evaporation
process. By this technique \cite{nakamura:96:acme},
which they called anodisation controlled
miniaturisation enhancement (ACME), they were able to raise the
operating temperature of individual samples, and observed a modulation
of the source-drain current with the gate voltage up to temperatures of
30\,K. In these experiments, the resistances of the SET were monitored
in situ and could be increased by about two orders of magnitude. A
drawback with this technique, as with all anodisation techniques, is
that preexisting asymmetries are enhanced, causing most of the samples
to show a Coulomb staircase in their current-voltage characteristics.\par
\section{Microanodised niobium resistors}
  \label{sect:microanodi}
\subsection{Physical setup}
\begin{figure}[tpb]\centering
\epsfig{file=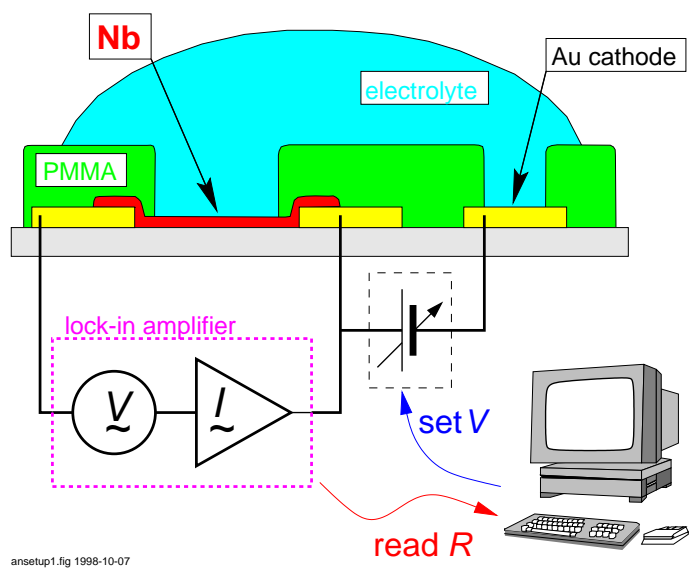,width=0.8\textwidth}
\caption[Niobium microanodisation setup]{Niobium microanodisation 
setup (schematic, not to scale). In reality, the leads to the resistance
monitoring and voltage control electronics were situated on the chip
perimeter, away from the electrolyte droplet.
}
\label{fig:ansetupx}
\end{figure}
Figure \ref{fig:ansetupx} is a sketch of the experimental setup for
resistor fabrication by anodisation with in situ resistance
monitoring. It corresponds very much to the scheme given in Western
Electric's 1959 patent \cite{westernelectric:60:resipat}, except that
our length scales are much smaller, that we specify the nature of the
monitoring device, and that we include the cathode on the sample (see
\ref{subsubsec:cathode}). \par
The areas of niobium that are to be anodised are exposed to the
electrolyte (see \ref{subsubsec:electrolyte}) by windows in the
anodisation mask (see \ref{subsubsec:anodimask}). The cell voltage is
controlled according to the resistance measurement (see
\ref{subsec:resimon}). \par
\subsubsection{Anodisation mask}
  \label{subsubsec:anodimask}
Had only the size resolution been a criterion, photolithography would
have been completely sufficient for the creation of the anodisation
window mask. We wanted, however, to have an absolute window placement
accuracy better than 1\,$\mu$m, which was impossible with our
photolithographic equipment. It appeared therefore most feasible to
exploit the electron beam lithography system's fiducial detection and
pattern alignment capabilities for the definition of the anodisation
mask. In addition, the electron scattering in an electron beam resist
material would automatically give the undercut profile required for the
deposition of top gates with the same mask.\par
As material for the anodisation mask, we chose a highly viscous solution
of PMMA, spin coating the wafer to a thickness of ca. 1.8\,$\mu$m (for
more
technical details, see the recipe in \ref{rec:windowmask} on page
\pageref{rec:windowmask}). This thickness was necessary to prevent
dielectric breakthrough of the PMMA under the anodisation voltages of up
to 25\,V.\par
The exposure and development of the anodisation mask were
straightforward, except maybe for the very long development time of
15~minutes that was necessary to completely open the windows. 
Images of anodisation mask windows can be seen in
figs.~\ref{fig:anomaopt} on page \pageref{fig:anomaopt} and
\ref{fig:tgatmont} on page \pageref{fig:tgatmont}.\par
\subsubsection{Electrolyte}
  \label{subsubsec:electrolyte}
As mentioned before, a lot of the development of electrolytes for
anodisation processes is poorly documented in the scientific
literature. Many recipes are based on experience rather than on a deep
understanding of the underlying process details. Demands for a good
electrolyte are stability and a low vapour pressure. The latter
eliminates the need for hermetically sealed encapsulations in
capacitors, and is a very welcome property for our application since any
significant change in the electrolyte composition during the anodisation
(with the electrolyte droplet being directly exposed to atmosphere) is
undesirable.\par
We could trace back the recipe for the electrolyte we used to a 1967
paper by Joynson et al. \cite{joynson:67:anodi}, but cannot rule out
that it might have been published elsewhere earlier. The electrolyte
consists of an aquaeous solution of ammonium pentaborate mixed with
ethylene glycol (see \ref{rec:electrolyte} for the details).\par
Originally intended to be used at 120$^\circ$C, this electrolyte can be
used at room temperature if it is regenerated by heating to about
100$^\circ$C under stirring for a few minutes not more than two days
prior to use. After two days, precipitations occur. The prepared
solution can be used for at least one year without any significant
change in properties. Niobium is a well-behaved metal in respect to
anodisation inasmuch as it works with many different electrolytes, and
is insensitive to contaminations \cite{guentherschulze:37:elko}.\par
The addition of ethylene glycol is an old, empirically founded
practice. In 1990, Bairachnyi et al. published a short article
\cite{bairachnyi:92:gly} in which they investigated the glycol's
role. They found that it improves the stoichiometry of the oxide film,
and thus the stability of the electric properties.\par

The long-time stability of our anodised wire samples might at least
partially be attributed to the beneficial effect of the glycol. In
Nb$_2$O$_5$-Nb bilayers, degradation is known to change the interface
between metal and oxide with time
\cite{boiko:93:degr}. Anodisation-trimmed resistors with very high sheet
resistances would be especially prone to ageing by interface
degradation. Luckily, in this application there are no field strengths
to be expected that would run as high as those across an electrolytic
capacitor dielectric, and that promote interface degradation 
\cite{palatnik:94:degr}.
Nevertheless, for real resistor applications, some
kind of protective coating might be necessary
\cite{waggener:65:ukpat}.\par 
\subsubsection{Cathode}
  \label{subsubsec:cathode}
The first anodisations were made with a tungsten wire as cathode,
inserted into the electrolyte droplet. It was later replaced by the
stainless steel needle of the syringe that was used to apply the
electrolyte droplet to the chip. This needle allowed to correct the
droplet size, and its wedged tip reduced the risk of accidental damage
to the sample or mask during repeated manual insertions of the
cathode.\par
Obviously, these designs would have required some kind of holder
mechanism for continuously monitored and executed anodisations. The
problem of droplet size reduction by evaporation during long
anodisations would have required a droplet size overhead in excess of a
(generous) safety distance between mask and electrode. These problems
were instantly remedied by integration of the cathode on the chip.\par
The cathode material does not appear to matter on this small scale. Of
course, a valve metal like niobium is a non-optimal choice since any
(unintentional) reversal of polarity would build an oxide layer on the
cathode. The simplest solution was to use one of the gold contact leads
on the prefabricated substrate chip as cathode. Such an on-chip cathode
can be seen on the left in fig.~\ref{fig:anomaopt}.\par
In any case, careful grounding of the electrolyte (syringe needle),
sample, and cathode during insertion and connection was found to be
essential. Due to the very small areas of niobium exposed by the
anodisation mask window, electrostatic charge would generate enough
current to drive the anodisation process far, even to pinchoff, in an
instant. \par
\subsection{Electrical setup}
  \label{subsec:resimon}
Our requirements for a resistance monitoring device were:
\begin{enumerate}
\item a certain precision, 
\item the ability to work in spite of the large potential difference
  between the sample and the cathode, which went up to almost 30\,V, and
\item a negligibly small distortion of the electric field along the
  sample. 
\end{enumerate}
The last requirement was of particular importance. Measuring the
resistance with a DC ohmmeter would introduce a considerable potential
drop along the sample, and since very small anodisation voltage
differences can result in large differences of the sheet resistance of
the anodised film, this approach would result in an uneven film
thickness. When we tried such a method initially, the same strip
resistance value was reached at significantly lower cell voltages than
with a ``proper'' resistance measurement. Obviously, the films had begun
to pinch off at the side with the higher anodisation voltage.\par
All three named requirements were met by using a lock-in technique. We
used a Stanford SR850DSP amplifier in current measuring mode. A sine
shaped voltage excitation with an rms amplitude of typically 4\,mV, and
up to 32\,mV for highly resistive samples, was applied. With a
sufficiently long time constant, an accuracy of the resistance
measurement of a few percent was achieved. Since the wiring of the
sample holder was quite open and pickup therefore non-negligible at low
frequencies, we used a lock-in frequency of 3\,kHz in the vast majority
of cases.\par
The voltage and resistance readings were fetched by a control computer
and logged. A control programme with a graphical user interface allowed
the experimentalist to modify the anodisation voltage ramping rate and,
with a little smooth touch, tune the resistance to a desired value. The
anodisation currents were not accessible to measurement since the
anodised areas were too small and the currents thus too weak.\par
\subsection{Anodisation dynamics}
  \label{subsec:anodidyn}
As expected, the sample resistance increased irreversibly as the
anodisation voltage was applied. For a given voltage and ramping rate,
the rate of resistance increase is a complicated function of voltage,
metal thickness, and time itself.
\begin{figure}\centering
\epsfig{file=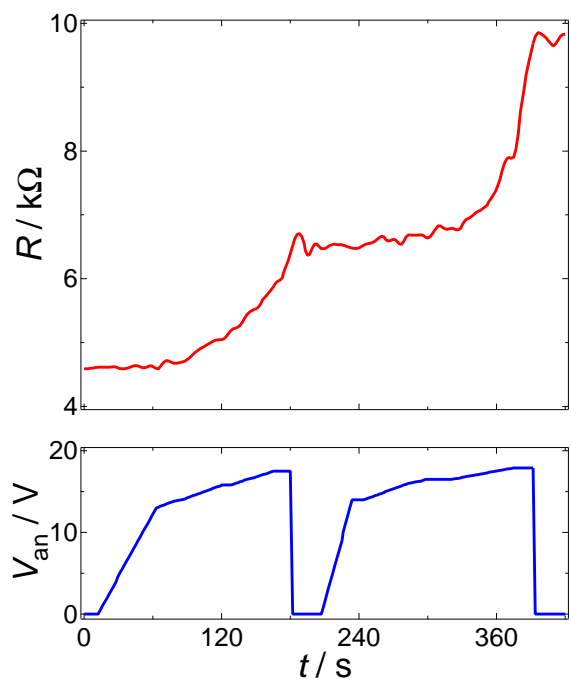,width=0.7\textwidth}
\caption[Anodisation voltage and sample
resistance]{%
Anodisation voltage and sample resistance. The resistance $R$ along the
wire sample, measured in situ, increases irreversibly. By adjusting
the cell voltage $V_\mathrm{an}$, the rate of increase of $R$ can be
controlled. Zeroing $V_\mathrm{an}$ holds $R$ at its current value.
Note that $dR/dT$ is not only a function of $V_\mathrm{an}$, but of the
time $t$ as well.}
\label{fig:twostpan}
\end{figure}
In fig.~\ref{fig:twostpan}, the time evolution of the resistance and the
voltage causing it are plotted as  functions of time. The resistance
values are rather low here since this was a sample in single electron
transistor-like geometry (see \ref{sect:setlikesamples}). Typical
features seen in fig.~\ref{fig:twostpan} are:
\begin{itemize}
\item It takes a certain voltage to cause any measurable impact on the
  resistance at all. Once that voltage is reached, the resistance
  reading climbs at an increasing pace. 
\item At high resistance values (small remaining film thicknesses),
  small differences in voltage cause big differences in resistance.
\item Zeroing the voltage promptly and safely holds the resistance at
  its maximum value.
\end{itemize}
As mentioned before, the popular notion of an anodisation constant,
i.\,e. a strictly linear dependence between the anodisation voltage and
the amount of oxide created, does not hold.
\begin{figure}\centering
\epsfig{file=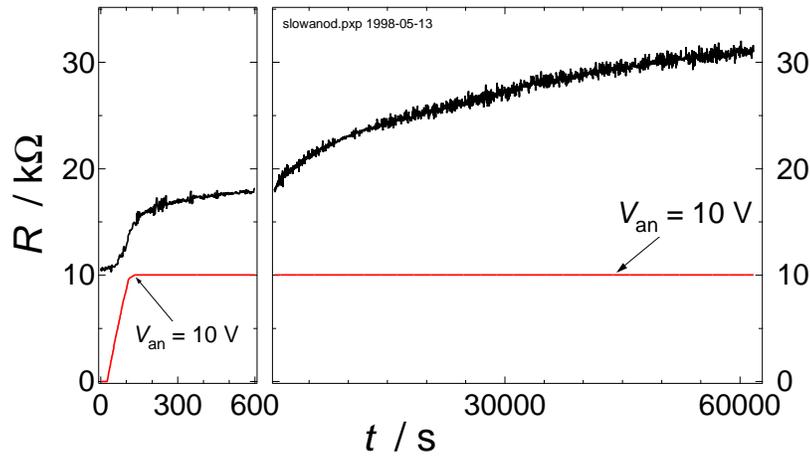,width=0.96\textwidth}
\caption[Slow anodisation process at constant voltage]{%
Slow anodisation process at constant anodisation voltage. The resistance
of the sample (a 10\,$\mu$m long wire) continued to increase with time.}
\label{fig:slowanod}
\end{figure}
This is directly visible in fig.~\ref{fig:slowanod}. In the anodisation
shown here, the voltage was ramped up quickly to 10\,V and then held
constant for more than sixteen hours. During the whole time, the
resistance continued to increase measurably. The resistance rate
decreased, but nevertheless, under these 
constant voltage conditions the oxide would
continue to grow until the metal were completely anodised.\par
This effect can be exploited for the high precision tuning of resistors:
by stopping the voltage ramping well before the desired resistance value
has been reached, and by simply waiting, one can reach the target resistance
with a precision only limited by the resistance measurement, and then
zero the voltage once that value has been reached. The slow dynamics
allow for improved precision by using longer integration times on the
lock-in amplifier.\par
\section{Measurement setup and procedures}
  \label{sect:lowtempsetup}
\subsection{Cryogenics}
  \label{subsec:cryogenics}
The first measurements were performed in a noncommercial dilution
refrigerator that reached temperatures down to 95\,mK. The temperature
could be measured with a  resistance thermometer calibrated over the
whole accessible temperature range. Most measurements documented in this
report were, however, done in a commercial dilution refrigerator of type
TLE\,200 from Oxford Instruments. A 
germanium resistance thermometer was
calibrated down to 45\,mK, and the base temperature of the cryostat was
below 20\,mK, as estimated from a preliminary measurement with nuclear
orientation thermometry. A magnetic field up to 5\,Tesla could be
applied perpendicular to the sample. With a ramp rate of
0.1\,T/min, the sample was warmed to approximately 40\,mK by eddy
currents.
\subsection{Amplifier electronics}
From the sample, the DC leads went via 
filters (described below)
and multiply thermally anchored wires to an amplifier box on
top of the cryostat at room temperature. The bias voltage was
symmetrised with respect to ground and fed to the samples via high ohmic
resistors in the mentioned amplifier box. Voltage drops over the sample
and over the bias resistors (proportional to the current) were picked up
by low noise amplifiers, and the amplified voltages sent outside the
shielded room for registration. Details about the measurement
electronics can be found in  Delsing's PhD thesis
\cite{delsing:90:thesis}. \par
Signals were measured with digital voltmeters,
initially with DMM of type Tektronix  DM5520  with a buffer capacity of 500
points, later with Keithley 2000 DMM storing 1024 data points. The
measurement times were synchronised with a Keithley 213 voltage source
providing the bias voltage, which was stepped rather than swept
continuously. Gate voltages were either generated with a second port on
this Keithley 213 source, or with a Stanford 
Research Systems DS\,345 signal
generator.\par
Unless explicitly mentioned otherwise, the sweep of the bias or the gate
voltage was always bidirectional, starting and ending at one of the
edges of the swept voltage region.\par
All data were registered electronically with the help of a GPIB equipped
Macintosh. For historical and practical reasons, the
measurement software was written in various versions of
LabVIEW (see appendix \ref{app:mdatadp}).\par
The special amplifier used for the noise measurements that are the
subject of chapter \ref{chap:noise} will be described separately in 
\ref{sect:bjoernsamplifier}.\par
\subsection{Shielding and filtering}
  \label{subsec:shieldfilter}
The cryostat was placed in a steel enclosure
(`shielded room'). Inside this enclosure, all
electronics were analog. Leads into the shielded
room were passed through filters in its wall. Inside the cryostat, the
leads were multiply filtered against high frequency radiation.
\begin{figure}
\centering
\epsfig{file=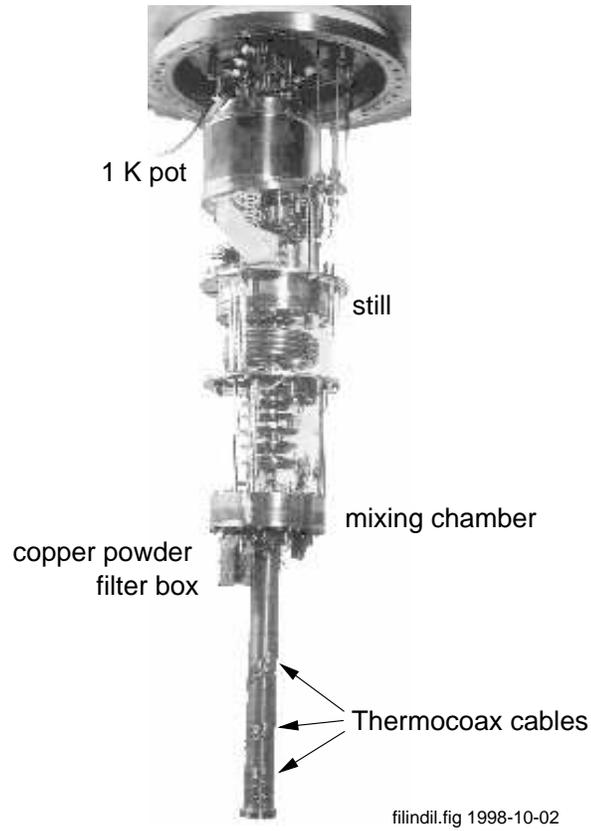,width=0.7\textwidth}
\caption[Dilution refrigerator with filters]{%
Dilution refrigerator insert with low temperature filters. DC leads are
formed as 500\,mm Thermocoax cables near the sample at dilution
refrigerator temperatures, and passed through a copper powder filter box.
}
\label{fig:filindil}
\end{figure} 
Figure~\ref{fig:filindil} shows the dilution refrigerator insert
with two types of filters mounted: the leads coming from room
temperature are heat sunk at the various stages and passed through a
copper powder filter \cite{martinis:87:diffprb} at the mixing chamber
stage. From there to the connector at the bottom of the insert, the DC
leads are formed by 500\,mm Thermocoax \cite{zorin:95:thermocoax},
a commercial coaxial cable with very good damping properties
\cite{fukushima:97:cpem}.
More information
about the cryostat and the filter design and properties can be found in
\cite{haviland:96:jvst}. 
\section{Resistor samples}
  \label{sect:resistorsamples}
\subsection{Sample geometry and characterisation}
\begin{figure}\centering
\epsfig{file=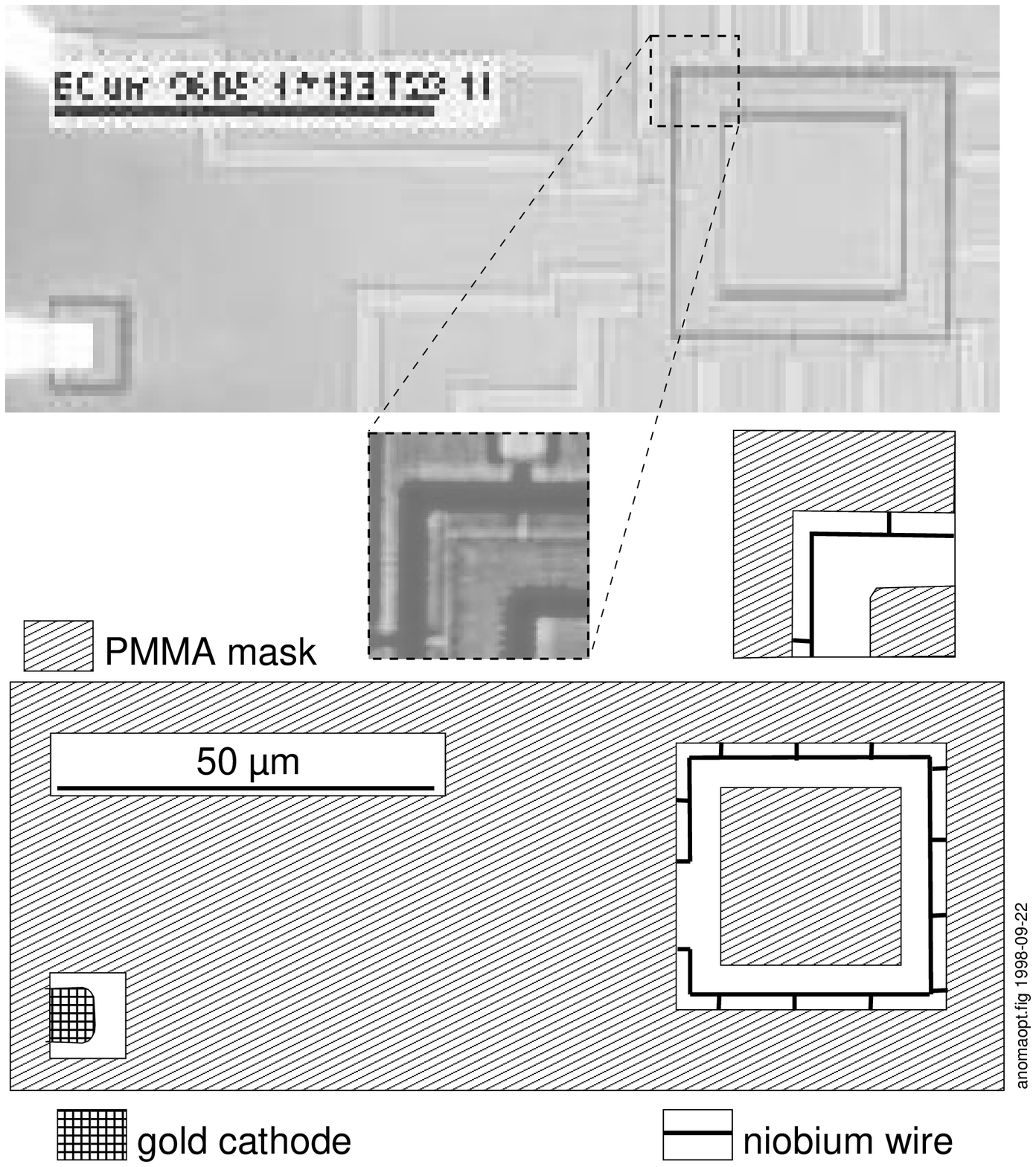,width=\textwidth}
\caption[Anodisation mask and geometry of resistor samples]{%
Optical micrograph of resistor samples and its artistic interpretation.
The Nb wire, divided into twelve sections of 10\,$\mu$m length each, is
barely visible through the window in the PMMA anodisation mask. The
small window on the left exposes a gold contact used as on-chip-cathode
to the electrolyte.}
\label{fig:anomaopt}
\end{figure}

The layout of the resistor samples is shown in
fig.~\ref{fig:anomaopt}. The standard geometry were strips of 10\,$\mu$m
length, 20\,nm thickness and a width limited by the lithography and
pattern transfer techniques. At the time these samples were made,
linewidths between 120\,nm and 180\,nm were the 
limit, caused by the pattern transfer to the germanium mask (this
improved later, when a better RIE machine was taken into operation).
The wires were either single, and attached to wider niobium pads for
four probe measurements, or grouped into a 120\,nm long wire as in
fig.\ref{fig:anomaopt}. These long wires had probe leads spaced at
10\,$\mu$m distance from each other, and allowed the measurement of the
resistance of each segment in a four probe configuration.\par
\begin{figure}\centering
\epsfig{file=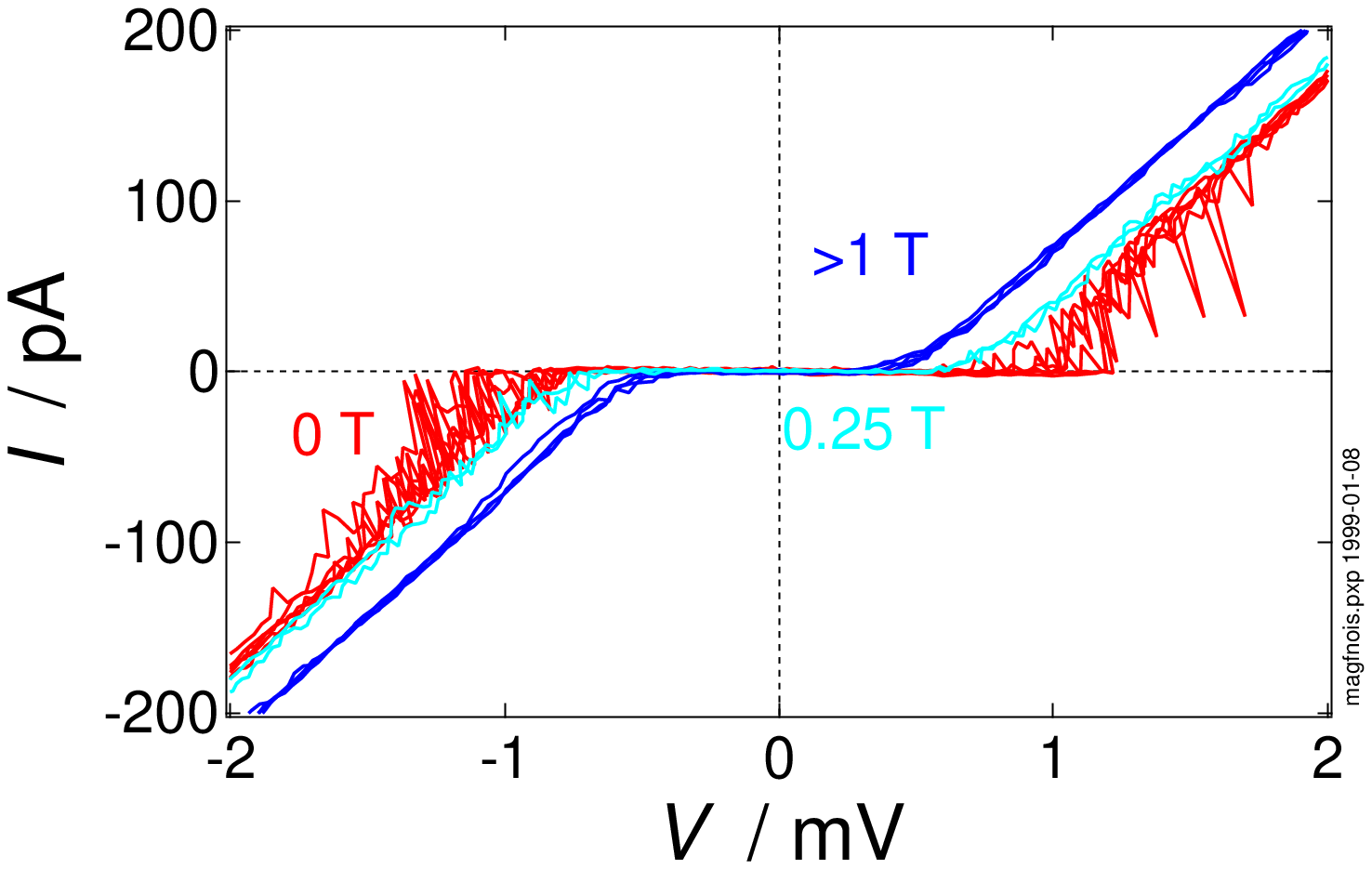,width=0.96\textwidth}
\caption[Magnetic field effect on resistor sample IVC]{%
Effect of an external magnetic field on the current-voltage
characteristics of a highly resistive samples. The threshold voltage is
reduced by about half. Telegraph noise affecting the IVC in the field
free state is smoothened out in the external field. Several traces have
been superimposed in each case.}
\label{fig:magfnois}
\end{figure}

The anodised samples were characterised by low temperature
magnetotransport measurements. For a highly resistive sample,
fig~\ref{fig:magfnois} shows the current-voltage characteristics in
different magnetic fields.
There is an obvious Coulomb blockade, whose threshold is reduced in an
external magnetic field. Another characteristic feature in the 0\,T
trace in fig.~\ref{fig:magfnois} is the telegraph noise pattern, that is
the switching of the IVC between two curves along the load line. This
telegraph noise was often observed in our samples, especially at high
resistances. The generally accepted explanation of this pattern is that
charge traps \cite{zimmerli:92:noiseapl} 
upon charging and discharging induce
charge differences on one grain that is important in the current path,
in the sense of acting like a single electron transistor coupled in
series with the rest of the film. The fluctuation of the charge in one
trap between the two states `charged' and `discharged' creates 
telegraph noise. Several such two level fluctuators (TLF) may be in
effect simultaneously.\par
Telegraph noise is also encountered in single electron transistors made
by conventional techniques. If it is seen, the sample is often dismissed
as ``infected by a TLF''. Sometimes the sample can be ``cured'' by
thermally cycling it and thereby changing the 
population of defects. 
Interestingly enough, applying a magnetic field on an ``infected'' SET
usually does not suppress the telegraph noise. Noise similar to that
seen in fig.~\ref{fig:magfnois} has been seen in one-dimensional regular
arrays of small junction SQUIDs 
\cite{haviland:96:jvst}. \par
We do not have a satisfactory explanation for the observed noise
pattern and can only state that here
it seems to be directly related to the
superconducting properties of the samples (see also
\ref{subsec:sitransitionexp}). To add to the confusion, one might add
that while here the sample is noisier in the superconducting state,
Wahlgren has observed that arrays were more noisy in the normal
conducting state \cite[ch.~7, fig.~4]{wahlgren:98:thesis}. Very similar
telegraph noise has been observed by Fujii et al. in completely normal
conducting composite films \cite{fujii:97:compofilms}.
\subsection{Superconductor-insulator transition}
  \label{subsec:sitransitionexp}
\begin{figure}\centering
\epsfig{file=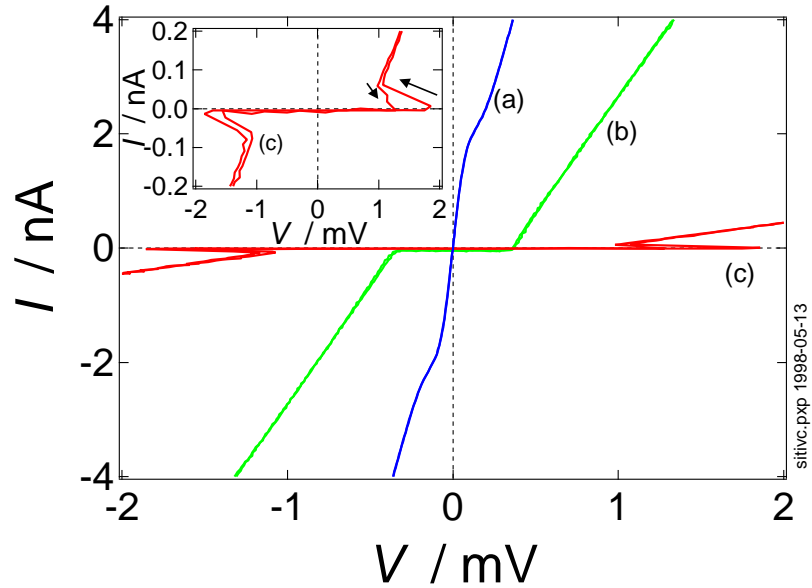,width=0.96\textwidth}
\caption[IVC of three resistor samples]{%
Current-voltage characteristics (IVC) of three resistor samples. With
increasing sheet resistance (measured at high bias and/or at high
temperature), the IVC changed from a supercurrent remnant ((a)
1.5\,k$\Omega/\Box$) to a sharp Coulomb blockade ((b)
8\,k$\Omega/\Box$). Samples with very high sheet resistance showed a
backbending IVC ((c) 40\,,k$\Omega/\Box$). Sample dimensions in each
case were $10\times 0.15\,\mu$m$^2$, the data were taken at 
temperatures of about 200\,mK (a) or at base temperature
below 50\,mK (b,c), respectively,
without applying an external magnetic field.}
\label{fig:sitivc}
\end{figure}
Figure~\ref{fig:sitivc} collects the types of current-voltage
characteristics we observed in the resistor samples. The data presented
here were taken on strips on three different chips. The strips had equal
dimensions, but were anodised for different times and to different final
voltages. The measurements were taken in a four probe configuration, at
temperatures between below 50\,mK and 200\,mK, and in the absence of an
externally applied magnetic field.\par
For the samples with lowest resistance, like in trace (a), we found a
remnant of the supercurrent, visible as a region of reduced differential
resistance for currents up to about 2\,nA. The sample whose IVC is shown
as (a) in fig.\ref{fig:sitivc} had a sheet resistance of approximately
1.5\,k$\Omega/\Box$. For samples anodised to greater depth, we observed
an increase of the differential resistance around zero bias that we have
already referred to as the ``Coulomb blockade'' (which will be finally
justified in \ref{sect:setlikesamples}). \par
Trace (b) in fig.~\ref{fig:sitivc} is a typical example of a sharp
Coulomb blockade with a well-defined threshold voltage. This sample had
a sheet resistance of about 8\,k$\Omega/\Box$. In samples with very high
sheet resistances, we observed not only a sharp blockade, but even a
backbending of the IVC (trace (c) and inset in
fig.~\ref{fig:sitivc}).\par
The backbending in the sweep
direction from low bias outwards 
could be explained by heating of the sample at the
onset of current flow, where the relatively high voltage leads to power
dissipation even at low currents. On the other hand, 
this would not explain a backbending in both sweep directions.
The backbending IVC
show a remarkable similarity with the IVC observed in arrays of
ultrasmall Josephson junctions by Geerligs et
al. \cite{geerligs:89:jjaprl} and Chen et
al. \cite{chen:92:ps}.\par 
Even the observed telegraph noise fits into the picture of the resistor
strips behaving as an array of superconducting junctions. Haviland et
al. \cite{haviland:96:jvst} have observed it accompanying the
superconductor-insulator transition in one-dimensional 
SQUID arrays, and gave
an explanation for the observed hysteresis by analogy to the
resistively shunted model for Josephson junctions 
\cite{haviland:97:phasdomabstract}. \par
\begin{figure}\centering
\epsfig{file=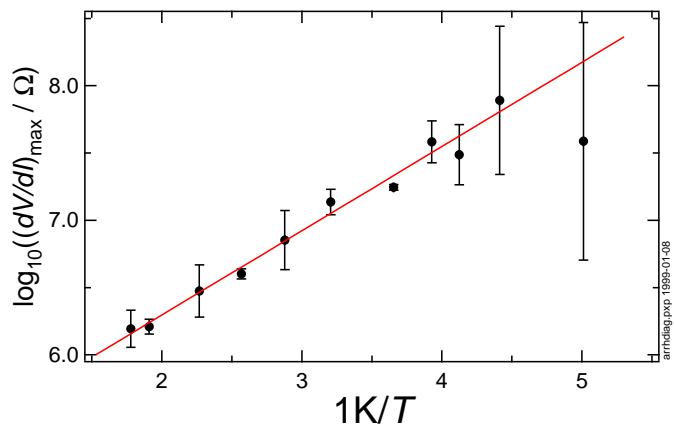,width=0.8\textwidth}
\caption[Arrhenius dependence of the zero bias differential
resistance]{Temperature dependence of the zero bias differential
resistance. The Arrhenius type dependence indicates a thermal
activation behaviour. Values have been determined from numerical
differentiation of measured current-voltage characteristics. No external
field was applied during these measurements.}
\label{fig:arrhdiag}
\end{figure}

The temperature dependence of the resistance is a usual criterion for
classifying a material as insulating or undergoing a superconducting
transition. Since our samples had such nonlinear IVC, one cannot assign
a global resistance value but looks instead, for example, at the
differential resistance at zero bias. A dedicated measurement would have
to involve a carefully devised biasing scheme and a sensitive detection,
preferably involving a lock-in technique \cite{chen:94:thesis}. Even
from the measured raw IVC one can, however, extract some information about
the zero bias differential resistance. For one sample, values gathered
from the numerical differentiation of IVC are plotted as a function of
inverse temperature in fig.~\ref{fig:arrhdiag}.
The errors indicated here were estimated from the curvature around the
maximum in the differential resistance. For the temperature range of
fig.~\ref{fig:arrhdiag}, before the sample went into a full Coulomb
blockade, we see that the zero bias resistance follows an Arrhenius law
\begin{equation}
  R_0(T)=R^\ast\cdot \mathrm{e}^
  {\frac{\displaystyle E_\mathrm{a}}{\displaystyle k_\mathrm{B}T}},
\end{equation}
which suggests a thermally activated behaviour
\cite{chen:94:thesis}. From fig.~\ref{fig:arrhdiag}, we extract an
activation energy corresponding to a temperature of about 0.6\,K, or to
a voltage of 50\,$\mu$V. This is of the same order of magnitude as the
voltage swing and the temperature of its disappearing in our samples in
the SET-like geometry (see \ref{sect:setlikesamples}). More quantitative
statements are complicated since the sample might have been
inhomogeneous; this one had a sheet resistance of about
4\,k$\Omega/\Box$. 
\subsection{Onset of the Coulomb blockade}
Defining a quantitative measure of the Coulomb blockade is an old
problem. Threshold voltages are hardly ever well-defined due to rounded
current-voltage characteristics, and since these IVC are also nonlinear
over many decades in bias voltage, extrapolating their tangent at the
end of an arbitrarily chosen voltage range would produce ambiguous
results. A solution to this problem was found by Wahlgren
\cite{wahlgren:95:prb,wahlgren:98:envprb}, who has shown the usefulness
of the \emph{offset voltage} analysis when studying environmental
effects on the Coulomb blockade.\par
\begin{figure}\centering
\begin{minipage}[c]{0.5\textwidth}
\epsfig{file=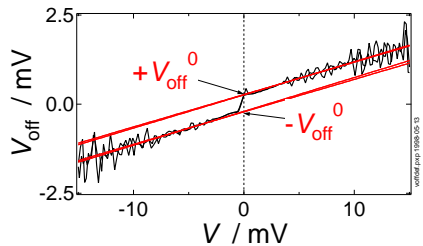,width=\textwidth}
\end{minipage}
\begin{minipage}[c]{0.48\textwidth}
\caption[Definition of $V_\mathrm{off}^0$]{Definition 
of $V_\mathrm{off}^0$. Extrapolation of $V_\mathrm{off}(V)$,
calculated from the tangent to the $I(V)$ curve, to zero bias gives 
$V_\mathrm{off}^0$, a measure of the Coulomb blockade.}
\label{fig:voffdef}
\end{minipage}
\end{figure}

The local offset voltage $V_\mathrm{off}(V)$ is computed numerically by
extrapolating the tangent to the $I(V)$ curve and taking its
intersection with the voltage axis:
\begin{equation}
  V_\mathrm{off}(V)=V-I(V)\;
 \left.\frac{dV^\prime}{dI}\right|_{V}.
\end{equation}
Figure~\ref{fig:voffdef} shows a calculation of $V_\mathrm{off}(V)$ for
a resistor sample. We see that $V_\mathrm{off}(V)$ has a jump at zero
bias and varies with approximately constant slopes in the vicinity. The
intersection of the extrapolation here with the $V_\mathrm{off}$-axis,
which we call $V_\mathrm{off}^0$, is therefore well-defined (see
fig.~\ref{fig:voffdef}). \par
This value has been shown to be of particular interest in the case of
the Coulomb blockade in double junctions, where it gives the limit for
the blockade in the so-called \emph{global rule} for low environment
impedances, where the whole electromagnetic environment influences the
Coulomb blockade \cite{wahlgren:95:prb}. It is not surprising that
$V_\mathrm{off}^0$ is useful in the case of our resistor samples. Since
we assume that we have a very large number of grains, probably several
thousand in a 10\,$\mu$m strip, the voltage drop per intergranular
junction is small, and according to the horizon picture of tunnelling
\cite[and references therein]{wahlgren:98:thesis}, the global rule
should then apply.\par
\begin{figure}\centering
\epsfig{file=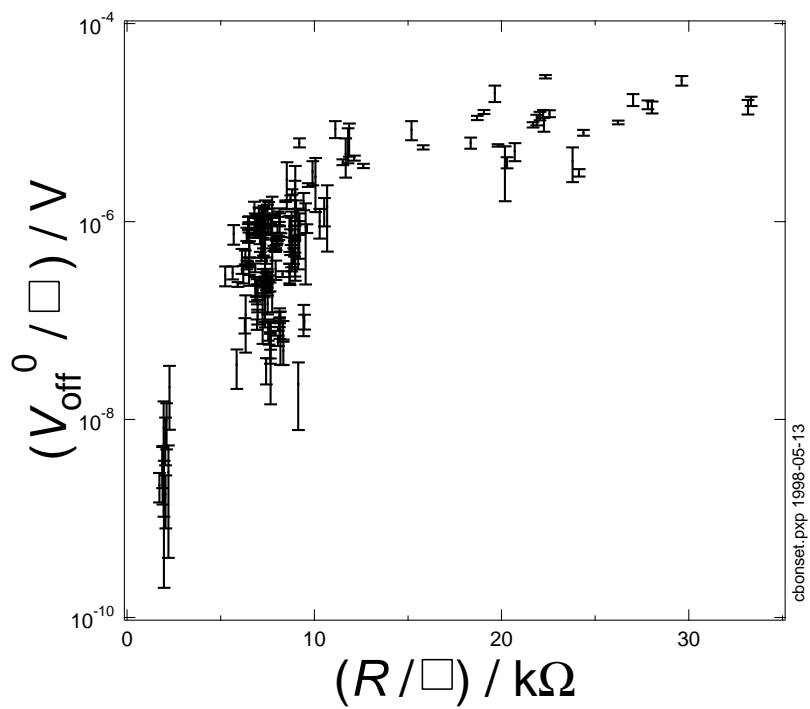,width=0.96\textwidth} 
\caption[Onset of CB in resistor samples]{%
Onset of the Coulomb blockade (CB) in resistor samples. Both
$V_\mathrm{off}^0$ and the resistance $R$ are normalised to the number
of squares. The CB is suppressed below 6\,k$\Omega/\Box$. Errors were
estimated from the standard deviation of
four independent extrapolations per data set (bidirectionally
swept IVC, extrapolated below and above zero, respectively).}
\label{fig:cbonset}
\end{figure}

For each IVC, which was taken bidirectionally by ramping the bias up and
down, we made four independent extrapolations of $V_\mathrm{off}$.
Figure~\ref{fig:cbonset} shows the results of this analysis for a set of
187 strips from four different batches with lengths between 10\,$\mu$m
and 120\,$\mu$m.
The errors have been estimated as the standard deviations of the four
$V_\mathrm{off}^0$ values per IVC. The correlation between extrapolated
zero bias offset voltage $V_\mathrm{off}^0$ and sample resistance $R$ is
obvious here where both quantities have been normalised to the number of
squares in the films.\par
Inhomogeneities may be responsible for the significant spread in the
data, but the trend is clear: the Coulomb blockade is appreciable for
all samples with a sheet resistance of more than 10\,k$\Omega/\Box$, and
at least three orders of magnitude less for all samples with less than
5\,k$\Omega/\Box$. In the intermediate region, there is  some
clutter due to measurement uncertainties. There is also an uncertainty
in the determination of the number of squares. The sample widths were
estimated from scanning electron microscope inspection of samples
produced under nominally identical conditions, and the error in the
measurement of the film area (not indicated in fig.~\ref{fig:cbonset})
may very well be up to 20\%.
\par
Within this measurement accuracy, the superconductor-insulator
transition we observed here in an unconventional way set in at a sheet
resistance that is compatible with $h/(4e^2)\approx 6.4$\,k$\Omega$. It
has been suggested from studies on ultrathin metal films
\cite{liu:92:ultrathinprl,belitz:94:rmp} that there should be a
universal sheet resistance for the superconductor-insulator transition
and that it might well be just $h/(4e^2)$. Values that agree with this,
by a factor of two or three at worst, have not only been found in
studies of quench-condensed films
\cite{jaeger:86:gallprb,haviland:89:2dscprl,jaeger:89:scprb}, but also in
regular arrays of ultrasmall Josephson junctions 
\cite{geerligs:89:jjaprl,chen:92:ps}. \par
Our resistor samples are of course not regular arrays of Josephson
junctions, and definitely not homogeneous, but granular. The average
grain size is supposedly a few nanometres (see
\ref{comm:grainsize}). The superconducting coherence length, on the
other hand, should be around 40\,nm for  niobium \cite{auer:73:prb},
and since this is much larger than the grain size, the distinction
between an homogeneous and a granular film should vanish
\cite{liu:93:sitprb}. 
\subsection{No gate effect}
  \label{subsec:gatefailures}
At this point in the investigations, it remained to show that the
observed feature that we have boldly called ``Coulomb blockade'' so far
actually was caused by charging effects. To prove this, one has to be
able to modify the current-voltage characteristics with the voltage on a
capacitively coupled gate. In Chandrasekhar's experiment
\cite{chandrasekhar:94:jltp}, side gates close to their short wires were
sufficient to produce a gate effect. The indium oxide they used showed
``the presence of large grains'' \cite{chandrasekhar:91:prl}, and they
noted that ``only one or perhaps two segments (...) [were] present'' in
their samples. \par
\begin{figure}\centering
\epsfig{file=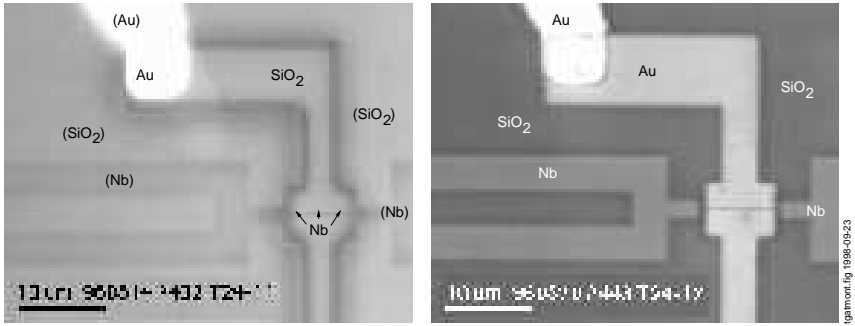,width=\textwidth}
\caption[Anodised Nb wire with top gate (no gate effect)]{%
Anodised wire with top gate, one of the geometries 
that we tried and that did not
give a gate effect. The optical micrographs show a 
sample after preparation
of the anodisation mask (left) and after evaporation of the gate and
liftoff (right). Materials in parentheses are seen through the PMMA mask.
}
\label{fig:tgatmont}
\end{figure}

Since  we expect to have many more segments, due to a smaller initial
grain size and the additional serration of the interface between niobium
and the oxide, it is not surprising that we did not see a gate effect
with side gates. Even top gates, with a much better capacitive coupling
to the strips, did not give a gate effect. Figure~\ref{fig:tgatmont}
shows a top gated sample before anodisation and after evaporation of the
gates.\par
\begin{figure}\centering
\epsfig{file=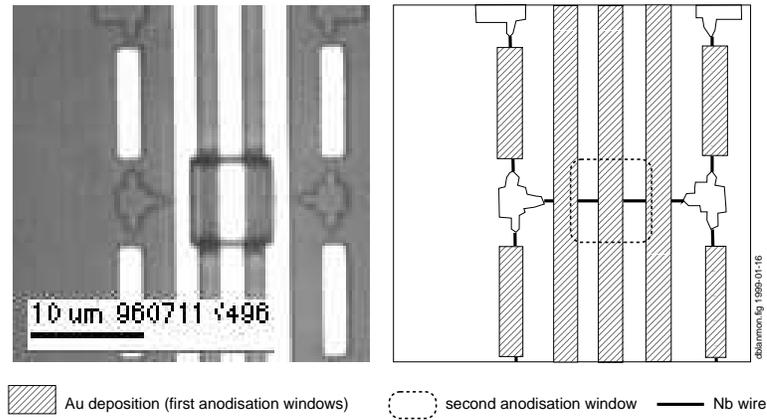,width=0.9\textwidth}
\caption[Double anodisation sample (no gate effect)]{%
Sample with a double anodisation mask, another device developed
in pursuit of a gate effect (yet another failure). In a first step, 
the niobium wires
had been anodised where the subsequently deposited gold marks the
windows. After removal of the first mask, a second mask was created and
opened as shown, for anodisation of the two pieces of niobium
wire exposed in the window and between the gold strips. The sample's
current-voltage characteristics could \textbf{not} be influenced with a gate
voltage applied to the central gold strip.
}
\label{fig:dblanmon}
\end{figure}

We also tried an approach with two anodisation processes, which can be 
seen as a step towards the samples in SET-like geometry. Even at the
length scales of a few micrometres, which can be seen in
fig.~\ref{fig:dblanmon}, there was no gate effect detectable.\par
\section{Samples in SET-like geometry}
  \label{sect:setlikesamples}
\subsection{Fabrication technique}
  \label{subsec:shadowmin}
\begin{figure}\centering
\epsfig{file=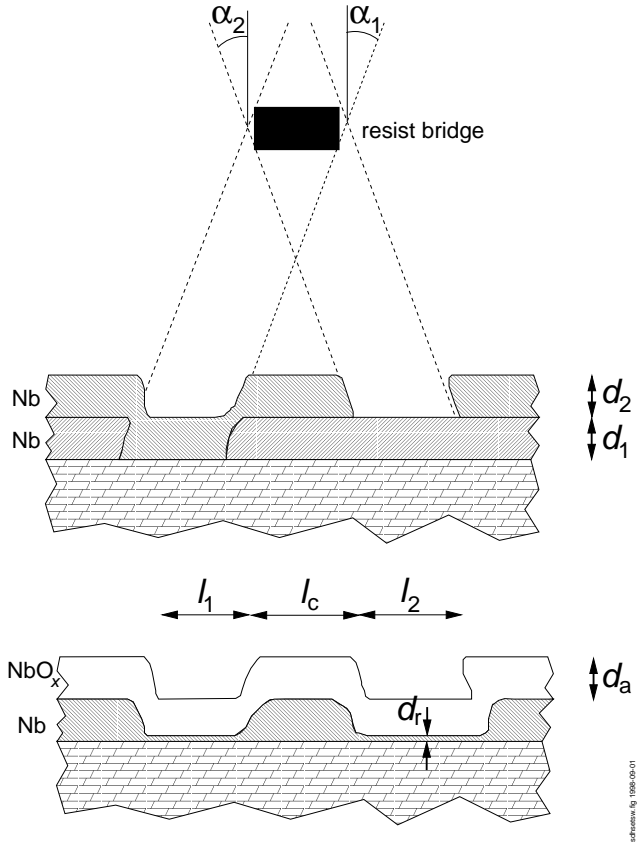,width=0.75\textwidth}
\caption[Shadow evaporation and anodisation technique for
the fabrication of SET-like structures]{%
Shadow evaporation and anodisation technique for
the fabrication of SET-like structures. Anodisation creates an oxide
layer of thickness $d_\mathrm{a}$, thinning the weak links to a
thickness $d_\mathrm{r}$. Neglected in this sketch are the deposition of
material on the bridge during the two evaporations, and the swelling of
the oxide film during anodisation.}
\label{fig:shdsetsw}
\end{figure}
\begin{figure}\centering
\epsfig{file=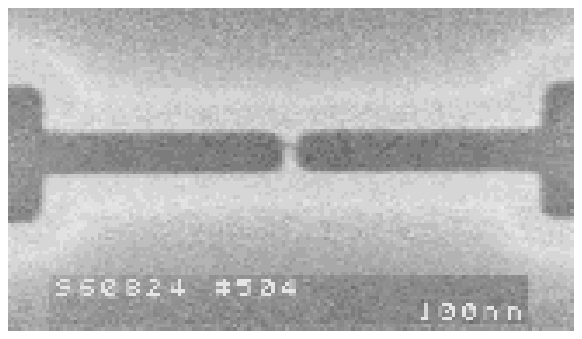,width=0.9\textwidth}
\caption[Mask for fabrication of SET-like devices by angular
evaporation]{%
Mask for fabrication of samples in SET-like geometry,
after pattern transfer. The darkest areas
are the substrate, oxidised silicon. Adjacent light areas show where the
organic resist support layers have been underetched
by oxygen RIE. The suspended germanium
bridge in the centre is here damaged by a tiny crack.
}
\label{fig:wealimsk}
\end{figure}
\begin{figure}\centering
\epsfig{file=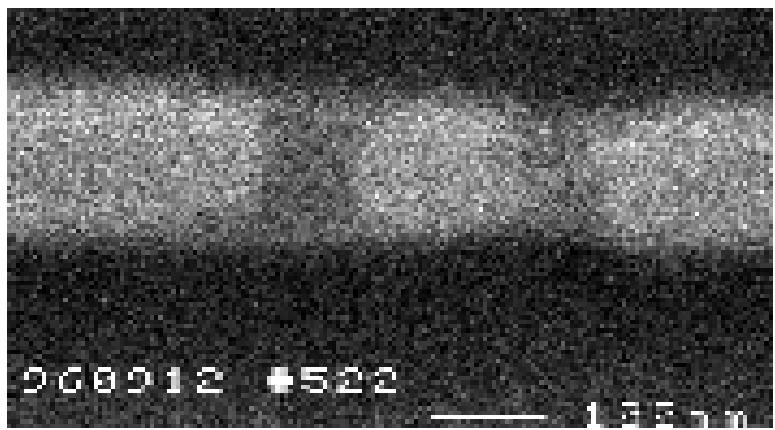,width=0.9\textwidth}
\caption[Variable thickness weak links in Nb wire]{%
Variable thickness weak links in a niobium wire made by double angle
evaporation. Scanning electron micrograph taken after Nb deposition and
liftoff and before further processing (anodisation).
}
\label{fig:thnsptnb}
\end{figure}

In order to reduce the number of grains participating in the electrical
transport and giving rise to the Coulomb blockade, we used the combined
shadow evaporation and anodisation technique illustrated in
fig.~\ref{fig:shdsetsw}. The corresponding mask is shown in
fig.~\ref{fig:wealimsk}, and a niobium film patterned by this technique
in fig.~\ref{fig:thnsptnb}.\par
The lengths $l_1\approx l_2$ of the weak links are defined by the
lithography as the width of the suspended bridge, while the island
length $l_\mathrm{c}$ is determined by the pattern overlap, and can be
made very small. In two subsequent niobium evaporations, film
thicknesses $d_1\approx d_2$ are deposited. It is critical to have the
difference between $d_1$ and $d_2$ as small as possible, so that both
thin spots are further thinned out to the same residual film thickness
$d_\mathrm{r}$. In the simplified sketch of fig.~\ref{fig:shdsetsw}, we
have neglected the deposition of material that leads to an asymmetry in
the shape of both weak links, and the swelling of the film during
anodisation (in reality, $d_\mathrm{a}\approx
3\cdot(d_1+d_2-d_\mathrm{r})$).\par 
The scanning electron micrograph of the mask (fig.~\ref{fig:wealimsk})
shows a problem encountered occasionally with four layer resist, namely
tensile stress in the germanium layer that can lead to cracks if only a
few individual bridges are created at large distances from each
other. If the pattern design allows it, additional holes can be made in
the mask near the intended bridge to release some of the stress. Also,
increasing the Ge thickness can remedy a lot of these problems.\par
Figure~\ref{fig:thnsptnb} shows the two niobium thin spots before
anodisation and deposition of the gate electrode on top of the
structure.\par 
The samples were rinsed carefully with deionised water immediately after
the anodisation. Ultrasonic excitation at this stage appeared to destroy
many samples. A gentle surface ashing made with oxygen RIE provided
sufficient adhesion of the gold gates to the samples. 50\,nm of gold
were evaporated to guarantee a continuous film even where it ran over
the strip edges. These gates had a two terminal resistance of about
60\,$\Omega$ at  a length of 160\,$\mu$m and a width of 8\,$\mu$m,
narrowing down to 3\,$\mu$m in the immediate vicinity of the anodised
area.\par
\begin{figure}
\begin{minipage}[c]{0.5\textwidth}
\epsfig{file=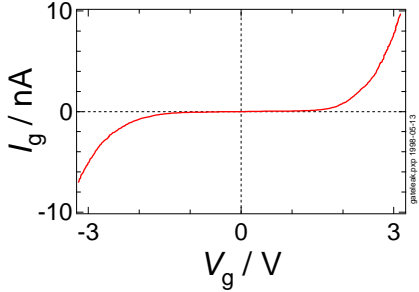,width=\textwidth}
\end{minipage}
\begin{minipage}[c]{0.48\textwidth}
\caption[IVC for gate leakage]{%
Current-voltage characteristic for the leakage current between an
anodised
double weak link structure and its top gate, taken at about 30\,mK.
}
\label{fig:gateleak}
\end{minipage}
\end{figure}

We tested the gate insulation by measuring the current-voltage
characteristics between gate and source, as shown in
fig.~\ref{fig:gateleak}. 
In most samples, the insulation was found to be this good. Up to a gate
voltage of about 1.5\,V, the gate current was too small to be measured,
which translates to an insulation resistance of at least
30\,G$\Omega$. For higher gate voltages, a measurable leak current set
in. In some samples, this current flow started at voltages as low as
0.5\,V. The leakage was easily detected by the shift of the measured
IVC, and these samples were dismissed.\par
\subsection{IVC and gate effect}
\begin{figure}\centering
\epsfig{file=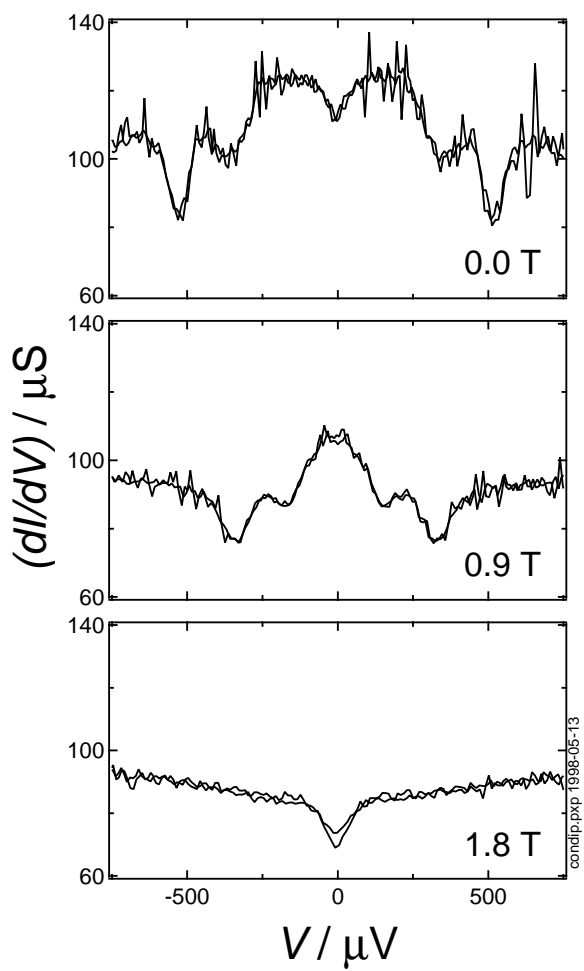,width=0.7\textwidth}
\caption[Coulomb blockade in an SET-like anodised
structure]{%
Coulomb blockade in an SET-like anodised structure. As superconductivity
was suppressed by an external magnetic field, the off-zero-bias
conductance peaks disappeared, and a Coulomb blockade for single
electrons remained. Data were taken at $T\approx 40$\,mK.}
\label{fig:condip}
\end{figure}

All sufficiently anodised samples showed a Coulomb blockade at
millikelvin temperatures. Sometimes it manifested itself just as a
hardly perceptible dip in the differential conductivity at zero bias. In
other samples, we observed a rather sharp blockade, in several samples
even at 4.2\,K. In neither of these cases, very weak or very strong
blockade, did we manage to modulate the IVC with a gate voltage. Only
when the IVC in the case without external field had a more complicated
structure, as in the top panel of fig.~\ref{fig:condip}, was there a
measurable gate effect.
As an external magnetic field was ramped up, the series of conductivity
dips and peaks moved towards zero bias and finally merged into a single
conductance dip there, obviously resulting from the Coulomb blockade of
single electron tunnelling. The off-zero-bias structures are probably
related to the superconducting properties, and the zero bias conductance
dip in the field-free case may be indicating the Coulomb blockade of
Cooper pair tunnelling \cite{haviland:91:zfpb}.\par
To measure the gate response of the sample's IVC, we biased it at a
number of practically constant currents and swept the gate voltage up
and down with a frequency of about 8\,mHz. The voltage between drain and
source was registered in the usual way with the low noise amplifiers in
the cryostat top box. The gate voltage was applied via a voltage
divider, and diagnostic leads made it possible to verify that the gate
voltage was actually present on the sample.\par
\begin{figure}\centering
\epsfig{file=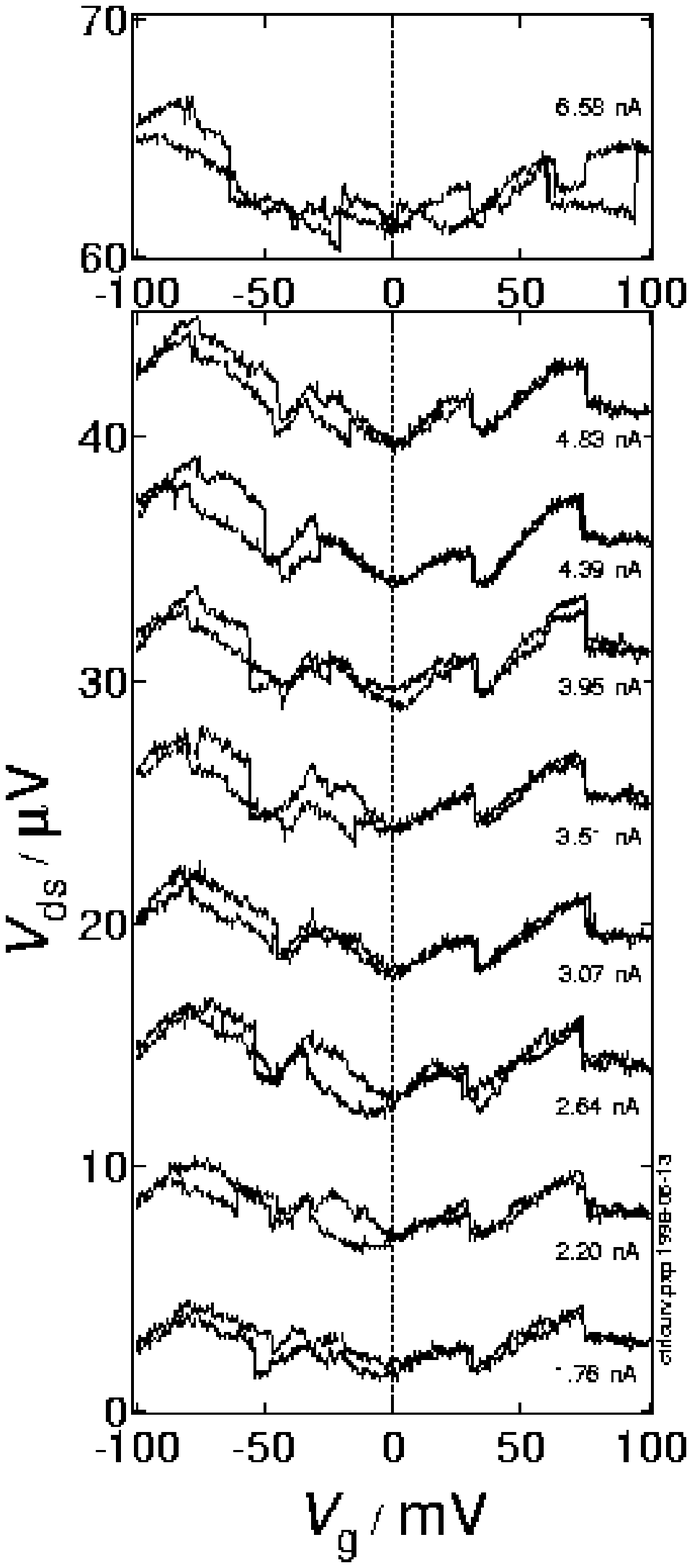,width=0.5\textwidth}
\caption[Control curves for an SET-like anodised
structure]{
Control curves for an SET-like anodised structure. The drain-source
voltage $V_\mathrm{ds}$ at different current bias points oscillated as
the voltage $V_\mathrm{g}$ applied to the gate on top of the structure
was ramped up and down (once per each bias point). 
Data were taken at dilution refrigerator base temperature below
30\,mK, and superconductivity was suppressed by an external field of
2\,T.}
\label{fig:ctrlcurv}
\end{figure}

In the superconducting state, no gate modulation was discernible from
all noise without further analysis. For the measurement shown in
fig.~\ref{fig:ctrlcurv}, we suppressed superconductivity completely by
applying an external field of 2\,T.
The causal influence of the gate voltage is obvious. These control
curves are far from the almost perfect sine curves a single electron
transistor gives, but are typical for systems of multiple tunnel
junctions (MTJ). The large period of the oscillations of about 50\,mV in
gate voltage is also typical for an MTJ system. If we just had a two
junction SET, this period would indicate an unreasonably low total
island capacitance of only 3\,aF.\par
\begin{figure}\centering
\epsfig{file=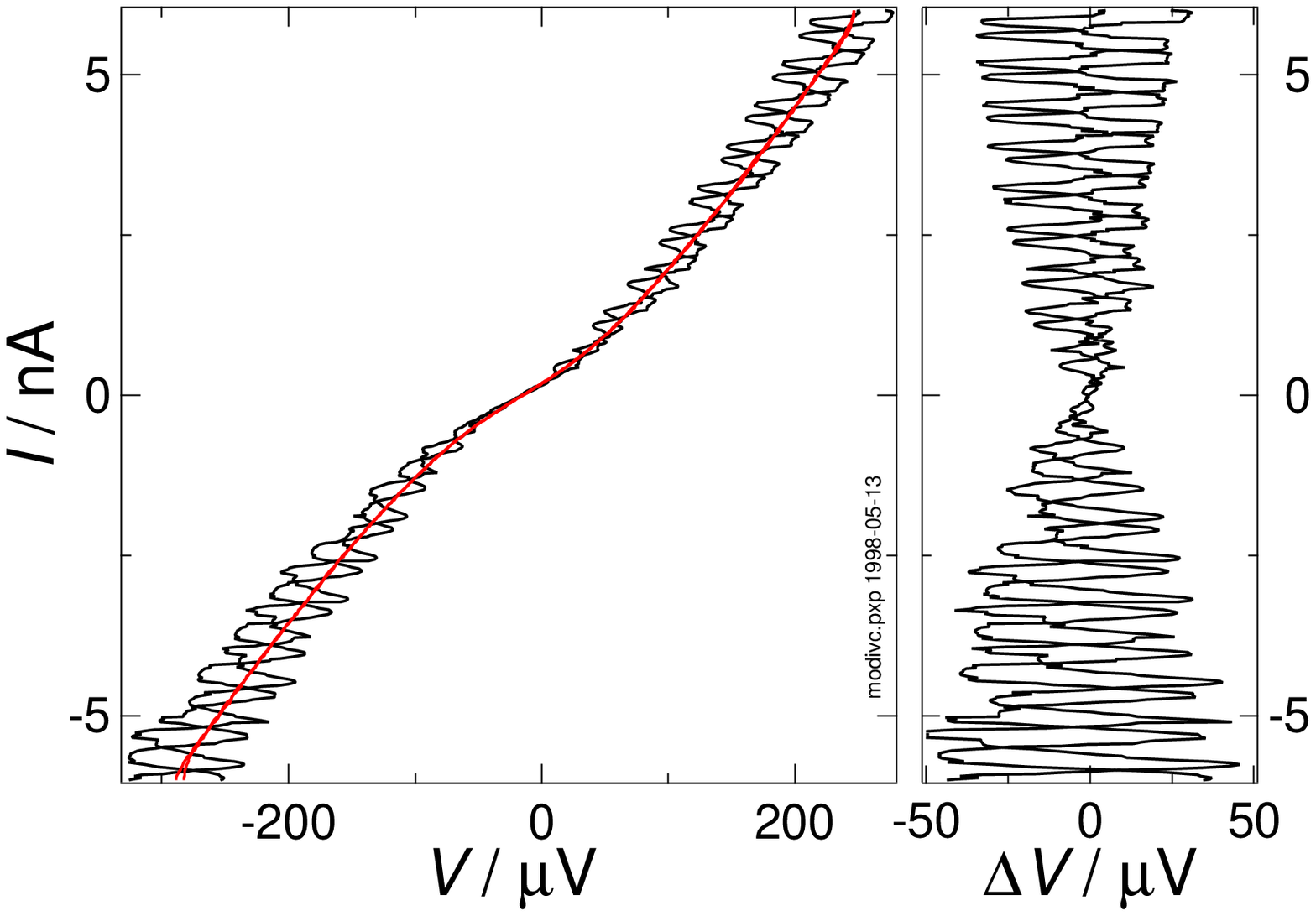,width=0.96\textwidth}
\caption[IVC of an SET-like anodised structure with gate
modulation]{%
IVC of an SET-like anodised structure with gate
modulation. The rms value of the sine-modulated gate voltage was 120\,mV
and its frequency 40\,times that of the bias sweep. The right panel
shows the deviation of $V$ from the unmodulated IVC in the left
panel. Data were taken at a temperature below 50\,mK, and no external
magnetic field was applied.}
\label{fig:modivc}
\end{figure}
Instead of showing the gate effect by recording modulation curves at
fixed bias points, one can map out the whole modulation range of the
current-voltage characteristics at once \cite{fulton:87:charprl}
by recording the IVC while simultaneously oscillating the gate voltage
at a higher frequency. This is very useful for a quick verification, and
for a gate leak test. If the gate is leaking current, the IVC will be
shifted as a whole by the oscillating gate voltage, whereas a true gate
effect is indicated by a behaviour as in fig.~\ref{fig:modivc}.
The voltage swing $\Delta V$ vanishes at zero bias. In the right panel,
$\Delta V$ is plotted separately. It has been calculated as the
difference between the modulated IVC in the left panel and an IVC
recorded in the same configuration with the gate voltage amplitude
zeroed.\par 
\subsection{Temperature dependence of the gate effect}
\begin{figure}\centering
\epsfig{file=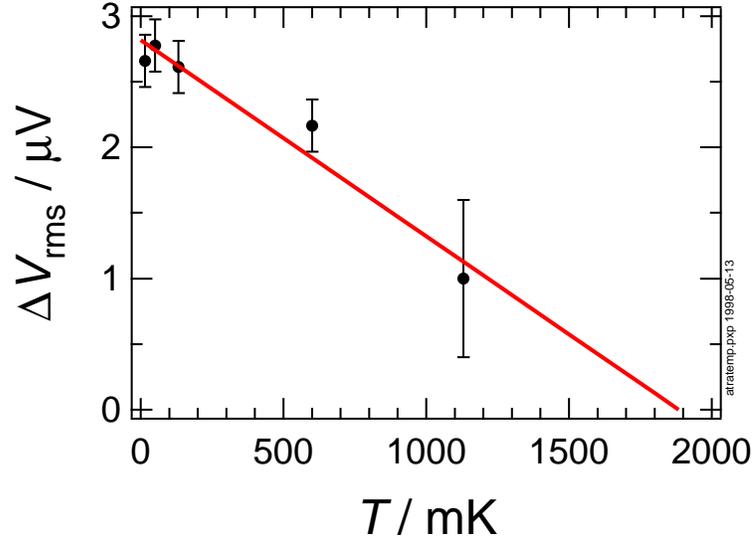,width=0.9\textwidth}
\caption[Temperature dependence of gate modulation]{%
Temperature dependence of the voltage swing in an 
anodised weak link sample with SET-like geometry.
} 
\label{fig:atratemp}
\end{figure}
An important parameter for a single electron device is the temperature
$T^\ast$ at which the gate effect vanishes. It should give information
on the energy scale at which the Coulomb blockade occurs. For single
electron transistors made by angular evaporation, Wahlgren
\cite{wahlgren:94:xjobb} gives the approximate relation between $T^\ast$
and the maximum voltage swing $\Delta V_\mathrm{max}$
\begin{equation}
  \label{eq:swingvanish}
  e\Delta V_\mathrm{max}\approx 4k_\mathrm{B}T^\ast.
\end{equation}
This relation seems to hold even for the very small SET made by Nakamura
et al. \cite{nakamura:96:set100k} that had a $T^\ast$ of more than
100\,K. In both these cases, the junction resistances were considerably
above $R_\mathrm{K}$.\par
The temperature dependence of the voltage swing in one of our samples in
SET geometry is shown in fig.~\ref{fig:atratemp}. For the low
temperature values up to 600\,mK, the swing $\Delta V$ between gate
modulated and unmodulated IVC was analysed in the bias region between $-
7.5\,\mathrm{nA}$ and $7.5\,\mathrm{nA}$, and the root-mean-square value
is plotted together with the estimated uncertainty. The high temperature
value at 1130\,mK was determined by comparison of the voltage swings at
$\pm 7.5\,\mathrm{nA}$ in two measurements with a square shaped gate
signal, and by normalisation to the low temperature amplitude.\par
As expected, the amplitude of the modulation decreased with
temperature. A simple extrapolation of the few data points in
fig.~\ref{fig:atratemp} gives a $T^\ast$ somewhere between 2\,K and
3\,K.\par 
Instead of a maximum voltage swing of about 100\,$\mu$V 
at low temperature, we
would have expected a value several times higher for the measured
$T^\ast$ according to eq.~(\ref{eq:swingvanish}). The comparatively low
resistance of these samples may be responsible for the suppression of
the IVC modulation amplitude.\par
\section{Conclusion: anodised niobium nanostructures}
We have demonstrated that anodic oxidation of nanofabricated niobium
thin film wires can be used to produce resistors of several hundred
kiloohms on a length of ten micrometres. This technique is intrinsically
limited by the onset of a Coulomb blockade when the sheet resistance
per square
exceeds the quantum resistance 6\,k$\Omega/\Box$. The anodised wires show
transport properties typical of an array of ultrasmall Josephson
junctions, where superconducting effects coexist with charging
effects. The low temperature 
current-voltage characteristics
of these samples show a
superconductor-insulator transition where the degree of anodisation is
the tuning parameter.\par
Placing short anodised areas with the aid of lithographic techniques,
one can fabricate transistor-like samples whose 
current-voltage characteristics can be modulated by
a gate voltage. This gate effect disappears, however, already at a
temperature of a few Kelvin, which might turn out to be the limiting
factor for potential applications.\par
\chapter{Noise in a single electron transistor}
\label{chap:noise}
\section{Noise model}
\label{sect:noisemodel}
\begin{figure}[tbp]\centering
\epsfig{file=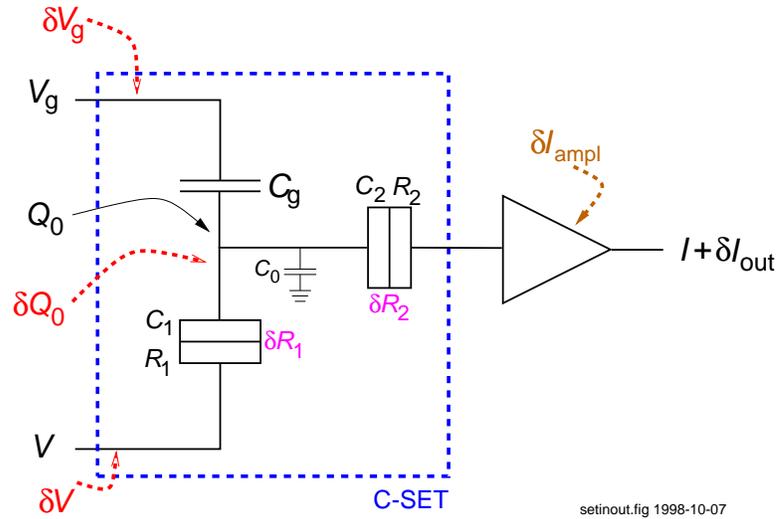,width=0.9\textwidth}
\caption[C-SET: Input and output quantities, parameters, noise sources]{
Schematic diagram of a 
capacitively coupled single electron transistor
(C-SET), consisting of two
junctions with resistances $R_1$, $R_2$ and (very low) capacitances
$C_1$, $C_2$,
and of a gate capacitor with $C_\mathrm{g}$. 
In a voltage biased configuration, the input quantities
voltage $V$, gate voltage $V_\mathrm{g}$ and offset charge
$Q_0$ due to background charges
result in a current $I$ at the output of the amplifier.
Its noise $\delta I$ is the result of noise at the input of the SET,
especially fluctuations in the background charge $\delta Q_0$, of the
amplifier noise $\delta I_\mathrm{ampl}$, and of fluctuations in the
junction resistances $\delta R_i$.
In this simplification, the triangular symbol stands for any generic
amplifier. The output signal can actually be a voltage that is
proportional to the current being measured, and in that case, the noise
added by the amplifier will be a voltage noise.}
\label{fig:setinout}
\end{figure}

The model we use for the single electron transistor's input and output
noise is laid out in fig.~\ref{fig:setinout}.
The capacitively coupled SET has three input quantities, all of which
might be affected by noise: the bias voltage $V$, the gate voltage
$V_\mathrm{g}$, and the background charge $Q_0$. However, due to careful
filtering of the leads (see~\ref{subsec:shieldfilter}), we can neglect
the noise in both bias and gate voltages, which leaves the background
charge as the only noisy input quantity.\par
Background charge noise is generally associated with charge
traps \cite{tavkhelidze:97:isec,tavkhelidze:98:lfnjap}. 
Their exact nature and location is still disputed, they may 
actually be at some distance from the island
\cite{zorin:96:bgnoiseprb,zimmerman:97:dist}.
Whenever these traps, located sufficiently close to the C-SET's
island, change their charge state, the offset charge of
the island is changed, and the transport characteristics of the SET are
modified. This is analogous to the situation in SQUIDs, where background
fluxons frozen in the superconductor are responsible for a large portion
of the low frequency noise. Like in the SQUID analog, the background
charge noise in SET appears to be largely a materials problem. However,
no significant improvements have been made yet by varying the materials
for SET \cite{ji:94:traps,bouchiat:cpem96}.
Whereas in SQUID, the low frequency noise contribution from
background fluxons can be eliminated by a bias reversal scheme
\cite{foglietti:86:apl}, no such scheme has yet been devised for SET,
making improvements on the materials and circuitry more urgent than
ever. \par
The second contribution to LFN in SQUIDs comes from fluctuations of the
junction resistances $R_{1,2}$. It appears obvious that such a
fluctuation could also occur in an SET and contribute as an output noise
source. The observation of LFN caused by resistance fluctuations has
been reported recently \cite{krupenin:98:stackjap}. Speculative to date is
a correlation between background charge noise and junction resistance
fluctuations \cite{starmark:98:condmat}, which might be caused by
charged defects in the tunnel barriers, affecting both the barrier
transparency and the island charge upon charging.\par
Any combination of bias voltage $V$, gate voltage $V_\mathrm{g}$ and
offset charge $Q_0$ defines an operating point of the SET
with a certain output current $I$ and gain $\partial I/\partial
Q_\mathrm{g}$. Background charge fluctuations with spectral noise
density $S_{Q_\mathrm{g}}(\omega)$ and resistance fluctuations with
spectral noise densities $S_{R_i}(\omega)$ will then add up, taking the
former's contribution up to the second order in the gain and neglecting
correlation, to a current noise
\begin{equation}
  \label{eq:noisemodel}
  S_{I_{Q,R}}(\omega)=
  \left(
  \left(\frac{\partial I}{\partial Q_\mathrm{g}}\right)^2 +
  \frac{\alpha}{4} \left( e
  \frac{\partial^2 I}{\partial Q_\mathrm{g}^2}\right)^2\right)
  S_{Q_\mathrm{g}}(\omega)+
  \sum_{i=1}^2 \left(\frac{\partial I}{\partial R_i}\right)^2
  S_{R_i}(\omega),
\end{equation} 
where $\alpha$ is an expansion coefficient describing contributions in
second order in gain that we will come back to below 
(see page \pageref{calc:alpha}).\par
Another fundamental noise source is shot noise \cite{schottky:18:schrot}
generated in the tunnel junctions. In our case, it is negligible up to
bias voltages well above the Coulomb blockade threshold, so that the
noise generated by both junctions can be regarded as uncorrelated
\cite{korotkov:96:sensiapl}. For linear arrays with $n \gg 2$ junctions, the
shot noise suppression resulting from this uncorrelation has recently
been measured \cite{wahlgren:98:thesis}. The shot noise of our
single electron transistor at high bias is then
\begin{equation}
  S_{I_\mathrm{e}}(I)=\frac{2eI}{2}=eI. 
\end{equation}
Measurements on an SET require an amplifier that inevitably adds noise
to the output signal. The total noise measured at the output of the
amplifier is the sum of the amplifier noise, the amplified shot noise,
the amplified SET output noise caused by resistance fluctuations, and
the amplified background charge input noise of the SET.  \par
Mind the consistently inconsistent use of the term \emph{charge noise} in
the literature. We will try to be stringent here and either speak of
\emph{background charge noise}, alias \emph{input charge noise}, 
or of \emph{input
referred noise}, alias \emph{charge equivalent noise}. The latter is
defined in analogy to \emph{flux noise} in SQUID, which is just the
spectral density of input flux fluctuations that would cause the
observed noise at the system's output under the assumption that no other
noise contributions arise from the SQUID itself or the amplifier
electronics, in other words, for an ideal system. 
\par
Similarly, charge
equivalent noise is the measured noise at the output of the system (SET
plus amplifier), divided by the amplifier gain and by the SET's gain.
Since an SET is useless without an amplifier, there is a certain
justification for using this quantity as a figure of merit. It is
customary to compare the noise  of SET systems at the frequency
$f=10$\,Hz, where the LFN is pronounced in almost all samples, yet
measurement times are still resonably short. \par
It appears that the evaluation at 10\,Hz has no drawbacks over the
approach regarding integrated quantities in the frequency band (50\dots
100)\,Hz \cite{starmark:98:condmat}.\par
Having the lowest charge
equivalent noise at 10\,Hz is a matter of a chase between single
electronics laboratories. The current record is held by the 
PTB-MSU collaboration
that achieved $2.5\cdot
10^{-5}\,e/\sqrt{\mathrm{Hz}}$ 
\cite{krupenin:98:stackjap} in
aluminium devices of stacked design that minimised the contact area
between substrate and island. 
The previous record of $7\cdot 10^{-5}\,e/\sqrt{\mathrm{Hz}}$ 
had also been set in a multilayer device \cite{visscher:95:setapl}.
\par
Other characteristic figures one might like to compare are the corner
frequency of the LFN and the white noise figure. These quantities,
however, say more about the amplifier than about the SET, and since we
regard the SET as the device under test and the amplifier as an
unavoidable complication, we will not work with these figures.\par
In the evaluation of the measurements presented below, we introduce the
\emph{differential charge equivalent noise}
(see~\ref{sect:gaindependence}), which is a quantity describing only
properties of the SET without involving amplifier characteristics at
all.\par 
\section{Sample fabrication and characterisation}
  \label{sect:noisefabrication}
\subsection{Fabrication}
\begin{figure}[bpt]\centering
\epsfig{file=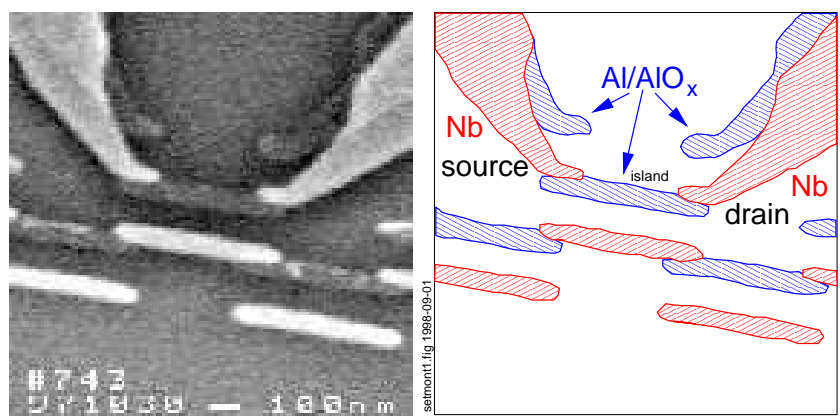,width=\textwidth}
\caption[SET with Nb leads and Al island]{%
Single electron transistor with niobium leads and aluminium
islands. Left: scanning electron micrograph, right: artistic
interpretation.}
\label{fig:setfoto1}
\end{figure}

The sample, whose scanning electron micrograph is shown in
fig.~\ref{fig:setfoto1}, was made by the shadow evaporation technique
(see \ref{subsec:niemdol}), using the four layer resist described in
\ref{subsec:fourlayerresist}. Details about the fabrication process can
also be found in the recipe appendix (\ref{rec:fourlayerprep}). \par
Substrate material was silicon from a two inch wafer, thermally
oxidised to a depth of approximately 1\,$\mu$m. The base material for
the junction structure was aluminium, which here formed the island of
the transistor. The aluminium can be seen constituting the structures
with relatively poor contrast in fig.~\ref{fig:setfoto1}. The film
thickness was 20\,nm, and the deposition was carried out under an angle
of $-21^\circ$ to the substrate normal.\par
\begin{figure}\centering
\epsfig{file=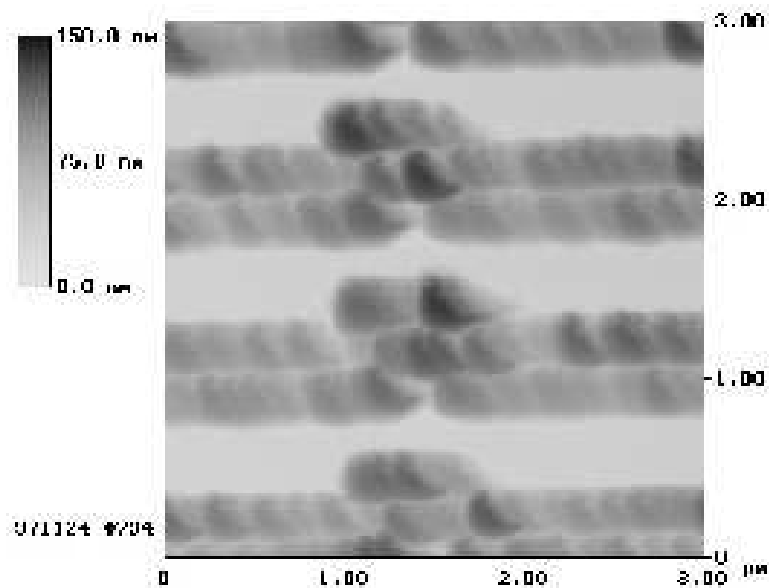,width=0.9\textwidth}
\caption[AFM image of an array of SET test structures]{%
Atomic force micrograph (AFM) of an array of SET test structures,
demonstrating the coarse grains of the aluminium film
(upper strips) compared to the smoother niobium (lower) strips.
}
\label{fig:setarafm}
\end{figure}

The primary reason for choosing aluminium as the base electrode material
was the evaporation source available at this time, an effusion
cell. This cell only delivered an evaporation rate of 3\,nm per minute,
so that the evaporation of the bottom metal layer took more than six
minutes. This resulted in a rather coarse grained structure of the
aluminium film, as demonstrated by the atomic force micrograph of
fig.~\ref{fig:setarafm}. 
It would have been impossible to cover a niobium bottom layer with a
sufficiently thin, yet homogeneous, aluminium layer as the fabrication
of Nb-AlO$_x$-Nb junctions would have required.\par
The aluminium layer, evaporated from a melt
whose nominal purity was 5N, was oxidised after the first
evaporation in the load lock of the UHV evaporation system, using
non-dehumidified air at a pressure of 8.8\,Pa for 20\,minutes, and after
two hours of pumping, the niobium counterelectrodes were deposited by
electron gun evaporation from a melt of nominal purity 2N8.\par
In retrospect, it must be regarded as unfortunate that we evaporated the
niobium right away without any pre-evaporations against the closed
shutter. Such practice \cite{harada:94:nbset} might have improved the
quality of the film by keeping the background pressure lower, due to the
chemisorption pumping (\emph{gettering}) by the fresh niobium layer. While
the evaporation system had a standard background pressure of
$3\cdot 10^{-7}\,\mathrm{Pa}$ without pre-evaporations, it went as low
as $5\cdot 10^{-8}\,\mathrm{Pa}$ after the deposition of a fresh niobium
layer and about an hour of cooling. During the niobium evaporations, the
pressure indicator showed an increase to approximately one hundred times
the base pressure. A further complication was that the niobium was not
deposited in one continuous operation, but instead opening the shutter
for two seconds and then closing it for eight seconds a total of five
times, giving a total niobium film thickness of approximately 20\,nm
(with an accuracy of several nm due to the rate uncertainty and
difficulty in exact shutter timing). This pulsed evaporation procedure
was intended to minimise the heat load to the substrate and the resist,
and thereby to minimise grain size and prevent resist damage. We had
seen thermally damaged resist structures earlier in this series of
experiments, but later found that we could attribute these problems to
incorrect resist fabrication caused by a faulty baking temperature
control (see \ref{subsec:fourlayerresist} and \ref{rec:fourlayerprep} 
for details on correct resist fabrication). The pulsed evaporation
procedure was later abandoned. \par
\par
Figure~\ref{fig:setfoto1} shows the niobium film, deposited under an
angle of $+21^\circ$ to the substrate normal, as the structures giving
much better contrast than aluminium in the scanning electron
microscope. The excess island created by the shadow evaporation here
happens to form part of a linear array of tunnel junctions not related to
the investigations described here (as a matter of fact, it was
broken). The gate electrode is situated far outside the imaged
detail. \par 
Since SEM inspection regularly damages ultrasmall, high ohmic tunnel
junctions, fig.~\ref{fig:setfoto1} in fact shows a transistor nominally
identical to the sample characterised below. Two chips (with four SET
structures each) on a contiguous piece of substrate were processed
simultaneously, and the SEM image shows the structure on the second chip
corresponding to the sample under consideration on the first chip. \par
Apart from samples with niobium electrodes and aluminium islands as
shown in fig.~\ref{fig:setfoto1}, even such with aluminium electrodes and
niobium islands were made. We have, however, no useful noise measurement
data from these samples.
\subsection{DC characterisation}
Unless stated otherwise, the DC characterisation measurements of the
sample were carried out in the dilution refrigerator mentioned in
\ref{subsec:cryogenics}, at the base temperature of $(30\pm
5)\,\mathrm{mK}$ which it reached at the time of these measurements. 
Contrary to the general statement in \ref{subsec:shieldfilter}, the
shielded room was partially dismounted during these measurements.
\subsubsection{Normal conducting state}
The transistor's serial resistance, i.\,e. the sum of both junctions'
resistances 
\begin{equation}
  R_\mathrm{T}=R_1+R_2,
\end{equation}
was measured between room temperature and liquid helium temperature one
week after the fabrication was completed (i.\,e. after removal of the
sample from the UHV system). From the room temperature value of
$(125\pm 5)\,$k$\Omega$, $R_\mathrm{T}$ rose upon cooling to 4.2\,K to
$(165\pm 8)\,$k$\Omega$ at high bias. Around zero bias, the differential
resistance increased to about 215\,k$\Omega$ due to the Coulomb
blockade.\par
The sample appeared to be rather stable against ageing and thermal
cycling. A first set of measurements was started when the sample was one
month old, the second measurement campaign, from which all data
presented here stem, another two months later. During this time, no
significant changes of the electrical characteristics were observed.\par
\begin{figure}\centering
\epsfig{file=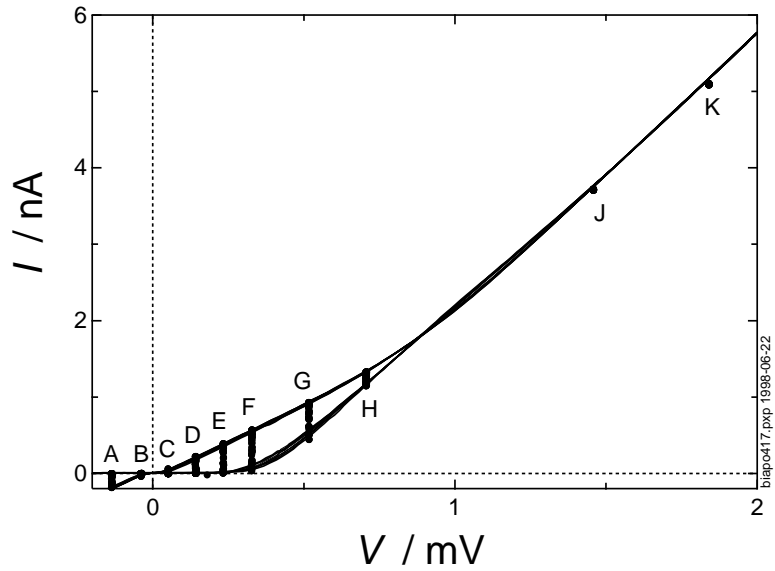,width=\textwidth}
\caption[IVC and noise measurement points]{%
Current-voltage characteristics of the single electron transistor at
base temperature ($\approx 30$\,mK) in the normal conducting state, for
minimum and maximum Coulomb blockade. The letters indicate the voltage
bias points for the noise measurements.
}
\label{fig:biapoinv}
\end{figure}

The $I$-$V$ characteristics in the normal conducting state, as shown in
fig.~\ref{fig:biapoinv}, were taken in an externally applied magnetic
field of 5\,T.
Absence of the Coulomb staircase in these characteristics indicates that
the two junctions were quite similar. This is corroborated by the
observation that the spread in transistor resistances among the four
double junctions on the chip was less than 20\%,
even though they may have had slightly different dimensions due to the
proximity effect and non-identical surroundings (the junction patterns
themselves were nominally identical), and because two of them had
aluminium islands, while the other two had niobium islands. \par
The total capacitance of the island, i.\,e. the sum of the junction
capacitances $C_{1,2}$, the gate capacitance $C_\mathrm{g}$ and the
capacitance to ground $C_0$,
\begin{equation}
  C_\Sigma=C_1+C_2+C_\mathrm{g}+C_0,
\end{equation}
was determined at dilution refrigerator base temperature using the
offset voltage analysis method described by Wahlgren et al.
\cite{wahlgren:95:prb,wahlgren:98:envprb}. From an extrapolated offset
voltage at zero bias of $V_\mathrm{off}^0=(325\pm 15)\,\mu$V, we found
an island capacitance
\begin{equation}
  C_\Sigma=\frac{e}{V_\mathrm{off}^0}=
  (0.49\pm 0.02)\,\mathrm{fF}.
\end{equation}
The gate capacitance was rather low because of the large distance to the
gate (since a dedicated gate electrode was damaged, another contact had
to substitute it). From a gate voltage period of the current at fixed
bias voltage of
$V_\mathrm{p}=(0.512\pm 0.008)\,$V, we found a gate capacitance
\begin{equation}
  C_\mathrm{g}=\frac{e}{V_\mathrm{p}}=
  (0.313\pm 0.003)\,\mathrm{aF}.
\end{equation}
See fig.~\ref{fig:biasmods}, top right, where the original data from
this measurement were used to illustrate the principle of SET operation.
\subsubsection{Superconducting state}
  \label{subsubsec:dccharsup}
In the superconducting state, at base temperature without any external
magnetic field applied, twice the superconducting energy gaps of both
the island and the source/drain materials are added to the Coulomb
gap. An analysis is best done by looking at the peaks in the
differential conductance plotted versus bias voltage. Upon variation of
the gate voltage, we found a minimum distance of the peaks of
$2\Delta_\mathrm{Al}+2\Delta_\mathrm{Nb}=(850\pm 20)\,\mu$eV. 
Assuming
$\Delta_\mathrm{Al}=(190\pm 10)\,\mu$eV, this means
$\Delta_\mathrm{Nb}=(235\pm 15)\,\mu$eV. While clearly suboptimal, this
nevertheless constitutes an improvement in gap value
over an aluminium top layer of
25\%,
even under the non-optimised 
deposition conditions as described above.\par
In this tunnelling experiment, we probed the niobium's superconducting
energy gap $\Delta$ in the electrodes near the junctions. We did not
measure the critical temperature 
of this sample directly since we did not 
have a suitable setup for measurements between approximately 1\,K
and 4.2\,K, i.\,e. between the ranges of a dilution refrigerators and an
unpumped Helium-4 bath. Though the critical temperature would have been
a nice figure of merit, the quantity really relevant for the performance
of a superconducting Coulomb blockade device is just the energy gap, and
it is known that very disordered niobium films no longer show the BCS
dependence of $T_\mathrm{c}$ on $\Delta$. For high disorder, the ratio
$2\Delta(T=0)/(k_\mathrm{B}T_\mathrm{c})$ falls below the BCS value of
3.53 \cite{camerlingo:85:disorder}. One should thus keep in mind that
our mentioning of $\Delta$ is somewhat more humble and relevant than
quoting a critical temperature.
\section{Noise measurement setup}
  \label{sect:bjoernsamplifier}
For the measurements of the current and its spectral noise density, we
used an amplifier designed and built by Bj\"orm Starmark. Detailed
\cite{starmark:97:isec} and very detailed \cite{starmark:98:lic}
information about this amplifier is available, and we will restrict
ourselves to summarising the amplifier properties here 
as far as they are
relevant to our investigations.\par
\subsection{Transimpedance amplifier}
\begin{figure}\centering
\epsfig{file=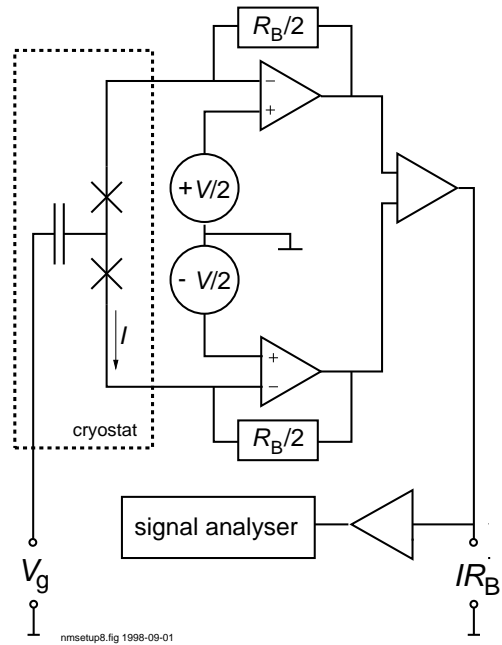,width=0.6\textwidth}
\caption[Setup for SET noise measurements]{%
Setup for noise measurements on a single electron transistor
(schematic, simplified).
The sample is voltage biased symmetrically with respect to ground, and
the resulting current measured with
help of a transimpedance amplifier placed on
top of the cryostat, inside a shielded enclosure. A signal analyser
performs a real time fast Fourier transform on the current signal
and calculates the noise spectral density.}
\label{fig:nmsetupx}
\end{figure}

The basic principle is that the amplifier translates an input current
into a relatively simply measurable output voltage. The amplifier's gain
has the dimension of an impedance, hence the name transimpedance
amplifier. Figure~\ref{fig:nmsetupx} shows a simplified schematic of the
noise measurement setup.
The sample, situated inside the cryostat, is voltage biased via two
operational amplifiers. Using the sketched biasing scheme, namely
symmetrical bias with respect to ground, reduced the effect of pickup
from external noise sources in the leads near the amplifier
\cite[3.5.6]{starmark:98:lic}. The output of the differential
preamplifier is actually the sum of the bias voltage $V$ and the
amplified current, which in turn is in very good approximation
$R_\mathrm{B}I$. \par
All amplifier parts were located at room temperature on top of the
cryostat, inside the (not completely) shielded room. The current signal,
i.\,e. the amplifier output from which the bias voltage has been
subtracted by virtue of a circuit element omitted from
fig.~\ref{fig:nmsetupx}, was fed to a digital voltmeter outside the
measurement cabin and to an HP\,35565 dynamic signal analyser. Unlike
in all the DC measurements described so far, this signal was 
not fed through shielded room feedthrough filters, but directly with a
coaxial cable.\par
The HP\,35565 performed a real time FFT and calculated the noise
spectral density of the current signal. We covered the frequency range
from 1\,Hz to 300\,Hz, averaging 25 spectra ending at 10\,Hz and 100
spectra each ending at 100\,Hz and 1\,kHz (evaluated up to 300\,Hz),
respectively, trading off accuracy against speed in the lowest frequency
region.\par
We made noise measurements at the bias voltage points labelled
\textsf{A} to \textsf{K} in fig.~\ref{fig:biapoinv}. At each bias point,
a series of 21 different gate voltages was applied, spanning about 1.6
modulation periods or elementary charges induced on the island by the
gate (see the top right graph of fig.~\ref{fig:biasmods} on page
\pageref{fig:biasmods}). At any bias and gate voltage pair, the noise
measurements took about five minutes, and though it would have been
better to have more data, the total measurement time was limited to
about eighteen hours by the lifetime of amplifier batteries.\par
\subsection{Amplifier noise}
  \label{subsec:amplinoise}
Over the relevant frequency range from 1\,Hz to 300\,Hz, the amplifier
noise was due to two contributions, the thermal noise of the feedback
resistors $R_\mathrm{B}$ at (room) temperature $T_\mathrm{B}$, and the
input equivalent voltage noise $e_\mathrm{n}$ of the operational
amplifiers. With $r_0=dV/dI$, the output impedance of the SET, the
amplifier noise reads
\begin{equation}
  i_\mathrm{n,ampl}(T,f)=\sqrt{2k_\mathrm{B}
      \frac{T_\mathrm{B}}{R_\mathrm{B}}+
      \frac{e_\mathrm{n}^2(f)}{2r_0^2} }.
\end{equation}
The Analog Devices AD\,743 ultralow noise BIFET
op-amp had an $e_\mathrm{n}$ of
5.5\,nV/$\sqrt{\mathrm{Hz}}$ (at 10\,Hz)
\cite{ad:ad743specs}, and thus we could, thanks to
the relatively high impedance of this particular sample, neglect the
second contribution. This left us with an amplifier noise entirely due
to the thermal noise in the feedback resistors, for which we used
$i_\mathrm{n,ampl}=(28\pm 2)\,\mathrm{fA}/\sqrt{\mathrm{Hz}}$ over the
whole frequency range in the following.
\section{Measurement results}
  \label{sect:gaindependence}
\subsection{Gain determination}
The single electron transistor's 
small signal gain can be expressed via its
transconductance and gate capacitance as
\begin{equation}
  \frac{dI}{dQ_\mathrm{g}}=
  \frac{dI}{dV_\mathrm{g}} \frac{dV_\mathrm{g}}{dQ_\mathrm{g}}=
  \frac{dI}{dV_\mathrm{g}} \frac{1}{C_\mathrm{g}}.
\end{equation}
We determined the transconductance $dI/dV_\mathrm{g}$ by numerical
differentiation of current and gate voltage data taken simultaneously
with the noise measurements. The sparsity of the gate voltage points (21
per bias point) introduced the dominating uncertainty in the
determination of the gain. Nevertheless, we preferred this random
uncertainty over introducing a systematic error by fitting some
analytical expression to the data. \par
In the early first measurement campaign mentioned earlier, we had
attempted to measure the gain directly. For this purpose, a small AC
component had been superimposed on the gate voltage, and the resulting
AC component of the current signal had been read out with a lock-in
amplifier. However, in order to attain a useful signal level, we had to
increase the gate AC component so much that the spectra were blurred by
numerous peaks at harmonics, subharmonics, and beat frequencies of the
AC signal and of 50\,Hz mains. Crosstalk between the leads in the
cryostat appears to be responsible, abetted by the low gate
capacitance. \par
\subsection{Current noise spectral density}
\begin{figure}\centering
\epsfig{file=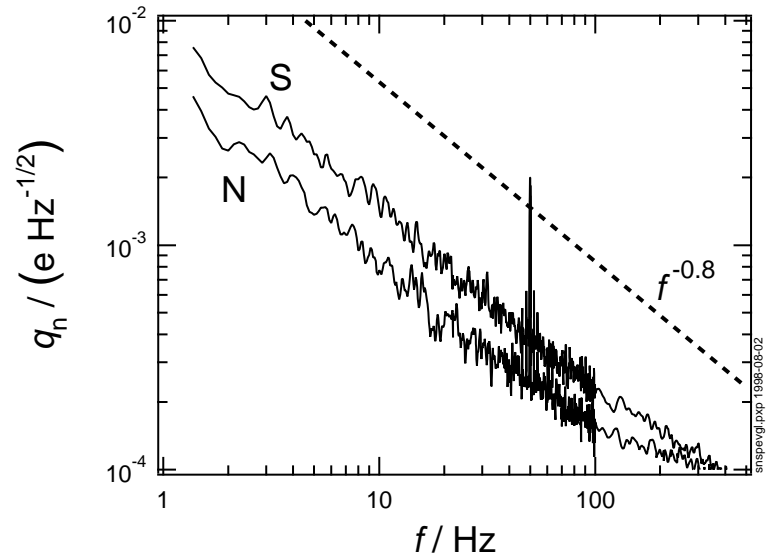,width=0.95\textwidth}
\caption[Input referred noise at maximum gain]{%
Input referred (charge equivalent) noise, at the bias and gate voltage
points giving maximum gain, for the normal conducting state (\textsf{N},
cf. fig.~\ref{fig:biapoinv}, point \textsf{F}) and superconducting state
(\textsf{S}), respectively. The dashed line indicates the frequency
dependence $i_\mathrm{n}\propto f^{-0.8}$.
}
\label{fig:snspevgl}
\end{figure}
In fig.~\ref{fig:snspevgl}, the input referred (charge equivalent) noise
of the SET is shown as a function of frequency at those two (bias and
gate voltage) points that gave the highest gains in the superconducting
and normal state, respectively. In both cases, the total current noise
at the output $i_\mathrm{n}$, including the amplifier noise
$i_\mathrm{n,ampl}$ (see \ref{subsec:amplinoise}), has been divided by
the respective gains:
\begin{equation}
  q_\mathrm{n}=i_\mathrm{n} \left(\frac{dI}{dQ_\mathrm{g}}
    \right)^{-1}.
\end{equation}
Though the maximum gain in the superconducting state was twice as high as
in the normal conducting state (3.4\,nA/e and 1.7\,nA/e, respectively),
the input referred noise does not differ significantly between the two
states. This is a first rough indication for the dominating influence of
input noise on the overall noise, which we shall quantify
below. \par
We further see the same frequency dependence $q_\mathrm{n}\propto
f^{-0.8}$, corresponding to a dependence $S_q\propto f^{-1.6}$ of the
spectral power density, in both the superconducting and normal
state. This same frequency dependence had been observed in 
a number of all-aluminium
SET on substrates from the same batch \cite{starmark:98:condmat}. \par
At 300\,Hz, we see the crossover from input dominated to output
dominated, namely amplifier (white), noise. Since the same amplifier
noise is divided by the higher gain in the superconducting case, the
input referred noise for the superconducting sample is  lower
than for a normal conducting one; this is  a consequence of the
input referral. The engineer's justification is that a higher gain
means a better signal-to-noise ratio, and this is exactly what the lower
input referred noise indicates.\par
In the following, we will concentrate on evaluating the noise at the
frequency $f=10\,$Hz. Instead of simply reading the spot value for the
noise spectral density from the measured raw data, we improved the
accuracy by producing linear fits to the bilogarithmic noise density
and frequency data in the ranges 3\,Hz to 10\,Hz and 10\,Hz to 30\,Hz,
respectively, and taking the average of the two fits at 10\,Hz (after a
plausibility check) as the measured noise value. The ranges were chosen
to stay above the region of data inaccuracy due to the limited
measurement time, and below the 50\,Hz mains peak.\par
When speaking of the net current noise
$i_\mathrm{n,net}$, as we will do in the following,
we mean the measured current noise $i_\mathrm{n}(f)$ from which the
amplifier noise $i_\mathrm{n,ampl}$ (flat) and the shot noise
$i_\mathrm{n,e}=\sqrt{eI}$ have been subtracted, the latter
only being significant for the two highest bias points well above the
Coulomb blockade threshold:
\begin{equation}
  i_\mathrm{n,net}=\sqrt{\,i_\mathrm{n}^2-i_\mathrm{n,ampl}^2-
  eI}.
\end{equation}
\subsection{Gain dependence of the current noise}
\begin{figure}\centering
\epsfig{file=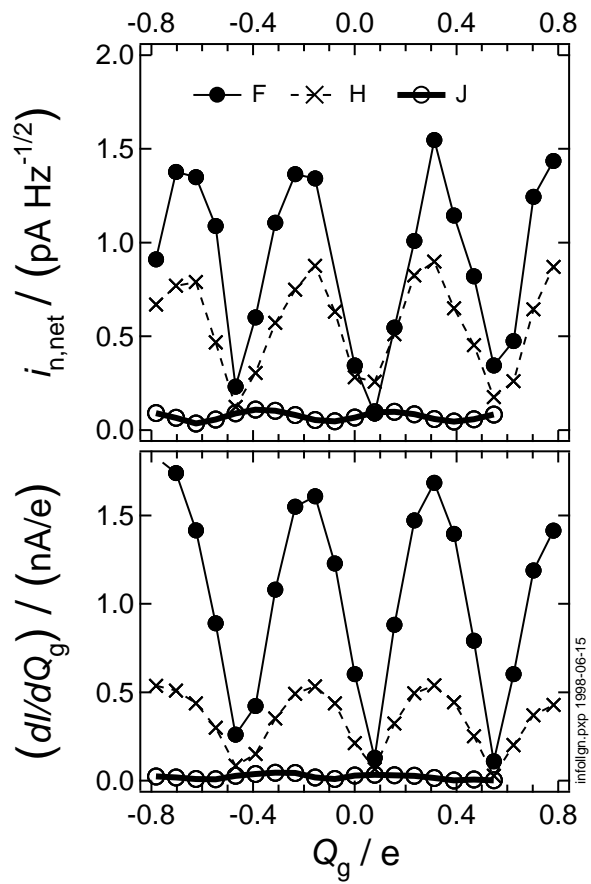,width=0.75\textwidth}
\caption[Current noise and gain as a function of gate charge]{%
Net current noise at 10\,Hz $i_\mathrm{n,net}$ and gain 
$dI/dQ_\mathrm{g}$ as a function of the charge induced on the gate of
the SET. Amplifier noise and shot noise have been subtracted from the
measured noise to calculate the net noise. Bias points are labelled as
in fig.~\ref{fig:biapoinv}. Measurements were taken at 30\,mK with the
sample driven into the normal conducting state.
}
\label{fig:infollgn}
\end{figure}
In fig.~\ref{fig:infollgn}, the net current noise at 10\,Hz (top panel)
and the SET's gain (bottom) are plotted, respectively, as a function of
the gate charge, for three bias points (cf. fig.~\ref{fig:biapoinv}; an
amplifier battery ran empty 
near the end of the measurement \textsf{J}). It
is immediately obvious (and comes as no surprise
\cite{starmark:98:condmat}) that the noise follows the gain, or in other
words, that the noise has dominantly input noise character. \par
\begin{figure}\centering
\epsfig{file=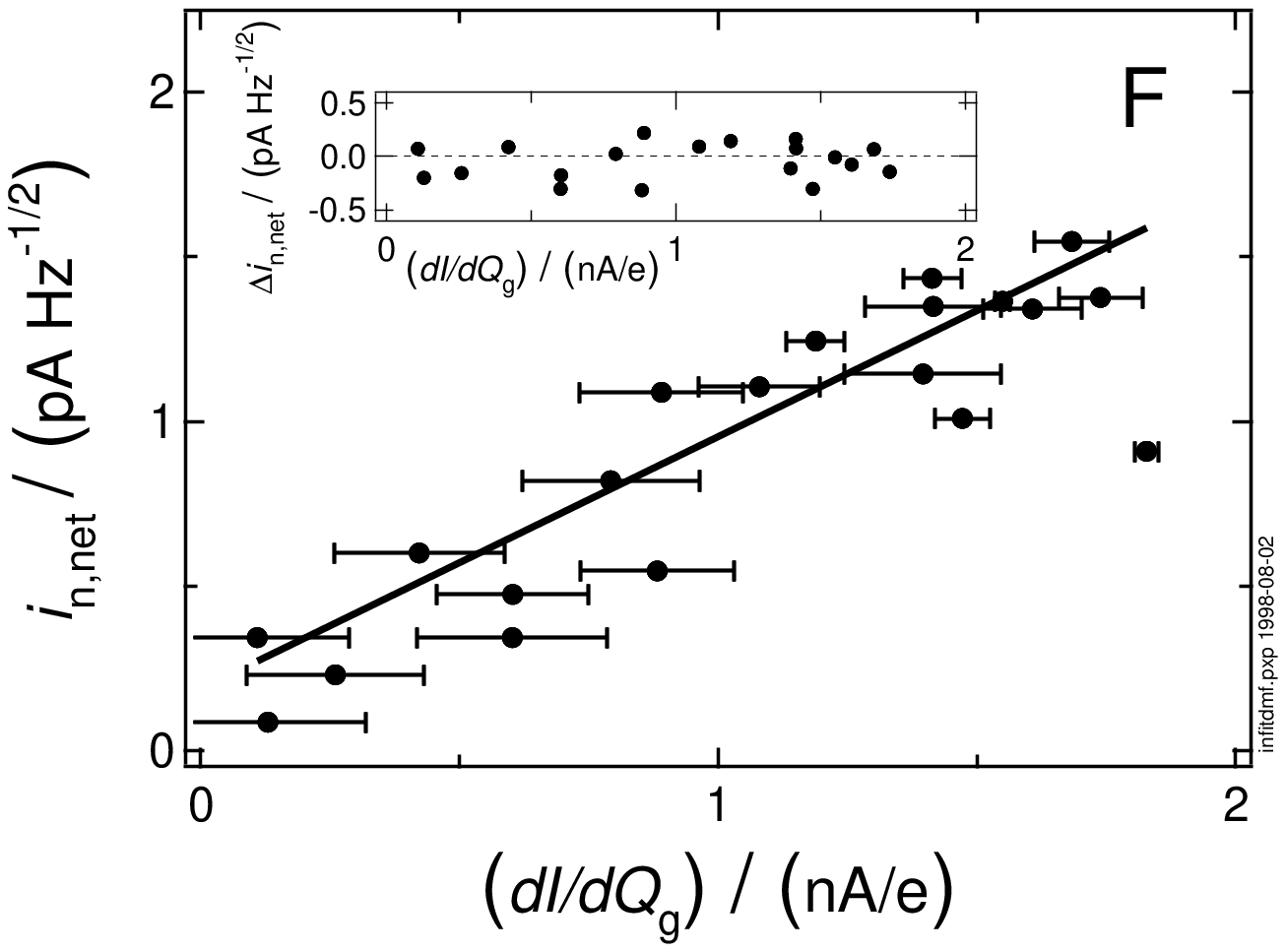,width=\textwidth}
\caption[Current noise as function of gain]{%
Dependence of the net current noise at 10\,Hz on the gain. The thick
line shows a linear least squares fit to the data points, and the fit
residuals are shown in the inset. The error margin on the gain is
relatively large due to the small number of gate voltage points per bias
point. These data were taken in bias point \textsf{F}
(cf. fig.~\ref{fig:biapoinv}), at 30\,mK and in the normal conducting state.}
\label{fig:infitdmf}
\end{figure}
For a more quantitative analysis, we plotted the net current noise as a
function of the gain in fig.~\ref{fig:infitdmf}. At each bias point, the
gain dependence of the net current noise could well be described by a
linear relation, as the graph in fig.~\ref{fig:infitdmf} exemplifies for
bias point \textsf{F}. The figure's inset shows that the residuals to a
least squares linear fit spread non-systematically. This means that
within our measurement accuracy, we cannot identify any other gain
dependent noise component apart from charge input noise. Specifically,
we do not see evidence for charge-resistance correlation noise, for
which the square of the current noise,
$S_{I_{R,\mathrm{corr}}}=i_\mathrm{n_{R,\mathrm{corr}}}^2$,
should depend linearly on the gain \cite{starmark:98:condmat}. \par
We will refer to the slope
\begin{equation}
  q_\mathrm{n}^\mathrm{fit}=\left\langle
  di_\mathrm{n}/d(dI/dQ_\mathrm{g}) \right\rangle
\end{equation}
in the diagrams exemplified by fig.~\ref{fig:infitdmf} as
\emph{differential charge equivalent noise}. To our knowledge, this
quantity has not been studied systematically before. \par
An upper limit for the contribution to the noise in second order in the
gain, described by the term proportional to  $\alpha$ in
eq.\,(\ref{eq:noisemodel}), can be given. 
\label{calc:alpha}
The coefficient $\alpha$ can
be evaluated as \cite{starmark:98:condmat}
\begin{equation}
  \alpha(f)=\frac{1}{e^2 S_{Q_\mathrm{g}}(f)}
  \int_{-\infty}^{+\infty}S_{Q_\mathrm{g}}(f^\prime)
  S_{Q_\mathrm{g}}(f-f^\prime)\,df^\prime.
\end{equation}
We found that $\alpha\lesssim 10^{-4}$, practically independent of
frequency, and that thus we could neglect the noise effect of second
order input charge noise.
\subsection{Resistance fluctuations}
An upper estimate for the resistance fluctuations can be given within
the phenomenological low frequency noise model
\cite{starmark:98:condmat}. Details on the calculations can be found in
\cite[section 2.3.3]{starmark:98:lic}. 
A number of rather strong assumptions
have to be made:
\begin{enumerate}
\item The estimation is only applicable for bias voltages well above the
  blockade (threshold) voltage, where the dynamic junction resistance is
  approximately constant.
\item The junctions are identical in their DC as well as in their noise
  properties. 
\item The background charge noise $S_{Q_\mathrm{g}}$ is not dependent on
  anything other than possibly the bias voltage.
\end{enumerate}
In the frequency band between 50\,Hz and 100\,Hz, the total SET
resistance $R_1+R_2=(165\pm 8)\,$k$\Omega$ did, under the above
assumptions, not fluctuate by more than 31\,$\Omega$.\par
\section{Deviations from ideal charge noise behaviour}
  \label{sect:nonidealnoise}
\subsection{Bias dependence}
In the model presented in \ref{sect:noisemodel}, a constant charge input
noise independent of any transistor back action would produce a constant
differential charge equivalent noise, independent of the bias voltage
for example. This is, however, not the case we find experimentally.\par
\begin{figure}\centering
\epsfig{file=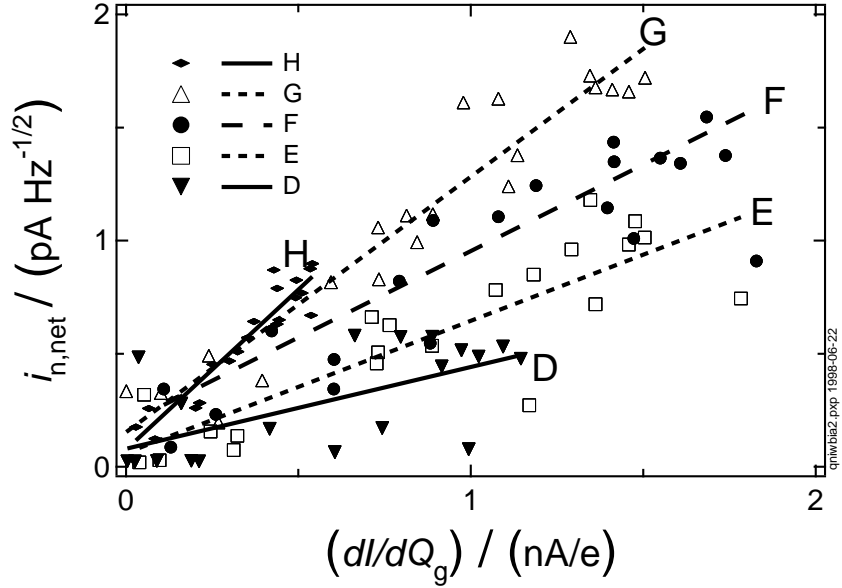,width=\textwidth}
\caption[Current noise as function of gain
for several bias points]{%
Gain dependence of the net current noise (amplifier noise and shot noise
have been subtracted) at 10\,Hz (base temperature, normal conducting
state). With increasing bias (\textsf{D}\dots\textsf{H},
cf. fig.~\ref{fig:biapoinv}), the slope of the noise-gain relation
increases, from $0.36\cdot 10^{-3}\,e/\sqrt{\mathrm{Hz}}$ at
bias point \textsf{D} to 
$1.42\cdot 10^{-3}\,e/\sqrt{\mathrm{Hz}}$ at point \textsf{H}.
These slopes have been calculated by a least squares fit to the data as
illustrated in fig.~\ref{fig:infitdmf}, error bars have been omitted to
reduce clutter.
}
\label{fig:qniwbia2}
\end{figure}
For the five bias points around that with maximum gain, the plots used
for the determination of the differential charge equivalent noise (as
demonstrated in fig.~\ref{fig:infitdmf}) have been superimposed in
fig.~\ref{fig:qniwbia2}. It is immediately evident that the differential
charge equivalent noise increases systematically with the bias voltage,
roughly by a factor of four when the bias voltages increases
fivefold.\par
This bias dependence of the noise at low temperatures has immediate
consequences for the optimal operating point of an SET
electrometer. While maximising the gain maximises the signal, an
optimisation of the signal-to-noise ratio may be achieved by 
sacrificing
some gain against reduced noise by biasing on the lower flank of the
gain-vs.-bias curve. \par
\subsection{Temperature dependence}
\label{subsec:noisetempdep}
The simplest back action mechanism by which the bias voltage could
influence the noise sources is heating of the electrons 
or phonons in the barriers, the island and leads, and the surfaces in
their vicinity, by the dissipation near the junctions. Wolf et
al. \cite{wolf:97:cpem} observed a weak dependence of the low frequency
noise on the transport current, with a proportionality to the fourth
root of the current, and suggested self-heating as an explanation. For a
single two level fluctuator, Kenyon et al. \cite{kenyon:98:tlftempprepr}
observed a bias dependence of the noise. Under the general assumption,
that $1/f$-like low frequency noise results from the superposition of
many such TLF, these experiments corroborate 
the self-heating hypothesis.\par
\begin{figure}\centering
\epsfig{file=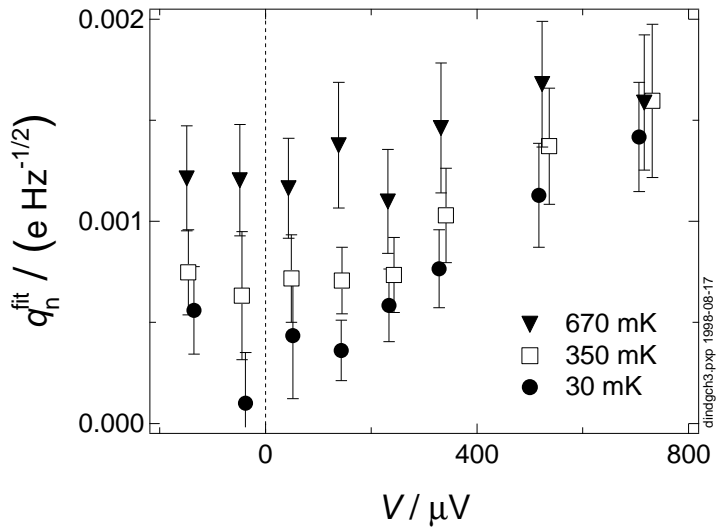,width=0.9\textwidth}
\caption[Differential charge equivalent noise as function of bias
for three different temperatures]{%
Differential charge equivalent noise (proportionality constant relating
gain increase to increase in net current noise), as a function of bias
voltage and at different temperatures. 
Data values were determined as the slopes of the linear fit curves in
the noise current versus gain diagrams (cf. fig.~\ref{fig:infitdmf}).
The error margins were estimated
from the average amplitude of the fit residuals.
While the increase of the differential charge equivalent noise with bias
is evident at 30\,mK, it is masked by temperatures of about half a
Kelvin. 
}
\label{fig:dindgch3}
\end{figure}
To test this hypothesis, we repeated the set of measurements described
so far at elevated temperatures of 350\,mK and 670\,mK. The evaluation
results for the differential charge equivalent noise as a function of
bias are plotted in fig.~\ref{fig:dindgch3}. \par
Within the error margin, which we estimated from the average amplitude
of the current noise-vs.-gain fits, the noise at low bias rose
significantly with temperature, while at higher bias of about twice the
threshold voltage, noise were independent of the ambient
temperature. At even higher bias, gains were so low at all temperatures
that zero gain noise (see \ref{subsec:zerogainnoise}) masked the
differential charge equivalent noise. Nevertheless, the trend in
fig.~\ref{fig:dindgch3} is clear: at temperatures of approximately a
little less than two thirds of a Kelvin, the bias dependence of the
differential charge equivalent noise vanished. \par
The implications for high sensitivity electrometry are obvious. If one
is forced to operate an SET electrometer at temperatures  
approaching that corresponding to the charging energy,
the signal-to-noise ratio
can  be optimised by simply maximising the signal, which in turn means
biasing the SET at the point giving the global maximum in gain. \par 
\subsection{Model calculation of the self-heating}
If it is self-heating of the transistor that activates the noise
sources, we should expect an equilibrium temperature of about half a
Kelvin, since this is the temperature at which the bias
dependence of the noise saturated (see \ref{subsec:noisetempdep}). \par
At very low temperatures, the thermal coupling between different
subsystems in the solid state becomes weak. Energy is transported into
the electronic system of the SET island via the junctions. Due to their
relatively high resistance, we can neglect their thermal conductivity
and thus their cooling effect 
\cite{korotkov:94:overheating}. Heat transfer
from the electrons in the island goes only to the phonon system of the
island. In equilibrium, the temperatures of the electron system
$T_\mathrm{el}$ and of the island phonon system $T_\mathrm{ph}$ are
related by \cite{korotkov:94:overheating}
\begin{equation}
  T_\mathrm{el}=\sqrt[5]{T_\mathrm{ph}^5+
  \frac{P}{\Sigma\Omega}},
\end{equation}
where $\Omega$ is the island's volume, 
$P$ the power dissipated in the island of the SET, 
and $\Sigma\approx 1\cdot 10^9$\,Wm$^{-3}$K$^{-5}$
\cite{kauppinen:96:elphprb} a
coefficient describing the electron-phonon coupling. 
Values for $\Sigma$ in the literature vary by about one order of
magnitude between $0.3 \cdot 10^9$\,Wm$^{-3}$K$^{-5}$
\cite{verbrugh:95:islandsize} and $2.4 \cdot 10^9$\,Wm$^{-3}$K$^{-5}$
\cite{wellstood:94:hotelprb}.\par
Determining the island's phonon temperature is not as
straightforward. In thick films one could use the model of the island's
phonon system coupled to the substrate's phonon system via the area $A$
with the \emph{Kapitza resistance} $\alpha$. For thick film samples at
very low temperatures, this can lead to significant temperature
differences between the different phonon systems. In thin films,
however, this model does not apply at very low temperatures
\cite{wellstood:94:hotelprb}, and the
phonon population of the film and the substrate can no longer be clearly
separated. \par
If we calculate the temperature of the phonon system under the
assumption that the usual relation \cite{korotkov:94:overheating}
\begin{equation}
  \label{eq:sucker}
  T_\mathrm{ph}=\sqrt[4]{\frac{P}{A\alpha}+T_\mathrm{substrate}^4}
\end{equation}
holds, where $\alpha=100$\,W\,m$^{-2}$\,K$^{-4}$
\cite{korotkov:94:overheating}, we find a $T_\mathrm{ph}$ of the order
of 500\,mK. For a film thickness of 40\,nm, this is just about the
temperature below which eq.~(\ref{eq:sucker}) loses relevance
\cite{wellstood:94:hotelprb}. \par
The electron temperature is not affected much by the phonon temperature,
and at the upper edge of our bias range (800\,$\mu$V), we find values
for $T_\mathrm{el}$ between 800\,mK (neglecting any heating of the
phonon system) and 1100\,mK (assuming that eq.~(\ref{eq:sucker}) is fully
applicable). \par
Stretching the model represented by eq.~(\ref{eq:sucker}) to its limits,
we can motivate a heating of the phonons of the island to about half a
Kelvin. Thus, all of the noise increase with bias could be attributed to
ohmic self-heating.\par
Another
bias dependence could have been caused by the noise sources being
activated by fluctuations of the island potential, that should increase
with the potential differences to the electrodes. If this mechanism had
been more dominant, we would have found more bias dependence of the
noise than had been possible to explain with self heating, but since we
obtain temperatures of about 0.5\,K from the model calculations with
conservative parameter estimates, such a contribution by island
potential fluctuations cannot be detected within the accuracy of these
measurements. 
\subsection{Zero gain noise}
\label{subsec:zerogainnoise}
Another deviation from the ideal input charge noise behaviour is the
gain independent component that we call \emph{zero gain noise}. It
readily presents itself as the offset along the noise axis in the fit
procedure used to determine the differential charge equivalent
noise. Zero gain noise also spooks around the literature as \emph{excess
noise} \cite{starmark:98:condmat}, which is unfortunate inasmuch this
term is a historical synonym 
\cite{rogers:87:nbnoise} for low frequency noise (see
\ref{subsec:lfnbackground}). \par
\begin{figure}\centering
\epsfig{file=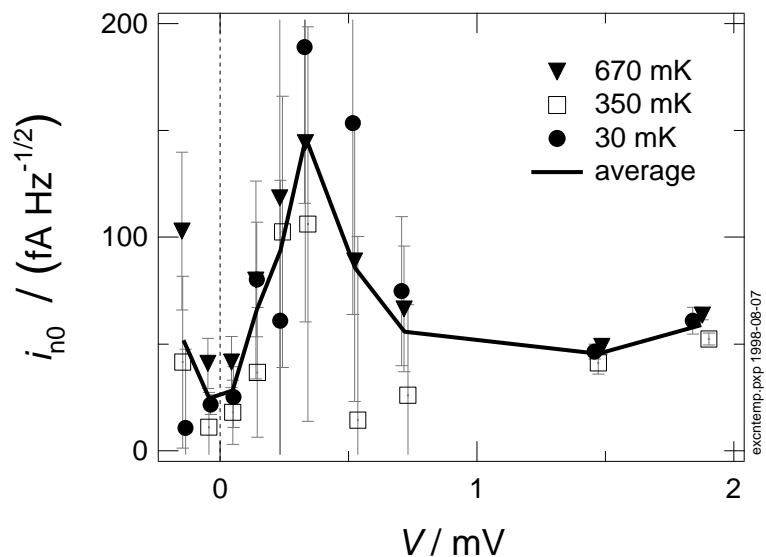,width=0.9\textwidth}
\caption[Zero gain (extrapolated) noise]{%
Zero gain noise at 10\,Hz (normal conducting state) as a function of
bias voltage, for the same temperatures as in
fig.~\ref{fig:dindgch3}. The values were calculated from the vertical
axis intersection in the fit procedure illustrated in
fig.~\ref{fig:infitdmf}, the errors have been estimated from the ratio
between the average amplitude of the fit residuals and the gain range.
Within this accuracy, no temperature dependence is visible. The zero
gain noise has a maximum close to the threshold voltage of the SET.
}
\label{fig:excntemp}
\end{figure}
Figure~\ref{fig:excntemp} shows that the zero gain noise clearly was a
function of bias, with a dependence roughly resembling that of the
maximum gain on the bias, i.\,e. peaking slightly above the threshold
voltage. While the maximum gain, however, dropped by about one half upon
raising the temperature from 30\,mK to 670\,mK, the zero gain noise was
temperature independent within our measurement accuracy. We do not have
a model explaining the observed zero gain noise at the present time, and
it is apparent that more numerous and especially more precise data are
needed here.
\section{Conclusion: noise in a single electron transistor}
We studied the low frequency current noise of a voltage biased single
electron transistor in the normal conducting state and found it
dominated by input charge noise. Introducing a technique that analyses
the differential charge equivalent noise, that is the coefficient
relating an increase in bias to the resulting increase in noise, we
found a second order effect, namely a bias and temperature dependence of
the noise. Both dependences can be reduced to a temperature dependence
when the self-heating of the electron gas in the transistor is taken
into account. Within our measurement accuracy, a bias dependence of the
noise due to island potential fluctuations cannot be detected.
The details of the
mechanism and the exact location of the noise sources, however, remain
unknown. 
\chapter{Conclusion}
\section{What's new? What's useful? What's not?}
In this thesis, we have investigated new fabrication methods for
resistors and charging effect devices, and a new evaluation technique
for low frequency noise measurements of single electron transistors.\par
Anodisation-made resistors allow us to pack large resistances on fairly
short lengths. The limitation is not set by fabrication issues, but is
rather inherent due to the onset of the Coulomb blockade at high sheet
resistances. It can be imagined that useful applications for such
resistors should exist.\par
Exploiting the Coulomb blockade for devices is not equally
straightforward. The process suffers from asymmetry in the residual
metal thicknesses, which is a problem common to all anodisation
miniaturisation schemes. The fine grained structure gives us a multiple
tunnel junction system with a gate effect vanishing at only a few
Kelvin. Given that practically the only application area envisioned
today for (random)multiple tunnel junctions is memory, these anodised
niobium systems are facing a tough competition.\par
Noise, namely low frequency noise, has not been considered enough in
the modelling of single electronics devices, especially regarding high
temperature operation. Our measurements indicate that low frequency
noise may become worse with higher temperatures. These measurements have
of course only scratched the surface of all that is to be discovered
about low frequency noise. Now that the equipment and an evaluation
technique are at hand, extensive measurements should be done soon to
establish causal dependencies between fabrication and sample parameters
on one hand and noise data on the other hand. It should then be possible
to obtain reasonable estimates on how detrimental low frequency noise
will be on the road towards room temperature single electronics.\par
\section{Suggestions for future research}
\hspace*{\fill}
\parbox{0.4\textwidth}{\textit{%
(...) [W]e can do that, too, some time.\\
\hspace*{\fill}Monica S. Lewinsky\cite{lewinsky}
}}\par\vspace*{2ex}
Although experiments reported by other groups indicate that there is
no dramatic improvement of the low frequency noise properties 
of C-SET to be
expected from the use of other substrate materials, some more systematic
research on the materials influence might be worthwhile.
\par
Korotkov's model for the low frequency noise in SET
\cite{starmark:98:condmat} came after the measurements presented here
were made. With this model in hand, we would have concentrated on 
more (and that means better)
measurements at low gain. This should allow for improved conclusions
about many aspects of the LFN, be it resistance fluctuations,
charge-resistance-correlation noise, and the nature of the zero gain
noise. 
\par
With very small high-ohmic resistors and junctions with oxide barriers,
we have the two key components of a resistively coupled single electron
transistor. It should be possible to combine both components on a single
chip, in just one angular evaporation step. Prototypes exist, but all
dozen or so of them had at least one of the junctions destroyed or a
pinchoff in the resistor strip. The idea of integrating the Nb based
junctions into an R-SET was the main reason to limiting the film
thicknesses to 20\,nm that are so unfavourable for the superconducting
properties. By adding some halide contaminants to the anodisation
electrolyte, it should be possible to anodise only the niobium line to
the SET island into a resistor, while simultaneously destroying the
parallel line of aluminium inevitably generated in the shadow
evaporation deposition. 
\cleardoublepage
\addcontentsline{toc}{chapter}{Bibliography}
\bibliography{da,database}
\begin{appendix}
\chapter{Symbols and notation}
\label{chap:symbols}
\setlongtables
\begin{longtable}{|p{0.08\textwidth}|p{0.56\textwidth}|p{0.24\textwidth}|}
\hline \endhead
\hline \endfoot
\hline ~\vphantom{{\Large (}} & %
   meaning (numerical value) \cite{cohen:93:const} & %
   SI unit \\ \hline \endfirsthead
$\alpha$ \vphantom{{\Large (}} &
   expansion coefficient &
   1 \\
$\alpha$ \vphantom{{\Large (}} &
   Kapitza resistance &
   kg\,s$^{-3}$\,K$^{-4}$ \\
$\Delta$ \vphantom{{\Large (}} &
   superconducting energy gap &
   m$^2$\,kg\,s$^{-2}$ \\
$\Delta{}V$ \vphantom{{\Large (}} &
   voltage swing (modulation) &
   m$^2$\,kg\,s$^{-3}$\,A$^{-1}$ \\
$\gamma$ \vphantom{{\Large (}} &
   phase difference over Josephson junction &
   1 \\
$\eta$ \vphantom{{\Large (}} &
   gain &
   {\tiny depends on ampl. type} \\
$\varepsilon$ \vphantom{{\Large (}} &
   relative dielectric permittivity &
   1 \\
$\varepsilon_0$ \vphantom{{\Large (}} &
   dielectric permittivity of vacuum \newline
   ($8.854\ldots\cdot 10^{-12}$)&
   m$^{-3}$\,kg$^{-1}$\,s$^{4}$\,A$^2$  \\
$\lambda$ \vphantom{{\Large (}} &
   London penetration depth &
   m \\
$\mu_0$ \vphantom{{\Large (}} &
   permeability of vacuum ($1.256\ldots\cdot 10^{-6}$) &
   m\,kg\,s$^{-2}$\,A$^{-2}$ \\
$\xi$ \vphantom{{\Large (}} &
   coherence length &
   m \\
$\varrho$ \vphantom{{\Large (}} &
   density &
   m$^{-3}$\,kg \\
$\Sigma$ \vphantom{{\Large (}} &
   electron-phonon interaction constant &
   m$^{-1}$\,kg\,s$^{-3}$\,K$^{-5}$ \\
$\tau$ \vphantom{{\Large (}} &
   lifetime &
   s \\
$\phi$ \vphantom{{\Large (}} &
   phase of multiparticle wavefunction &
   1 \\
$\Omega$  \vphantom{{\Large (}} &
   volume &
   m$^3$ \\
$A$ \vphantom{{\Large (}} &
   area &
   m$^2$ \\
$B$ \vphantom{{\Large (}} &
   magnetic flux density &
   kg\,s$^{-2}$\,A$^{-1}$ \\
$C$ \vphantom{{\Large (}} &
   capacitance &
   m$^{-2}$\,kg$^{-1}$\,s$^{4}$\,A$^{2}$ \\
$C_0$ \vphantom{{\Large (}} &
   stray capacitance &
   m$^{-2}$\,kg$^{-1}$\,s$^{4}$\,A$^{2}$ \\
$C_\mathrm{g}$ \vphantom{{\Large (}} &
   gate capacitance &
   m$^{-2}$\,kg$^{-1}$\,s$^{4}$\,A$^{2}$ \\
$C_\Sigma$ \vphantom{{\Large (}} &
   total island capacitance &
   m$^{-2}$\,kg$^{-1}$\,s$^{4}$\,A$^{2}$ \\
$d$ \vphantom{{\Large (}} & 
   thickness &
   m \\
$d_\mathrm{b}$ \vphantom{{\Large (}} & 
   barrier width &
   m \\
$e$ \vphantom{{\Large (}} & 
   elementary charge ($1.602\ldots\cdot 10^{-19}$) &
   s\,A \\
$e_\mathrm{n}$ \vphantom{{\Large (}} & 
   op amp input equivalent noise &
   m$^2$\,kg\,s$^{-3.5}$\,A$^{-1}$ \\
$E$ \vphantom{{\Large (}} &
   energy &
   m$^2$\,kg\,s$^{-2}$ \\
$E_\mathrm{a}$ \vphantom{{\Large (}} &
   activation energy &
   m$^2$\,kg\,s$^{-2}$ \\
$E_\mathrm{c}$ \vphantom{{\Large (}} &
   characteristic charging energy $e^2/(2C)$&
   m$^2$\,kg\,s$^{-2}$ \\
$E_\mathrm{ch}$ \vphantom{{\Large (}} &
   charging energy &
   m$^2$\,kg\,s$^{-2}$ \\
$E_\mathrm{F}$ \vphantom{{\Large (}} &
   Fermi energy &
   m$^2$\,kg\,s$^{-2}$ \\
$E_\mathrm{J}$ \vphantom{{\Large (}} &
   Josephson (coupling) energy &
   m$^2$\,kg\,s$^{-2}$ \\
$f$ \vphantom{{\Large (}} &
   frequency & s$^{-1}$ \\
$G$ \vphantom{{\Large (}} &
   conductivity &
   m$^{-2}$\,kg$^{-1}$\,s$^{3}$\,A$^{2}$ \\
$H$ \vphantom{{\Large (}} &
   magnetic field &
   m$^{-1}$\,A \\
$h$ \vphantom{{\Large (}} &
   Planck's constant ($6.626\ldots\cdot 10^{-34}$) &
   m$^2$\,kg\,s$^{-1}$ \\
$\hbar$ \vphantom{{\Large (}} &
   Planck's constant divided by $2\pi$ ($1.054\ldots\cdot 10^{-34}$) &
   m$^2$\,kg\,s$^{-1}$ \\
$i_\mathrm{n}$  \vphantom{{\Large (}} &
   current noise &
   s$^{-1/2}$\,A \\
$I$ \vphantom{{\Large (}} &
   current &
   A \\
$I_\mathrm{c}$ \vphantom{{\Large (}} &
   critical current &
   A \\
$I_\mathrm{c0}$ \vphantom{{\Large (}} &
   maximum critical current &
   A \\
$I_\mathrm{s}$ \vphantom{{\Large (}} &
   supercurrent &
   A \\
$k_\mathrm{B}$ \vphantom{{\Large (}} &
   Boltzmann constant ($1.380\ldots\cdot 10^{-23}$) &
   m$^2$\,kg\,s$^{-2}$\,K$^{-1}$ \\
$l$ \vphantom{{\Large (}} &
   length &
   m \\
$P$ \vphantom{{\Large (}} &
   power &
   m$^2$\,kg\,s$^{-3}$ \\
$q_\mathrm{n}$  \vphantom{{\Large (}} &
   charge noise &
   s$^{1/2}$\,A \\
$Q$ \vphantom{{\Large (}} & 
   charge  &
   s\,A \\
$Q_0$ \vphantom{{\Large (}} & 
   offset charge  &
   s\,A \\
$Q_\mathrm{g}$ \vphantom{{\Large (}} & 
   gate charge  &
   s\,A \\
$r_0$ \vphantom{{\Large (}} & 
   output impedance &
   m$^2$\,kg\,s$^{-3}$\,A$^{-2}$ \\
$R$ \vphantom{{\Large (}} & 
   resistance &
   m$^2$\,kg\,s$^{-3}$\,A$^{-2}$ \\
$R^\ast$ \vphantom{{\Large (}} & 
   resistance constant &
   m$^2$\,kg\,s$^{-3}$\,A$^{-2}$ \\
$R_0$ \vphantom{{\Large (}} & 
   zero bias (dynamic) resistance &
   m$^2$\,kg\,s$^{-3}$\,A$^{-2}$ \\
$R_I$  \vphantom{{\Large (}} & 
   current autocorrelation function &
   A$^2$ \\
$R_\mathrm{K}$ \vphantom{{\Large (}} & 
   quantum resistance &
   m$^2$\,kg\,s$^{-3}$\,A$^{-2}$ \\
$R_{\mbox{\tiny K-90}}$ \vphantom{{\Large (}} & 
   Klitzing resistance (25812.807) &
   m$^2$\,kg\,s$^{-3}$\,A$^{-2}$ \\
$R_\mathrm{Q}$ \vphantom{{\Large (}} & 
   `quantum resistance for pairs' $R_\mathrm{K}/4$ &
   m$^2$\,kg\,s$^{-3}$\,A$^{-2}$ \\
$R_\mathrm{T}$ \vphantom{{\Large (}} & 
   tunnelling resistance &
   m$^2$\,kg\,s$^{-3}$\,A$^{-2}$ \\
$S_I$ \vphantom{{\Large (}} & 
   current spectral noise density &
   s\,A$^2$ \\
$S_{I,\mathrm{e}}$ \vphantom{{\Large (}} & 
   current shot noise spectral density &
   s\,A$^2$ \\
$S_{I,\mathrm{th}}$ \vphantom{{\Large (}} & 
   current thermal noise spectral density &
   s\,A$^2$ \\
$S_{Q_\mathrm{g}}$ \vphantom{{\Large (}} & 
   background charge noise spectral density &
   s$^3$\,A$^2$ \\
$S_R$ \vphantom{{\Large (}} & 
   resistance noise spectral density &
   m$^4$\,kg$^2$\,s$^{-5}$\,A$^{-4}$ \\
$t$ \vphantom{{\Large (}} &
   time &
   s \\
$T$ \vphantom{{\Large (}} & 
   temperature &
   K \\
$T^\ast$ \vphantom{{\Large (}} & 
   temperature of vanishing gate effect &
   K \\
$T_\mathrm{c}$ \vphantom{{\Large (}} & 
   critical temperature &
   K \\
$T_\mathrm{el}$ \vphantom{{\Large (}} & 
   electron temperature &
   K \\
$T_\mathrm{ph}$ \vphantom{{\Large (}} & 
   phonon temperature &
   K \\
$V$ \vphantom{{\Large (}} & 
   voltage, bias voltage &
   m$^2$\,kg\,s$^{-3}$\,A$^{-1}$ \\
$V_\mathrm{an}$ \vphantom{{\Large (}} & 
   anodisation (cell) voltage &
   m$^2$\,kg\,s$^{-3}$\,A$^{-1}$ \\
$V_\mathrm{b}$ \vphantom{{\Large (}} & 
   (tunnel) barrier height &
   m$^2$\,kg\,s$^{-3}$\,A$^{-1}$ \\
$V_\mathrm{g}$ \vphantom{{\Large (}} & 
   gate voltage &
   m$^2$\,kg\,s$^{-3}$\,A$^{-1}$ \\
$V_\mathrm{off}$ \vphantom{{\Large (}} & 
   offset voltage &
   m$^2$\,kg\,s$^{-3}$\,A$^{-1}$ \\
$V_\mathrm{off}^0$ \vphantom{{\Large (}} & 
   (extrapolated) zero bias offset voltage &
   m$^2$\,kg\,s$^{-3}$\,A$^{-1}$ \\
$V_\mathrm{th}$ \vphantom{{\Large (}} & 
   threshold voltage &
   m$^2$\,kg\,s$^{-3}$\,A$^{-1}$ \\
$x$ \vphantom{{\Large (}} &
   spatial coordinate &
   m \\
\end{longtable}

\chapter{Glossary and abbreviations}
\label{glossary}
\begin{description}
\item[\AA{}ngstr\"om (\AA{}):] Outdated unit of
   length. 1\,\AA{}$=10^{-10}$\,m. 
\item[ACME:] Anodization controlled miniaturization enhancement
   \cite{nakamura:96:acme}
\item[ADC:] Analog-to-digital converter
\item[ADP:] Automated data processing
\item[AF:] Anodic film (more general than AOF, may incorporate
   inclusions from the electrolyte)
\item[AFM:] Atomic force microscope/microscopy/micrograph
\item[AIS:] Automated information system
\item[Ammonium pentaborate:] (NH$_4$)B$_5$O$_8\cdot x$H$_2$O
   \cite{apbdictionaryentry}, where $x$ 
   gives the amount of crystal water. 
   If undefined, we assume $x\approx 4$ (APB tetrahydrate).
\item[AOF:] Anodic oxide film
\item[APB:] Ammonium pentaborate
\item[ASCII:] American Standard Code for Information Interchange
\item[Aspect ratio:] Ratio between height and width of a
  structure. Fabrication of structures with high aspect ratios requires
  very anisotropic etching or deposition techniques.
\item[BCS:] Bardeen-Cooper-Schrieffer (theory of superconductivity)
  \cite[and references therein]{claeson:74:supcon} 
\item[BIFET:] Bipolar output FET input operational amplifier
\item[CAD:] Computer Aided Design
\item[CAM:] Computer Aided Manufacturing
\item[Cb:] Chemical symbol for Columbium
\item[CB:] Coulomb blockade
\item[CBCPT:] Coulomb blockade of Cooper pair tunnelling
\item[CHET:] Charging effect transistor
   \cite{amman:89:chet}
\item[Chip marks:] Alignment marks (JEOL EBL system) used for highest
   precision alignment. Three chip marks have to be situated within one
   field. In the work described here, wafer marks were used instead.
\item[Columbium (Cb):] old name for Niobium (Nb), used in the
   Angloamerican language space until about 1950 and in the American
   metallurgical community even later.
\item[Contrast:] Degree to which the physicochemical properties
   exploited in resist development differ in exposed areas compared to
   unexposed areas
\item[Corner frequency:] Frequency above which low frequency noise
  becomes negligible compared to white noise
\item[CP:] Copolymer P(MMA-MAA)
\item[DAQ:] Data acquisition
\item[DMM:] Digital multimeter
\item[DMS:] Dilute magnetic semiconductor
\item[DOS:] Density of states
\item[DSA:] Dynamic signal analyser
\item[DUT:] Device under test
\item[DVM:] Digital voltmeter
\item[DXF:] Drawing Exchange Format
\item[EBL:] Electron beam lithography
\item[EOS:] Electron optical system
\item[Field:] Area that can be written by the EBL without the need of
   moving the stage, $80\times 80\,\mu$m$^2$ in highest resolution mode. At
   the edges of the fields, stitching error occurs, so fine structures
   should not cross field boundaries.
\item[FFT:] Fast Fourier transform
\item[Forming:] The process of growing an anodic [oxide] film
\item[Gauss (G):] Hopelessly outdated unit of magnetic flux density in
   one of the various CGS systems. Corresponds to (but is not equal to)
   0.1\,mT.
\item[GDS-II:] Stream format, a.\,k.\,a. Calma Stream. The industry
   standard for lithographic pattern data.
\item[GLP:] Good Laboratory Practice, a set of
  rules that define the proper conduct and documentation of scientific
  experiments \cite{usgov:98:glp}.
\item[GPIB:] General purpose interface bus (IEEE-488)
\item[HEMT:] High electron mobility transistor
\item[IBE:] Ion beam etching (milling)
\item[Inch:] Outdated unit of length. 1\,in$=$25.4\,mm
\item[Igor:] A data evaluation and graphing programme for scientists
   and engineers from WaveMetrics
\item[IUPAC:] International Union of Pure and Applied Chemistry
\item[IVC:] Current-voltage characteristic
\item[JEOL:] Japanese manufacturer of electron optical research
   instrumentation 
\item[JJ:] Josephson junction
\item[JJA:] Josephson junction array
\item[Jobdeck file (JDF):] Text file in the JEOL EXPRESS
   system that describes which patterns (chips) are to be exposed and
   where they are to be positioned relative to each other, 
   and that contains the definition of the shot modulation
   and the alignment (mark detection) parameters.
   It also points to a calibration sequence of the electron optical
   system that will be carried out at the beginning of the exposure
   and eventually during the exposure.
\item[LabVIEW:] A graphical programming language for data acquisition
   and experiment control from National Instruments
\item[LFN:] Low frequency noise
\item[LIA:] Lock-in amplifier
\item[Linux:] The best operating system for small computers
\item[MEK:] Methylethylketone
\item[Microfabrication:] Fabrication of devices with typical linear
   dimensions below 1\,$\mu$m
\item[M-IT:] Metal-insulator transition
\item[ML:] Monolayer
\item[MSU:] Moscow State University
\item[NaN:] Not a number
\item[Nanofabrication:] The art and science of producing non-random
   structures with typical linear dimensions less than 100\,nm
\item[Negative resist:] Resist that is removed during development where
   it has \emph{not} been exposed. Example is SAL\,601 (e-beam resist).
\item[PMGI:] Poly(dimethyl glutarimide), a positive e-beam and deep UV
   resist 
\item[PMMA:] Polymethylmethacrylate, a positive e-beam resist
   \cite{haller:68:pmma}
\item[P(MMA-MAA):] Copolymer
   poly(methylmethacrylate-methacrylic acid), a positive e-beam resist,
   more sensitive than PMMA
\item[Positive resist:] Resist that is removed during development where
   it has been exposed. Examples are PMMA (e-beam resist) or S-1813
   (photoresist). 
\item[PROXECCO:] A commercial computer programme for proximity
   correction \cite{eisenmann:93:proxecco}.
\item[Proximity correction:] Increasing the exposure dose for narrow
   and/or isolated features to compensate for the proximity effect.
\item[Proximity effect:] Additional exposure of pixels with many
   neighbouring exposed pixels due to scattering of the electron beam in
   the resist and substrate and to secondary electrons.
\item[PTB:] Physikalisch-Technische Bundesanstalt
\item[QPT:] Quantum phase transition
\item[RAM:] Random access memory
\item[RIE:] Reactive ion etching
\item[Rotation:] Angular misorientation of the  sample relative to the
   sample holder and 
   consequently the whole electron beam lithography machine.
   Rotation has to be compensated by the EOS, increasing inaccuracies
   (stitching error) and pattern distortions. A limit on 
   allowed rotation is
   set in the internal configuration files of the JEOL system.
\item[RRR:] Residual resistance ratio, between the resistances at room
   temperature and just above the resistive transition or at 4.2\,K; a
   measure for the quality often used for Nb. 
\item[S-1813:] A positive photoresist
\item[SAIL:] Self-aligned in-line technique, developed in Jena
   \cite{bluethner:96:sailjdp}
\item[SAL\,101:] A developer for PMGI
\item[SAL\,601:] A negative e-beam resist
\item[Schedule file (SDF):] Text file in the JEOL EXPRESS
   system that describes where the arrangement of patterns defined in the
   jobdeck file is to be placed relative to the machine and what the
   reference dose for the shot modulation is.
   It also contains information on hardware settings and definitions
   for the alignment mark detection.
\item[SECO:] Step-edge cut-off technique \cite{altmeyer:97:thesis}
\item[Selectivity:] Ratio of the solubilities of different resists
   exposed simultaneously, important for the resolution in processes
   involving two layer resist systems
\item[SEM:] Scanning electron microscope
\item[Sensitivity:] Reciprocal of the irradiation dose required to
   produce the physicochemical modifications in a resist needed for
   development
\item[SET:] (``ess-eeh-tee'') Single electron tunnelling, 
   alt. single electron
   (tunnelling) transistor
\item[Shadow evaporation technique:] also known as Dolan technique,
   Niemeyer-Dolan technique, nonvertical evaporation technique etc.
   A method of forming very small overlap
   junctions in the shadowed area underneath a suspended bridge on the
   substrate. Self-aligning, involves only one lithography
   step. Introduced by Niemeyer \cite{niemeyer:74:mitt}, in its present
   form with resist mask by Dolan \cite{dolan:77:masks}.
\item[Shot modulation:] JEOL-specific implementation of handling
   the assignment of doses to pattern parts (to compensate the
   proximity effect). Each primitive is assigned a shot rank 
   (an integer number) that
   corresponds to a certain dose enhancement factor
   (a floating point number). This assignment
   is called the shot modulation.
\item[S-IT:] Superconductor-insulator transition
\item[SNAP:] Selective niobium anodization process
   \cite{kroger:81:anodiapl} 
\item[SnL:] Swedish Nanometre Laboratory, G\"oteborg.
\item[SPM:] Scanning probe microscopy. AFM and STM are both SPM
  techniques. 
\item[SQUID:] Superconducting quantum interference device
\item[Stitching error:] Misalignment of parts of the electron
   beam exposed pattern at the boundaries of fields and subfields.
   Stitching error increases with sample rotation.
\item[STM:] Scanning tunnelling microscopy
\item[Subfield:] Area that can be written by the EBL without switching
   digital-to-analog converters, $10\times 10\,\mu$m$^2$ in highest
   resolution mode. At subfield boundaries, slight stitching 
   error occurs, so the
   finest nanostructures should not cross them.
\item[Tear-off technique:] A special form of angular evaporation
   technique where some material is deposited on resist sidewalls and
   removed during liftoff. Requires good control over the undercut and
   the evaporation angles.
\item[TEM:] Transmission electron microscopy
\item[TLF:] Two level fluctuator
\item[UHV:] Ultra high vacuum, below $10^{-6}$\,Pa
\item[Vector scan:] EBL mode where the beam is swept only over the areas
   that are to be exposed, as opposed to raster scan, where it is swept
   over the whole sample and simply blanked from non-exposure
   areas. Requires faster electron optics and makes systems more
   expensive, but can save a lot of exposure time.
\item[VI:] Virtual Instrument, a LabVIEW programme
\item[VTB:] Variable thickness bridge
\item[Wafer marks:] Alignment marks (JEOL EBL system) that can be placed
   almost anywhere on the sample. Of course, precision of alignment
   improves when the marks are as close to the writing area as possible.
\item[WCE:] Wet chemical etching
\item[White noise:] Noise with a frequency independent spectral density,
  usually occurring at some intermediate frequency range
\item[ZEP\,520:] A positive e-beam resist
\item[ZEP\,7000B-97:] A positive e-beam resist with high sensitivity
\end{description}

\chapter{Recipes}
\label{chap:recipes}
All recipes assume that reactive ion etching (RIE) is done in a
Plasmatherm Batchtop 70 with a seven inch electrode (area
248\,cm$^2$), an electrode distance of 60\,mm and a working
frequency of 13.56\,MHz.
\par
The contact printer 
operates in the wavelength range 
(320\dots 420)\,nm$^2$.\par
Electron beam lithography was done with a JEOL JBX 5D-II system
with CeB$_6$  cathode.
\section{Substrate preparation}
\subsection{Photomask making (positive resist process)}
\label{rec:photomaskpositive}
\begin{enumerate}
\item Clean a Cr mask with RIE. Process gas oxygen, pressure 67\,Pa,
flow 36\,$\mu$mol/s, rf power 250\,W, time 120\,s.
\item Spin ZEP 7000B-97, solvent chlorobenzene, at 2000~rpm,
  to a thickness of approximately 500\,nm.
\item Bake in an oven for 20~min at 180$^\circ$C.
\item Expose using 4th lens, 3rd aperture, 10\,nA beam current,
  with a dose of 10\,$\mu$C/cm$^2$.
\item Develop for (6\dots 8)~min in a 1:1 mixture (by volume)
  of MEK and Ethylmalonate.
\item Rinse in Ethylmalonate.
\item Blow dry with nitrogen (not quite easy).
\item Ash the surface with RIE. Process gas oxygen, pressure 13\,Pa,
flow 36\,$\mu$mol/s, rf power 50\,W, time 30\,s.
\item Wet etch in Balzers Chrome Etch \#4 as long as necessary,
  about 2~min when the solution is fresh.
\item Rinse with deionised water.
\item Strip the resist with RIE. Process gas oxygen, pressure 67\,Pa,
flow 36\,$\mu$mol/s, rf power 250\,W, time 120\,s.
\end{enumerate}
\subsection{Photomask making (negative resist process)}
\label{rec:photomasknegative}
\begin{enumerate}
\item Rinse a Cr mask with deionised tap water.
\item Ash the surface with RIE. Process gas oxygen, pressure 33\,Pa,
flow 36\,$\mu$mol/s, rf power 50\,W, time 30\,s.
\item Spin Microposit Primer.
\item Spin Shipley SAL-601 at 4000\,rpm, giving a thickness of about
800\,nm.
\item Preexposure bake for 20\,min at 90\,$^\circ$C
   in an oven.
\item E-beam expose in the JEOL JBX 5D-II. Design dose
10\,$\mu$C/cm$^2$, acceleration voltage 50\,kV, fourth lens
   (working distance 39\,mm),
   third aperture
   (diameter 300\,$\mu$m), current 5\,nA.
\item Postexposure bake for 20\,min at 110\,$^\circ$C
   in an oven.
\item Develop in Microposit MF322 for about 6\,min, inspect in the
microscope.
\item Ash the surface in the RIE (see above) and immediately 
   thereafter
\item Etch in Balzers No.\,4 chromium etch 
(composition: 200\,g cerium ammonium nitrate, 35\,mL 98\% acetic acid,
filled with deionised water to 1000\,mL).
\item Remove the resist by stripping with RIE. Process gas oxygen, pressure
66\,Pa, flow 7\,$\mu$mol/s, rf power 250\,W, time 120\,s.
\end{enumerate}

\subsection{Gold pad photolithography (carrier chips)}
\label{sect:padrecipe}\label{rec:chiplitho}
\begin{enumerate}
\item Strip the surface 
of an oxidised two inch Si wafer with RIE.  Process gas oxygen, pressure
66\,Pa, flow 7\,$\mu$mol/s, rf power 250\,W, time 120\,s.
\item Spin Shipley S-1813 at 5500\,rpm, giving a thickness of about
1000\,nm.
\item Bake for 7:30\,min at 110\,$^\circ$C
   on a hotplate.
\item Expose for 12\,s at an intensity of 10\,mW/cm$^2$, correspondingly
longer or shorter for different intensities.
\item Develop in a 1:1 mixture (by volume) of Microposit Developer and
deionised water for 60\,s, rinse thoroughly with deionised water from
the tap. \textbf{Or:}
\item Develop in pure MF\,322 developer for 15\,s and rinse.
\item Ash the surface with RIE. Process gas oxygen, pressure 33\,Pa,
flow 36\,$\mu$mol/s, rf power 50\,W, time 30\,s. Immediately 
thereafter
\item evaporate 20\,nm of Ni$_{0.6}$Cr$_{0.4}$ 
at 0.1\,nm/s and
\item 80\,nm Au at 0.2\,nm/s.
\item Liftoff in slightly warmed acetone.
\item Presaw from the back to a depth of about 50\,$\mu$m. When cutting
alignment edges from the front side, try to preserve a C$_4$
symmetric circumference shape of the wafer; this facilitates later
resist preparation.
\end{enumerate}

\section{Niobium nanofabrication}
\subsection{Four layer resist preparation}
  \label{rec:fourlayerprep}
\begin{enumerate}
\item Ash the surface 
of a wafer with gold chip patterns
with RIE. Process gas oxygen, pressure 33\,Pa,
flow 36\,$\mu$mol/s, rf power 50\,W, time 30\,s.
\item Spin 350k PMMA (1.8\,\%, in xylene) at 2500\,rpm to a thickness
of about 50\,nm.
\item Bake for 12\,min at 170\,$^\circ$C
   on a hotplate.
\item Spin Shipley S-1813, diluted 1:1 by volume with Shipley
P-Thinner, at 3000\,rpm, giving a thickness of about 200\,nm.
\item Bake for 12\,min at 160\,$^\circ$C
   on a hotplate.
\item Evaporate 20\,nm Ge at 0.2\,nm/s.
\item Spin 350k PMMA (1.8\,\%, in xylene) at 2500\,rpm to a thickness
of about 50\,nm.
\item Bake for 10\,min at 150\,$^\circ$C
   on a hotplate.
\item Break into suitable chip sets for further handling.
\end{enumerate}

\subsection{Four layer resist exposure}
\label{sect:4lrexposure}
Acceleration voltage 50\,kV,
   first aperture
   (diameter 60\,$\mu$m), fifth lens
   (working distance 14\,mm), current 20\,pA for the fine patterns
   (1\,nA for the coarser leads).
   Area doses
\begin{itemize}
\item 1120\,$\mu$C/cm$^2$ for 20\,nm wide lines.
\item 400\,$\mu$C/cm$^2$ 
   for 100\,nm wide lines.
\item 280\,$\mu$C/cm$^2$ 
   for all wider lines and areas.
\end{itemize}

\subsection{Four layer resist proximity correction}
\label{sect:proxparapp}
(for PROXECCO:) double Gaussian with $\alpha=0.02\,\mu$m,
$\beta=10\,\mu$m and $\eta=0.5$. The low $\eta$,
the ratio between exposure by backscattered and by directly impacting
electrons,
is due to the Ge layer
that absorbs a large fraction of the backscattered electrons. Number of
doses 32, output quality fine, physical fracturing.

\subsection{Parameters for SET fabrication}
\begin{figure}
\begin{minipage}[b]{0.5\textwidth}
\epsfig{file=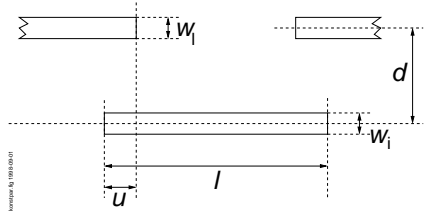,width=\textwidth}
\end{minipage}
\hspace*{0.1\textwidth}
\begin{minipage}[b]{0.4\textwidth}
\caption[Geometric design parameters for shadow evaporated SET]{%
\label{fig:konstpar}%
Geometric design parameters for SET made by shadow evaporation.}
\end{minipage}
\end{figure}
See fig.~\ref{fig:konstpar} for definition of geometric parameters.
Linewidths $w_\mathrm{l}\approx w_\mathrm{i}\approx 100$\,nm,
pattern shift $d=240$\,nm, island length $l=600$\,nm, overlap
$u=100$\,nm.

\subsection{Four layer resist processing}
\label{sect:flpprcapp}
\begin{enumerate}
\item Expose in the EBL machine (see \ref{sect:4lrexposure}).
\item Develop in a mixture of 10 volume parts isopropanole and 1
   volume part deionised water for 60\,s under ultrasonic
   excitation.
\item Reactive ion etching: pattern transfer to the Ge mask.
   Process gas CF$_4$, pressure 1.3\,Pa, flow 
   7.5\,$\mu$mol/s,
   rf power 14\,W, time 120\,s.
\item RIE of the support layers. Process gas O$_2$, pressure
   13\,Pa, 
   15\,$\mu$mol/s flow rate, rf power 20\,W, time 15\,min.
\item Evaporate Nb with e-gun heating. Deposition rate about
   0.5\,nm/s.
\item Liftoff in slightly warmed acetone, spraying chip centres
   directly with a syringe.
\end{enumerate}

\section{Niobium microanodisation}
\subsection{Anodisation window mask}
  \label{rec:windowmask}
\begin{enumerate}
\item Spin 950k PMMA (8\,\%, in 
chlorobenzene) at 5000 rpm, giving a thickness
of about 1.8\,$\mu$m.
\item Bake for 12\,min at 170\,$^\circ$C
   on a hotplate.
\item E-beam expose with an  area dose of
280\,$\mu$C/cm$^2$. Acceleration voltage 50\,kV,
   first aperture
   (diameter 60\,$\mu$m), fifth lens
   (working distance 14\,mm), current 1\,nA.
\item Develop in a mixture of 10 volume parts isopropanole and 1
volume part deionised water under ultrasonic excitation for 8\,min.
\end{enumerate}

\subsection{Electrolyte for Nb anodisation}
\label{rec:electrolyte}
Downscaled from the recipe of Joynson \cite{joynson:67:anodi}:
8.3\,g ammonium pentaborate, 60\,mL ethylene glycol and 40\,mL
distilled water to be stirred and heated to about 100$^\circ$\,C.
The solution has to be regenerated by heating and stirring before
using since the ammonium pentaborate precipitates.

\chapter{Measurement data handling}
  \label{app:mdatadp}
Modern physics is hardly imaginable without automatic data processing
(ADP). A disadvantage of ADP is that the handling of the data largely
takes place in volatile machine memory, and that the evaluation leaves
much less of a paper trail than it used to only about a decade
ago. Moreover, caused by the rapid changes in data storage, both in file
formats and in physical media, original data as well as evaluation
results can become inaccessible within a few years. While this may
basically be desirable, as a mechanism that prevents us with
overburdening ourselves with useless data garbage, it is at the brink of
violating one of the principles of our profession. Data acquisition and
evaluation must be reconstructable over a reasonable period of time;
this is a matter of Good Laboratory Practice.\par

This chapter outlines how the data gathered for and used in this thesis
were taken and stored, and how the electronic evaluation is
documented. We are now in the fortunate situation that storage
technologies have advanced so far that it has become feasible to keep
the entire body of original data on inexpensive media. With the
information in this appendix, it should be possible to understand and,
if necessary, use the original data that will accompany at least some of
the electronic versions of this work.\par

\section*{ADP infrastructure}
\begin{figure}\centering
\epsfig{file=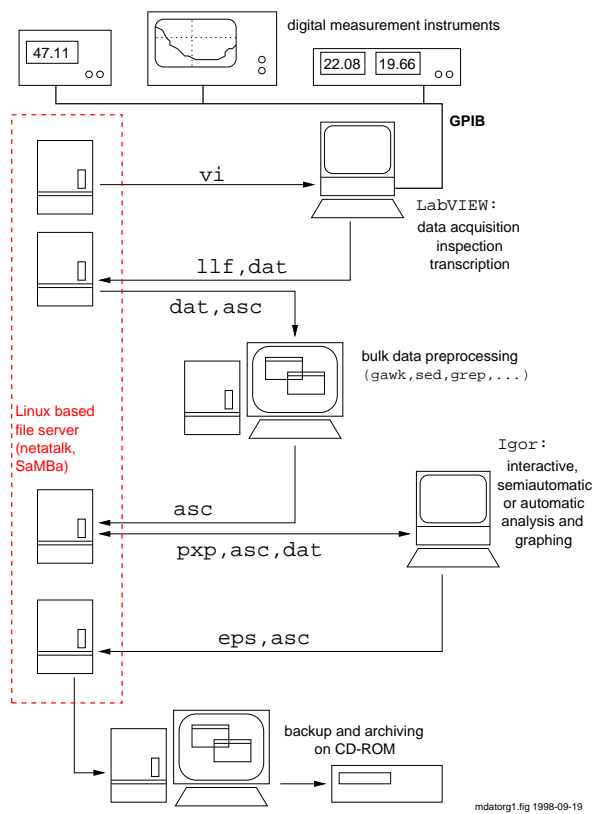,width=0.7\textwidth}
\caption[ADP organisation]{%
Automatic data processing: organisation and flow of measurement data.
All measurement instruments were digital and delivered raw data via GPIB to
a Macintosh computer running LabVIEW. Inspection of the data during the
measurement and the transcription to ASCII files were also done under
LabVIEW. The data files were organised and preprocessed on a Linux
workstation, and the final analysis and graphical presentation was
carried out on other Macintoshes and Windows NT workstations
using the Igor software. 
Files were kept on a Linux file server and backed up 
and archived via the Linux system.}
\label{fig:mdatorg}
\end{figure}
Figure \ref{fig:mdatorg} gives an overview over the flow of measurement
data. Digital measurement instruments, like multimeters, signal
analysers or lock-in amplifiers, were connected, usually via a GPIB, to
the measurement computer, which happened to be a succession of
Macintoshes over the years. Experiment control and data acquisition
programmes were written in LabVIEW, a
graphical programming language that in its current version is available
for and compatible between MacOS, various Windows versions and a few
commercial UNIX flavours. Data were stored in a LabVIEW binary format
that changed at least once during the course of this work. The data were
quickly transcribed to an ASCII form that is documented in detail in the
following section \ref{sect:fileformat}. All data evaluation was then
based on these ASCII files that are to be considered the original
data. The simple preprocessing of large amounts of data, such as
screening and extraction of specific quantities, was done with a number
of GNU tools like awk, sed and grep on a Linux workstation. \par

For the
final evaluation and graphical presentation of the data, the
(proprietary) software Igor from WaveMetrics 
was found most suitable, for a number of reasons.
Igor has a powerful scripting language that allows easy handling of
large sets of data, and file input/output. By breaking up long
evaluation sequences into short programmes, using ASCII files as
interface for the exchange of intermediate data, Igor can be employed
in a way that optimises traceability. All evaluation steps are
documented as instructions in the Igor procedures, which can be exported
and stored or printed as human-readable ASCII files. Like LabVIEW, the
latest version of Igor is available for Macintosh and 32\,bit-Windows
systems with Intel processors. Unfortunately, there is no UNIX/Linux
version.\par 

Data were kept on a Linux file server interfaced to the Macintosh world
with netatalk and to the Windows network with SaMBa. The measurement
data, photographs, and scanned documents are stored on compact discs.

\section*{Measurement data file formats}
\label{sect:fileformat}
\subsection*{Definitions}
A \textbf{logbook} is a book with a sturdy cover and paginated blank
pages in which circumstances of experiments are noted with a nondelible
pen immediately
as they occur. Entries made at a later date than the date of the
measurement, usually in the form of additional remarks, are clearly
marked as such.\par
A \textbf{measurement} is completed when data from an instrument have
been read either into AIS memory or by a human observer, independently
of whether these data have been recorded on storage media.\par
An \textbf{experiment} is an ensemble of measurements logically
belonging together, usually carried out with essentially unchanged
apparatus and a variation of typically one parameter. Every experiment
is identified by a three-digit numerus currens and 
must be described in a
logbook entry. The last digit of the year and the hexadecimal digit for
the month are prepended to this three digit number, giving a five
alphanumeric character long \textbf{experiment identification
number}.\par
Every experiment consists of a number of \textbf{logs}, that is a set
of measurements with unchanged parameters. Usually, one independent
variable is changed (`swept'), and recorded along with at least one
dependent variable and the relevant parameters. Logs within each
experiment are numbered with a three digit decimal number, starting from
000 for each experiment. The concatenation of experiment identification
number and this 
three digit log number gives the \textbf{log identification
number}. Individual logs need not have a corresponding entry in the
logbook, if parameters have been varied and recorded automatically; in
this case, a summary entry in the logbook is sufficient.\par
A \textbf{LabVIEW log file} records the data of exactly one experiment
in a National Instruments proprietary binary format.\par
\textbf{Transcription} is the process of translating the information
contained in a LabVIEW log file into ASCII files suitable for both human
and automatic reading and processing, and for archiving.\\
A \textbf{log file} is the ASCII translation of exactly one log. Its
name consists of the log identification number and the extension
\texttt{.dat} (note the compliance with ISO\,9660 and DOS restrictions
on filenames). Log files are only generated automatically from the
LabVIEW log file and never modified afterwards.

\subsection*{Types of information}
\label{informationseinteilung}
By semantical content, information contained in every log of an
experiment falls into one of the following categories.
\begin{description}
\item[Parameters] are recorded as pairs of parameter names and
parameter values. They give settings of experimental apparatus, and
setpoints and measured actual values of physical quantities.
\item[Data] consist of usually at least two, fairly large streams
of numerical data.
\item[Free formatted text] is text information given manually to
the programme writing the LabVIEW log files, describing hardware
settings, sample characteristics etc. It is used to back up the logbook
entries and to make information accessible to ADP.
\item[Other info] like timestamps either in name-value-pair format
like parameters, or as free formatted text.
\end{description}
According to into which of these categories information fall, their
formatting in the ASCII transcription will vary.

\subsection*{Implementation in LabVIEW}
Every log in a LabVIEW log file consists of
\begin{itemize}
\item a timestamp,
\item a free formatted comment text and
\item a number of (`an array of'
   in LabVIEW notation) clusters, each consisting in turn of
\begin{itemize}
\item a name string,
\item a parameter string and
\item a value array.
\end{itemize}
\end{itemize}
The information contained in each cluster falls into exactly one of 
two of the
categories described in \ref{informationseinteilung}:\par
For \textit{parameter clusters}, the name string is arbitrary yet
descriptive, and the parameter string contains exactly as many lines
(separated by newline characters dependent on the OS LabVIEW runs under)
as
there are values in the value array. Each line in the parameter name
string than gives the name of the parameter whose value is recorded in
the corresponding cell of the value array.\par
In \textit{data clusters}, the name string contains the name of the
variable, chosen consistently across as many experiments as possible
(see table \ref{table:variablennamen}). The parameter has either less
or more lines than there are values in the value array, usually less. In
the parameter string, keywords and parameters are recorded that the
measuring and recording programme might use to derive measured
quantities from raw data, for example a bias resistance value used to
compute the transport current from a voltage measurement.

\begin{table}
\caption{\label{table:variablennamen}%
Variable names (non-exhaustive compilation)
and generic ``wave'' names for Igor use. 
Unless indicated specifically, quantities are recorded
in the appropriate SI units, except for temperatures, which are recorded
in mK for historical reasons.}\vskip2ex
\centering
\begin{tabular}{|l|l|}\hline
variable name & description\\ 
\hline\hline
\texttt{i}\textit{logIDnr} & transport current\\
\texttt{isrbdc}\textit{logIDnr} & current times bias resistance (DC)\\
\texttt{v}\textit{logIDnr} & voltage\\
\texttt{vb}\textit{logIDnr} & bias voltage\\
\texttt{vg}\textit{logIDnr} & gate voltage\\
\texttt{vguac}\textit{logIDnr} & voltage before divider on gate (AC)\\
\texttt{vgudc}\textit{logIDnr} & voltage before divider on gate (DC)\\
\texttt{vsdc}\textit{logIDnr} & voltage over sample (DC)\\
\hline
\dots\texttt{mv}\dots & mean value\\
\dots\texttt{std}\dots & standard deviation \\
\hline
\texttt{inv}\dots & inverse (1/x) \\
\texttt{ln}\dots & natural logarithm of \dots \\
\texttt{log}\dots & decadic logarithm of \dots \\
\hline
\dots\texttt{mv} & \dots in mV \\
\dots\texttt{ua} & \dots in $\mu$A\\
\dots\texttt{ko} & \dots in k$\Omega$\\
\hline
\end{tabular}
\end{table}

\subsection*{Format of ASCII transcript}
In this section, the rules for the translation of LabVIEW log files into
log files are described and the characteristic patterns making these log
files accessible to ADP are defined. As an example, we show the
transcript of the first log (number 000) of the experiment with the
numerus currens 394, registered in December 1997. According to the
naming conventions outlined above, this file must have the name
\texttt{7c394000.dat}. The transcript starts with some information
intended for its identification in case the file should change its name
(like in a recovery action after a file system corruption):
\begin{verbatim}
MEASUREMENT DATA TRANSCRIPT FILE
COPYRIGHT henning@fy.chalmers.se
SOURCE_FILE_(LABVIEW_LOG) 7c394
FILENAME 7c394000.dat
\end{verbatim}
Then, the free formatted text is included, surrounded by characteristic
lines:
\begin{verbatim}
--- BEGIN_COMMENT ---
SAMPLE T47-34
completely Per's box
sweeping I-V
MAGN 05.000 T
base temperature
PINS 1-1-16-16
GATEPIN 3
stepping gate
--- END_COMMENT ---
\end{verbatim}
Even without consulting the logbook, one can now see that this file
contains a current-voltage characteristic of the device between pins 1
and 16 on chip T47-34, measured in a magnetic field of 5\,Tesla at
cryostat base temperature, and that the parameter varied in this
experiment is the gate voltage applied to pin 3 on the chip.\par
The next brief section of the log file is the time stamp, both in a
machine friendly and a human readable format:
\begin{verbatim}
TIMESTAMP 2964497225
DATE_TIME 1997-12-09    08:27
\end{verbatim}
This $I$-$V$ characteristic was logged at 08:27 hours local time on
9th December, 1997.\par
The following is the transcription result of three parameter clusters:
\begin{verbatim}
SWEEP cluster
BUFSIZ  1.024000E+3
LORAMP  -1.000000E+1
HIRAMP  1.000000E+1
SWPORT  3.000000E+0
TRPORT  2.000000E+0
SWMODE  1.000000E+0
ACCURACY cluster
DELAY   2.000000E+1
NPLC    1.000000E+0
AUTOCAL 0.000000E+0
DISPLAY 1.000000E+0
TEMPERATURE cluster
T_INIT  1.500000E+3
R_INIT  0.000000E+0
T_FINAL 1.500000E+3
R_FINAL 0.000000E+0
SENSOR  5.000000E+0
\end{verbatim}
Most of this should be self-explanatory. 
If applicable, ``1'' represents the Boolean ``true'' or ``on'', ``0''
represents ``false'' or ``off''.
\texttt{BUFSIZ} gives the
number of data points in a sweep, \texttt{SWMODE} equal 1 indicates
bidirectional sweep (0 would have been unidirectional). The implausible
values for initial and final temperature are due to a bug in the
temperature measurement software: since our thermometer is not
calibrated below what we believe to be 15\,mK, an appropriate output
would have been `NaN', as a synonym for `below 15\,mK'.\par
The next cluster is a data cluster for the variable named \texttt{i}:
\begin{verbatim}
--- BEGIN DATACOMMENTS i ---
i_R_BIAS 2E4
i_GAIN 1E3
--- END DATACOMMENTS i ---
--- BEGIN DATA i ---
i7c394000
-1.990785E-8
-1.984932E-8
     {1020 lines of text deleted here}
-1.979642E-8
-1.984588E-8
--- END DATA i ---
\end{verbatim}
The variable name is prepended to the lines of the parameter string,
separated by an underscore. Here, the current was computed by measuring
the voltage drop across the bias resistor of 20\,k$\Omega$, amplified
with a gain of 1000.
Similarly, the voltage across the sample \texttt{v} 
and the gate voltage \texttt{vg} are
recorded.\par
Immediately preceding the actual numerical data is one line that
contains a name uniquely identifying the data stream. The name consists
of the variable name and the log identification number. When handling
the data with Igor, this becomes the ``wave'' name.
\begin{verbatim}
--- BEGIN DATACOMMENTS v ---
v_GAIN 1E2
--- END DATACOMMENTS v ---
--- BEGIN DATA v ---
v7c394000
-2.734458E-3
-2.723223E-3
     {1020 lines of text deleted here}
-2.718614E-3
-2.729049E-3
--- END DATA v ---
--- BEGIN DATACOMMENTS vg ---
vg_GAIN 4E1
vg_FIX_RANGE 9 VOLT
--- END DATACOMMENTS vg ---
--- BEGIN DATA vg ---
vg7c394000
9.926481E-7
9.704308E-7
     {1020 lines of text deleted here}
1.125952E-6
1.092626E-6
--- END DATA vg ---
\end{verbatim}
Finally, the last cluster is a parameter cluster again, giving the gate
voltage setpoint. If the gate voltage had not been measured in every
point of the sweep, one measurement of the actual value would have been
included here:
\begin{verbatim}
SHELL_VOLTAGE cluster
V_SHELL_SETPOINT        0.000000E+0
V_SHELL_MEASURED        NaN
\end{verbatim}
The log file is concluded by a human readable line that also helps to
verify the completeness of the log file:
\begin{verbatim}
END OF MEASUREMENT DATA TRANSCRIPT FILE
\end{verbatim}

\subsection*{Applicability}
The above scheme was used for all 
low temperature and very low temperature
experiments from the beginning of
1996, starting with experiment number 62236
on 1996-02-20.
Data taking during the anodisation of samples were recorded directly in
ASCII in another, linewise constructed,
column oriented format, 
since these data had to be yanked to
the hard disk during the process.

\subsection*{Other file formats}
The most important of those
data formats, identified by the file name extension, that will be
encountered in the electronic material relating to this work, are given
in table~\ref{table:filetypes}.
\begin{table}
\caption{File name extensions and file types}
\label{table:filetypes}
\begin{center}
\begin{tabular}{|l|l|}\hline
extension & file type \\ \hline\hline
.txt & text (ASCII), generic \\
.asc & text (ASCII), intermediate data \\
.dat & text (ASCII), original data (not to be altered) \\
.pdf & Portable Document Format, often scanned material \\
.bmp & Windows bitmap, e.\,g. SEM original photos \\
.tif & TIFF images, e.\,g. optical microscope photographs \\
.jpg & JPEG/JFIF, compressed photographs \\
.vi & LabVIEW Virtual Instruments (programmes) \\
.bdat & LabVIEW log file, versions $\leq 3$ \\
.llf &  LabVIEW log file, versions $\geq 5$ \\
.pxp & Igor packed experiment \\
.eps & Encapsulated Postscript (graphics) \\
.tar & GNU tape archive \\
.gz & GNU gzip compressed file \\
.html & HTML (for navigation) \\
$\langle$none$\rangle$ & executable scripts \\
\hline
\end{tabular}
\end{center}
\end{table}

\chapter{Selected publications 1994--1998}
\section*{Superconductor physics and single electronics}
\begin{itemize}
\item
Torsten Henning, B.~Starmark, T.~Claeson, and P.~Delsing.
\newblock Bias and temperature dependence of the noise in a single electron
  transistor.
\newblock accepted by Eur. Phys. J. B 1998-10-12, 
\texttt{cond-mat/9810103}.
\item
B.~Starmark, Torsten Henning, A.~N. Korotkov, T.~Claeson, and P.~Delsing.
\newblock Gain dependence of the noise in the single electron transistor.
\newblock \texttt{cond-mat/9806354}.
\item
T.~Henning, B.~Starmark, and P.~Delsing.
\newblock {Stromrauschen eines Einzelelektronentransistors}.
\newblock Talk at Tagung Kryoelektronische 
Bau\-ele\-men\-te Kryo '98, Braunschweig,
  1998-10-11--13. Published in summary form only.
\item
Torsten Henning.
\newblock Coulomb blockade effects in anodised niobium nanostructures
  ({Coulomb}-blockad-effekter i anodiserade 
  {Nb}-nano\-struk\-turer).
\newblock Licentiate thesis, Institutionen f\"or Mikroelektronik och
  Nano\-ve\-ten\-skap, Chalmers Tekniska H\"ogskola AB 
  och G\"oteborgs Universitet,
  G\"oteborg, May 1997.
\newblock \texttt{cond-mat/9710037}.
\item
Torsten Henning, D.~B. Haviland, and P.~Delsing.
\newblock Coulomb blockade effects in anodized niobium nanostructures.
\newblock {\em Supercond. Sci. Technol.}, 10(9):727--732, September 1997.
\newblock \texttt{cond-mat/9706302}.
\item
Torsten Henning, D.~B. Haviland, and P.~Delsing.
\newblock Charging effects and superconductivity in anodised niobium
  nanostructures.
\newblock In H.~Koch and S.~Knappe, editors, {\em ISEC'97. 6th International
  Superconductive Electronics Conference. Extended Abstracts}, volume~2, pages
  227--229, Berlin, June 1997. Physikalisch-Technische Bundesanstalt.
\newblock ISBN 3-9805741-0-5.
\item
Torsten Henning, D.~B. Haviland, and P.~Delsing.
\newblock Fabrication of {Coulomb} blockade elements with an electrolytic
  anodization process.
\newblock {\em Electrochemical Society Meeting Abstracts}, 96-2:561, 1996.
\newblock Fall Meeting San Antonio, Texas, October 6-11. Published in summary
  form only.
\item
Torsten Henning, D.~B. Haviland, and P.~Delsing.
\newblock Transition from supercurrent to {Coulomb} blockade tuned by
  anodization of {Nb} wires.
\newblock {\em Czech. J. Phys.}, 46(Suppl. S4):2341--2342, 1996.
\newblock Proc. 21st Int. Conf. on Low Temperature Physics, Prague, August
  8--14, 1996.
\item
T.~Henning, D.~B. Haviland, P.~Delsing, and T.~Claeson.
\newblock Widerst{\"a}nde f{\"u}r {Einzelladungstunnelelemente}.
\newblock {\em Verhandl. DPG (VI)}, 30:1751, 1995.
\newblock DPG-Fr{\"u}hjahrstagung Berlin 1995. Published in summary form only.
\item
Torsten Henning, David~B. Haviland, and Per Delsing.
\newblock Nanofabrikation von {Widerst\"anden} aus anodisierten
  {Nb}-{Dr{\"a}hten}.
\newblock {\em Verhandl. DPG (VI)}, 31:1966--1967, 1996.
\newblock DPG-Fr{\"u}hjahrstagung Regensburg 1996. Published in summary form
  only.
\item
T.~Henning, D.~B. Haviland, P.~Delsing, and T.~Claeson.
\newblock {Nb} resistors for single electronics.
\newblock {\em Bull. APS (II)}, 40(1):205, March 1995.
\newblock APS March Meeting San Jose 1995. Published in summary form only.
\item
Torsten Henning, H.~Kliem, A.~Weyers, and W.~Bauhofer.
\newblock Characterization of high-temperature superconductor ceramics from
  their resistive transition.
\newblock {\em Supercond. Sci. Technol.}, 10(9):721--726, September 1997.
\newblock \texttt{cond-mat/9707165}.
\end{itemize}
\section*{Semiconductor physics}
\begin{itemize}
\item
P.~J. Klar, D.~Wolverson, J.~J. Davies, W.~Heimbrodt, M.~Happ, and T.~Henning.
\newblock Photoluminescence and photoluminescence excitation studies of lateral
  size effects in {Zn}$_{1-x}${Mn}$_x${Se}/{Zn}{Se} quantum disc samples of
  different radii.
\newblock {\em Phys. Rev. B}, 57(12):7114--7118, March 1998.
\newblock \texttt{cond-mat/9803208}.
\item
Ivan~J. Griffin, Peter~J. Klar, Daniel Wolverson, J.~John Davies, Bernard Lunn,
  Duncan~E. Ashenford, and Torsten Henning.
\newblock Magneto-photoluminescence studies of
  {Zn}$_{1-x}${Mn}$_x${Te}/{Zn}{Te} multiple quantum-well and quantum-dot
  structures.
\newblock {\em J. Crystal Growth}, 184/185:325--329, 1998.
\newblock \texttt{cond-mat/9805079}.
\item
P.~J. Klar, D.~Wolverson, D.~E. Ashenford, B.~Lunn, and Torsten Henning.
\newblock Comparison of {Zn}$_{1-x}${Mn}$_x${Te/ZnTe} multiple-quantum wells
  and quantum dots by below-bandgap photomodulated reflectivity.
\newblock {\em Semicond. Sci. and Tech.}, 11:1863--1872, 1996.
\newblock \texttt{cond-mat/9710008}.
\item
Peter~J. Klar, Daniel Wolverson, J.~John Davies, Bernard Lunn, Duncan~E.
  Ashenford, and Torsten Henning.
\newblock Spin-flip {Raman} scattering in quantum dots based on
  {Cd}$_{1-x}${Mn}$_x${Te/CdTe} quantum well structures.
\newblock In Matthias Scheffler and Roland Zimmermann, editors, {\em Proc. 23rd
  Int. Conf. on the Phys. of Semicond., Berlin, 21--26~July 1996}, volume~2,
  pages 1485--1488. World Scientific, June 1996.
\newblock ISBN 981-02-2777-9 (set), 981-02-2946-1 (vol. 2).
\end{itemize}

\end{appendix}
\end{document}